\documentclass[12pt,a4paper,reqno]{article}
\usepackage{amsmath,amsthm}

\numberwithin{equation}{section}

\def\Bg {{\cal B}} 
\def\Eg {{\cal E}} 
\def\Fg {{\cal F}}
\def\Hg {{\cal H}} 
\def\Mg {{\cal M}} 
\def\Ng {{\cal N}} 
\def\Sg {{\cal S}} 
\def\Xg {{\cal X}} 
\def\Yg {{\cal Y}} 
\def\Zg {{\cal Z}} 

\def\R{{\bf R}}

\def\C{{\bf C}}

\def\Qb{{\overline{Q}}}

\def\Cb{{\overline{\C}}}

\def\gb{{\overline{g}}}
\def\Bb{{\bf B}}

\def\a{\alpha}
\def\b{\beta}
\def\c{\gamma}
\def\Ga{\Gamma}
\def\d{\delta}

\def\p{\psi}
\def\psib{\overline{\psi}}
\def\f{\phi}
\def\ep{\varepsilon}

\def\th{\theta}

\def\k{\kappa}
\def\l{\lambda}

\def\m{\mu}
\def\n{\nu}
\def\r{\rho}
\def\s{\sigma}
\def\Si{\Sigma}
\def\t{\tau}
\def\w{\omega}
\def\W{\Omega}

\def\lbeq(#1){\label{eqn:#1}}
\def\refeq(#1){{\rm (\ref{eqn:#1})}}
\def\lbth(#1){\label{th:#1}}
\def\refth(#1){{\rm Theorem \ref{th:#1}}}
\def\lbprp(#1){\label{prp:#1}}
\def\refprp(#1){{\rm Proposition \ref{prp:#1}}}
\def\lblm(#1){\label{lm:#1}}
\def\reflm(#1){{\rm Lemma \ref{lm:#1}}}
\def\lbcor(#1){\label{cor:#1}}
\def\refcor(#1){{\rm Corollary \ref{cor:#1}}}
\def\lbrm(#1){\label{rm:#1}}
\def\refrm(#1){{\rm Remark \ref{rm:#1}}}
\def\lbass(#1){\label{ass:#1}}
\def\refass(#1){{\rm Assumption \ref{ass:#1}}}
\def\lbdf(#1){\label{df:#1}}
\def\refdf(#1){{\rm Definition \ref{df:#1}}}
\def\lbsec(#1){\label{s:#1}}
\def\refsec(#1){{\rm \S\ref{s:#1}}}
\def\lbsubsec(#1){\label{ss:#1}}
\def\refsubsec(#1){{\rm \S\ref{ss:#1}}}

\def\ben{\begin{enumerate}}
\def\een{\end{enumerate}}
\def\ep{\varepsilon}
\def\ph{\varphi}

\def\la{\langle}
\def\ra{\rangle}
\def\ds{\displaystyle}

\def\ax{{\la x \ra}}
\def\ay{{\la y \ra}}
\def\az{{\la z \ra}}

\def\br{\begin{array}}
\def\er{\end{array}}
\def\lap{\Delta}
\newcommand {\pa}{\partial}

\newtheorem{theorem}{Theorem}[section]
\newtheorem{lemma}[theorem]{Lemma}
\newtheorem{proposition}[theorem]{Proposition}
\newtheorem{definition}[theorem]{Definition}
\newtheorem{corollary}[theorem]{Corollary}

\theoremstyle{definition}
\newtheorem{remark}[theorem]{Remark}

\def\tL={{\tilde{L}}}

\newcommand{\bea}{\begin{eqnarray}}
\newcommand{\eea}{\end{eqnarray}}
\newcommand{\beq}{\begin{equation}}
\newcommand{\eeq}{\end{equation}}
\newcommand{\bdm}{\begin{displaymath}}
\newcommand{\edm}{\end{displaymath}}

\newcommand{\lf}{\left}
\newcommand{\ri}{\right}
\newcommand{\half}{\frac12}

\title{The $L^p$ boundedness of wave operators for 
Schr\"odinger operators with threshold singularities II. 
Even dimensional case}
\author{Domenico Finco\footnote{Department of Mathematics, Gakushuin University, 
1-5-1 Mejiro, Toshima-ku, Tokyo 171-8588, Japan. Present address: Institute f\"ur 
Angewandte Mathematik} \ and 
Kenji Yajima\thanks{
Department of Mathematics, Gakushuin University, 
1-5-1 Mejiro, Toshima-ku, Tokyo 171-8588, Japan.
Supported by JSPS grant in aid for scientific research 
No. 14340039 }}
\date{}
\begin{document}
\allowdisplaybreaks
\maketitle

\section{Introduction}

Let $H=-\lap + V(x)$ be a Schr\"odinger operator on $\R^m$, $m\geq 1$, 
with real potential $V(x)$ such that $|V(x)|\leq C \ax^{-\d}$, $\ax=(1+|x|^2)^{\frac12}$, 
for some $\d>2$. Then, $H$ with domain $D(H)=H^2(\R^m)$, the Sobolev space of order 2, 
is selfadjoint in the Hilbert space $\Hg=L^2(\R^m)$ and $C_0^\infty(\R^m)$ 
is a core. The spectrum $\s(H)$ of $H$ consists of absolutely continuous 
part $[0,\infty)$ and a finite number of non-positive eigenvalues $\{\l_j\}$ of finite 
multiplicities. The singular continuous spectrum and positive eigenvalues are absent 
from $H$. 
We denote the point, the continuous and the absolutely continuous subspaces for $H$ by 
$\Hg_{\rm p}$, $\Hg_{\rm c}$ and $\Hg_{\rm ac}$ respectively, and the orthogonal projections onto the 
respective subspaces by $P_{\rm p}$, $P_{\rm c}$ and $P_{\rm ac}$. We have $\Hg_{\rm ac}=\Hg_{\rm c}$ 
and $P_{\rm ac}=P_{\rm c}$; $H_0=-\lap$ is the free 
Schr\"odinger operator. 

The wave operators $W_\pm$ are defined by the following 
strong limits 
\[
W_\pm = \lim_{t \to \pm \infty} e^{itH}e^{-itH_0} 
\]
in $\Hg\equiv L^2(\R^m)$. 
It is well known that the limits exist and are complete in the sense that 
${\rm Image} \ W_\pm = \Hg_{\rm ac}$. The wave operators satisfy the so called 
intertwining property and the continuous part of $H$ is unitarily equivalent to 
$H_0$ via $W_\pm$: For any Borel functions on $\R$, we have 
\begin{equation}\lbeq(equiv)
f(H)P_{\rm ac}(H)= W_\pm f(H_0) W_\pm^\ast.  
\end{equation}
It follows that the mapping properties of $f(H)P_{\rm ac}(H)$ may be deduced from those of 
$f(H_0)$ once corresponding properties of $W_\pm$ are known. In this paper we shall prove the following 
theorem. We say that $H$ is of exceptional type if there exist no non-trivial 
solutions of $-\lap \f + V(x) \f =0$ which satisfies $|\f(x)|\leq C \ax^{2-m}$; 
$H$ is of generic type otherwise (see \refdf(gene) for an equivalent definition). 
We write $\Fg$ for the Fourier transform. Throughout this paper, 
we assume that $V$ satisfies the following condition: 
\begin{equation}\lbeq(cond-1)
\Fg (\ax^{2\s} V) \in L^{m_\ast}  \quad \mbox{for}\  \s>\frac1{m_\ast}\equiv \frac{m-2}{m-1}. 
\end{equation}
For integers $k\geq 0$, $W^{k,p}(\R^m)$ is the Sobolev space of order $k$. 
 
\begin{theorem}\lbth(even) Let $m\geq 6$ be even and $V$ satisfy \refeq(cond-1).  

\noindent  
{\rm (1)} Suppose, in addition, that $|V(x)|\leq C\ax^{-(m+2+\ep)}$ 
for some $C>0$ and $\ep>0$ and that $H$ is of generic type. 
Then, for all $1\leq p \leq \infty$, $W_\pm$ extend to bounded operators in $L^p(\R^m)$:  
\begin{equation}\lbeq(bounded) 
\| W_\pm u \|_{L^{p}} \leq C_p \|u\|_{L^{p}}, \quad \ u \in L^{p}(R^m) \cap L^{2}(\R^m). 
\end{equation} 
For $1<p<\infty$,  $W_\pm$ actually are bounded in $W^{k,p}(\R^m)$ 
for $0 \leq k \leq 2$. If derivatives $\pa^\a V(x)$ are bounded for $|\a|\leq \ell$ in addition, 
then  $W_\pm$ are bounded in $W^{k,p}(\R^m)$ for all $0 \leq k \leq \ell +2$ and $1<p<\infty$. 
For $p=1, \infty$ the same holds if $\pa^\a V(x)$,  $|\a|\leq \ell$, satisfy \refeq(cond-1) and 
$|\pa^\a V(x)|\leq C \ax^{-(m+2+\ep)}$ for some $C>0$ and $\ep>0$. 

\noindent 
{\rm (2)} Suppose, in addition, that 
$|V(x)|\leq C\ax^{-(m+4+\ep)}$ if $m=6$,  and $|V(x)|\leq C\ax^{-(m+3+\ep)}$ if $m\geq 8$ 
for some $C>0$ and $\ep>0$, and that $H$ is of exceptional type. Then, for $m/(m-2)<p<m/2$ 
and $0 \leq k \leq 2$, 
$W_\pm$ extend to bounded operators in $W^{k,p}(\R^m)$:
\begin{equation}\lbeq(bounded-1) 
\| W_\pm u \|_{W^{k,p}} \leq C_p \|u\|_{W^{k, p}}, \quad \ u \in W^{k,p}(R^m) \cap L^{2}(\R^m). 
\end{equation} 
If $\pa_x^\a V(x)$ are bounded for $|\a| \leq \ell$ in addition, then \refeq(bounded-1) holds 
for $0 \leq k \leq \ell+2$ and $W_\pm$ are bounded in $W^{k,p}(\R^m)$ for all $0 \leq k \leq \ell +2$.   
\end{theorem}

Some remaks are in order. 
\begin{remark} Some condition like \refeq(cond-1) is necessary for the theorem by virtue of 
the counter example due to \cite{GV} to the dispersive estimates for the corresponding time 
dependent 
Schr\"odinger equation, see below. 
\end{remark} 
\begin{remark} When $m \geq 3$ is odd, it is proved in a recent paper \cite{Y-odd} 
that $W_\pm$ are bounded in $L^p(\R^m)$ for all $1\leq p \leq \infty$ 
if $V$ satisfies \refeq(cond-1) and $|V(x)|\leq C\ax^{-(m+2+\ep)}$ and $H$ is of generic type; 
for $p$ between $\frac{m}{m-2}$ and $\frac{m}2$ if $V$ satisfies \refeq(cond-1) and  
$|V(x)|\leq C\ax^{-(m+3+\ep)}$ and $H$ is of exceptional type. The argument in Section 7 below implies 
that this result extends to the continuity of $W_\pm$ in $ W^{k,p}(R^m)$ as in \refth(even). 
The paper \cite{Y-odd} will be referred to as [I] in what follows. In \cite{Cu} an extension 
for some non-selfadjoint cases in $m=3$ and its application to nonlinear equations is presented. 
When $m=1$, it is 
recently shown (\cite{DF}) that $W_\pm$ are bounded in $L^p$ for $1<p<\infty$ 
(but not for $p=1$ or $p=\infty$) if $\int_{\R} \ax |V(x)| dx<\infty$ 
and $H$ is of generic type, or if $\int_{\R} \ax^2 |V(x)| dx<\infty$ and $H$ is of 
exceptional type (see \cite{Weder}, \cite{AY} for earlier results). 
\end{remark}
\begin{remark}
When $m \geq 4$ is even, it is long known (\cite{Y-d4}) that \refeq(bounded) is satisfied 
for all $1\leq p \leq \infty$ if $V$ satisfies 
\begin{equation}\lbeq(old)
\sum_{|\a|\leq k+ (m-2)/2}
\left(\int_{|x-y|\leq 1} |\pa_y^\a V(y)|^{p_0}dy \right)^{\frac1{p_0}} \leq C \ax^{-\left(\frac{3m}2+1+\ep\right)} 
\end{equation} 
for some $p_0 >\frac{m}2$ and $\ep>0$ and if {\it  $H$ is of generic type}. If $m \geq 6$, 
condition \refeq(old) implies that $\pa^\a V$, $|\a|\leq k$, satisfy both \refeq(cond-1) and 
$|\pa_x^\a V(x)|\leq C\ax^{-(m+2+\ep)}$ and \refth(even) (1) improves the result of \cite{Y-d4} 
for $m \geq 6$. When $m=2$, it is known 
(\cite{Y-d2}, \cite{JY}) that $W_\pm$ is bounded in $L^p(\R^2)$ for $1<p<\infty$ if $V$ satisfies 
$|V(x)|\leq C \ax^{-6-\ep}$ and if $H$ is of generic type.  
\end{remark} 
\begin{remark} If $m \geq 4$ and if $H$ is of exceptional type, $W_\pm$ is not bounded in $L^p(\R^m)$ 
when $1\leq p <\frac{m}{m-2}$ because this would contradict Murata's result (\cite{Mu}) 
on the decay in time in weighted $L^2$ spaces of solutions $e^{-itH}u$ of the 
corresponding time dependent Schr\"odinger equation. 
We strongly believe the same is true for $\frac{m}2<p \leq \infty$ 
though the proof is missing. Notice that when $m=4$, $\frac{m}{m-2}=\frac{m}2=2$. 
\end{remark} 
\begin{remark} By interpolating \refeq(bounded) for different $k's$ by the real interpolation method 
(\cite{BL}), estimates of \refth(even) can be extended to the ones between Besov spaces. 
\end{remark}  

When $f(\l)= e^{-it\l}$, \refeq(equiv) and \refeq(bounded) implies the so called 
$L^p$-$L^q$ estimates for the propagator of the corresponding time dependent Schr\"odinger equation  
\begin{equation}\lbeq(pq)  
\|e^{-itH}P_c u \|_p \leq C |t|^{-m\left(\frac12-\frac1{p}\right) }\|u\|_q 
\end{equation}
where $p,q$ are dual exponents, viz. $1/p+1/q=1$, and $2 \leq p \leq \infty$ if $H$ is of generic type 
and $ 2 \leq p <m/2 $ if $H$ is of generic case. When $1 \leq m \leq 3$ and if $H$ is of generic type, 
the $L^p$-$L^q$ estimate has been proven for $2 \leq p \leq \infty$ 
for much wider class of potentials by more direct methods (\cite{GS}, \cite{S}, \cite{G}); 
when $m=3$ and $H$ is of exceptional type it is proved that \refeq(pq) holds 
for $2 \leq p <3$ and 
\begin{equation}\lbeq(pqe) 
\|e^{-itH}P_c u \|_{L^{3,\infty}} \leq C_p t^{-\frac12}\|u\|_{L^{\frac32, 1}} 
\end{equation} 
replaces \refeq(pq) at the end point where $L^{p,q}$ are Lorentz spaces (\cite{ES}, \cite{Y-s}). 
However, when $m \geq 4$, the result obtained by using wave operators via \refth(even) (1) 
or \cite{Y-d4} gives the best estimates so far as far as the decay and smoothness 
assumption on the potentials is concerned. We should also emphasize that the $L^p$-$L^q$ estimate 
\refeq(pq) is proven for the first time when $m\geq 6$ and $H$ is exceptional type. 

The intertwing property and the boundedness results, \refeq(equiv) and \refeq(bounded), 
may be applied for various other functions $f(H)P_c$ and can provide useful estimates.  
We refer the readers to [I] as well as \cite{Y-d3} and \cite{Y-d4} 
for some more applications, and we shall be devoted to the proof of \refth(even) in the 
rest of the paper. 

We prove \refth(even) only for $W_-$, which we denote by $W$ for brevity. We shall mainly 
discuss the $L^p$ boundedness, as the extension to Sobolev spaces is immediate as will be shown 
in Section 7. We write $R(z)= (H-z)^{-1}$ and $R_0(z) = (H_0 -z)^{-1}$ 
for resolvents. 
We parametrize $z\in  \C \setminus [0,\infty)$ by $z=\l^2$ by $\l \in \C^+=\{z \in \C: \Im z >0 \}$ 
and define $G(\l) = R(\l^2)$ and $G_0(\l) = R_0(\l^2)$ for $\l \in \C^+$. 
They are $\Bb(\Hg)$-valued meromorphic functions of $\l \in \C^+$ and 
the limiting absorption principle (LAP for short) asserts that $G(\l)$ and $G_0(\l)$ 
when considered as $\Bb(\Hg_{\s}, \Hg_{-\s})$-valued functions have continuous extentions 
to $\Cb^+=\{z: \Im z \geq 0\}$, the closure of $\C^+$, where  $\Hg_{\c}= L^2(\R^m, \ax^{2\c}dx)$ 
is the weighted $L^2$ space and $\s>\frac12$. Our theory is based on the stationary 
representation of wave operators which expresses $W$ via boundary values 
of the resolvents (cf. \cite{KaS1}, \cite{Ku-0}):   
\begin{equation}\lbeq(stationary)
Wu= u -\frac{1}{\pi i}\int^\infty_0 G(\l)V(G_0(\l)-G_0(-\l))u \l d\l .
\end{equation} 
As in odd dimensional cases, we decompose $W$ into the high and the low energy parts 
$W=W_> + W_< \equiv W\Psi(H_0)^2 + W \Phi(H_0)^2$,  by using 
cut off functions $\Phi(\l)$ and $\Psi(\l)$ such that 
$\Phi(\l)^2 + \Psi(\l)^2 \equiv 1$, $\Phi(\l)=1$ near $\l=0$ and $\Phi(\l)=0$ for $|\l|>\l_0^2$ 
for a small constant $\l_0>0$ to be specified below. The operators $\Phi(H)$ and $\Phi(H_0)$ 
have continuous integral kernels bounded by $C_N\la x-y \ra^{-N}$ for any $N$ and they are 
are bounded in $L^p(\R^m)$ for any $1 \leq p \leq \infty $. 
By virtue of the intertwining property we have $W_> =\Psi(H)W\Psi(H_0)$ and 
$W_<=\Phi(H) W \Phi(H_0)$ and, combining this with \refeq(stationary)  
\begin{align} 
W_< = \Phi(H)\Phi(H_0) -\int^\infty_0 \Phi(H)G(\l)V(G_0(\l)-G_0(-\l))\Phi(H_0)\l \frac{d\l}{\pi i}, \lbeq(low)\\
W_> = \Psi(H)\Psi(H_0) -\int^\infty_0 \Psi(H)G(\l)V(G_0(\l)-G_0(-\l))\Psi(H_0) \l \frac{d\l}{\pi i}. \lbeq(high) 
\end{align}
We study the operators defined by the integrals 
in \refeq(low) and \refeq(high) separately. We use the following terminology. 
\begin{definition} We say that the integral kernel $K(x,y)$ is admissible if 
\begin{equation}\lbeq(schur)
\sup_{x} \int_{\R^m} |K(x,y)|dy + \sup_{y} \int_{\R^m} |K(x,y)|dx <\infty. 
\end{equation}
\end{definition} 
It is well known that integral operators with admissible integral kernels are bounded 
in $L^p$ for any $1\leq p \leq \infty$. 
\begin{definition}\lbdf(cond) The operator valued function $K(\l)$ 
of $\l \in (-\l_0, \l_0)$ which acts on functions on $\R^m$ is said to satisfy property   
$(K)_\r$, $\r>0$, if it satisfies the following two conditions: 
\ben 
\item[{\rm (1)}] For $0\leq \c \leq \frac{m-2}2$, $\l \mapsto \ax^{\r-\c}K(\l)\ax^{\r-\c} \in \Bb(\Hg)$ 
is of class $C^\c$.  
\item[{\rm (2)}] For $\frac{m}{2} \leq \c \leq \frac{m+2}2$, it is of class $C^{\c}$ 
for $\l \not=0$ and, for some $C>0$ and $N>0$, 
\begin{align}
\|\ax^{\r-\frac{m}2}K^{(\frac{m}2)}(\l)\ax^{\r-\frac{m}2}\|_{\Bb(\Hg)}\leq C \la\log \l\ra^N,  \lbeq(cond) \\
\|\ax^{\r-\frac{m+2}2}  K^{(\frac{m+2}2)}(\l)\ax^{\r-\frac{m+2}2}\|_{\Bb(\Hg)}\leq C|\l|^{-1}\la\log \l\ra^N.   
\lbeq(cond-a)
\end{align}
\een
\end{definition}

The plan of the paper is as follows. Section 2 is preparatory in nature. 
In subsection 2.1, we collect some 
results on the behavior of the free resolvent $G_0(\l)$ on the reals, near the threshold 
$\l=0$ in particular.  When $m$ is even, $G_0(\l)$ is convolution operator with 
the kernel 
\begin{equation}\lbeq(co-ker) 
G_0(\l,x) = \frac{C_m e^{i\l|x|}}{|x|^{m-2}}
\int^\infty_0 e^{-t}t^{\frac{m-3}2}\left(\frac{t}2-i\l|x|\right)^{\frac{m-3}2}dt  
\end{equation}
where $C_m={\ds\frac{ie^{-i(2\n+1)\pi/4}}{2(2\pi)^{\n +1}\Ga(\n+\frac12)}}$ and $G_0(\l)$ contains 
a term whose $(m-2)$-nd derivative becomes logarithmically singular at $\l=0$.  
We study mapping properties of the derivatives of such operators in detail (see \refprp(even-exp)). 
In subsection 2.2, we recall from \cite{Y-d3} the result  on the $L^p$ boundedness of Born approximations 
of wave operators. Using these results, we study 
the behavior of $(1+ G_0(\l)V)^{-1}$ in Section 3 and show that $V(1+ G_0(\l)V)^{-1}$ 
satisfies the property $(K)_\r$ for any $\r<\d-1$ if $H$ is of generic type; and 
when $H$ is of exceptional type that 
$(1+ G_0(\l)V)^{-1}$ has the expansion near $\l=0$ of the following form  
\begin{equation}\lbeq(expansion) 
(1+ G_0(\l)V)^{-1} = \frac{P_0 V}{\l^2} + \sum_{j=0}^2 \sum_{k=1}^2 \l^j \log^k \l D_{jk} + I + R_r (\l) ,
\end{equation} 
where $P_0$ is the orthogonal projection onto the $0$ eigenspace of $H$, 
$VD_{jk}$ are finite rank operators from $\Hg_{-(\d-3-\ep)}$ to $\Hg_{\d-3-\ep}$ for any $\ep>0$, 
and $VR_r(\l)$ satisfies the condition $(K)_\r$ for any $\r<\d-3$ if $m=6$ and $\r<\d-2$ if $m \geq 8$. 
When the dimension $m$ becomes 
the larger the formula \refeq(expansion) for $(1+ G_0(\l)V)^{-1}$ becomes the less complex 
thanks to the fact that $\l^{m-2}\log \l$ becomes less singular as $m$ increases.

We show in Section 4 and Section 5 that the low energy part $W_<$ is bounded in $L^p$ 
for all $1\leq p \leq \infty$ when $H$ is of generic type, and for $\frac{m-2}2 <p <\frac{m}2$ 
when $H$ is of exceptional type, respectively. 
For proving this we substitute $G_0(\l)V(1+ G_0(\l)V)^{-1}$ for $G(\l)V$ on the right 
of \refeq(low). In view of the fact that $V(1+ G_0(\l)V)^{-1}$ in generic case 
and $VR_r(\l)$ in exceptional case satisfy property $(K)_\r$ for some $\r>m+1$, we show in 
Section 4 that, if $K(\l)$ satisfies this property, then 
the operator $\W$ defined by 
\begin{equation}\lbeq(ome)
\W = \int^\infty_0 \Phi(H)G_0(\l) K(\l) (G_0(\l)-G_0(-\l))\Phi(H_0)\l \tilde \Phi(\l) d\l 
\end{equation} 
with an additional cut-off function 
$\tilde \Phi \in C_0^\infty(\R)$ such that $\tilde \Phi(\l) \Phi(\l) = \Phi(\l)$ is an integral operator with an admissible integral kernel (see \refprp(2)).  The basic idea of 
the proof is similar to the one used for similar purpose in odd dimensions, however, because of the 
more complex structure of $G_0(\l,x)$, the argument becomes a bit more involved and a rather 
unexpected cancellation between $G_0(\l)$ and $G_0(-\l)$ plays a crucial role. 

For studying $W_<$ when $H$ is of exceptional type, we substitue 
\refeq(expansion) for $(1+ G_0(\l)V)^{-1}$. Then, the identity $I$ produces the first 
Born approximation, which is 
bounded in $L^p$ for all $1\leq p \leq \infty$ (see \reflm(wr1)); \refprp(2) implies 
that $R_r(\l)$ produces an integral operator with admissible integral kernel; 
and we study operators produced by the singular terms 
$\l^{-2} P_0 V  + \sum_{j=0}^2 \sum_{k=1}^2 \l^j \log^k \l D_{jk}$ 
in Section 5. We shall deal with the one produced by 
the most singular term $\l^{-2}P_0 V$ in Subsection 5.1. 
Here again the basic idea is similar to the odd dimensional case: 
If $P_0 = \sum \f_j \otimes \f_j$, the operator under investigation  
$\int^\infty_0 G_0(\l) VP_0 V (G_0(\l)-G_0(-\l))\l^{-1} \Phi(\l) d\l$ 
is a linear combination of 
\[
Z_j u(x) = \int_{\R^m} \frac{(V\f_j)(x) F_ju (|x-y|)}{|x-y|^{m-2}}dy 
\]
where, with sperical average $M_ju(r)= |\Si|^{-1} \int_{\Si} (V\f_j \ast \check u)(r\w) d\w$,  
$\check u(x) = u(-x)$ and $\Si= S^{m-1}$ being the unit sphere of $\R^m$, $F_ju(\r)$ is given by 
\begin{multline*} 
F_ju(\r) = \int_0^\infty  \int_0^\infty  e^{-(t+s)}(ts)^{\frac{m-3}2} dt ds \\
\times \left\{\int^\infty_0 e^{-i\l\r} (s+2i\l\r)^{\frac{m-3}2} 
\left(\int_{\R} e^{i\l r}(t-2i\l r)^{\frac{m-3}2} r M_ju(r)dr\right) d\l \right\}. 
\end{multline*}
Observing that $F_ju(\r)$ and $M_ju(r)$ are one dimensional objects, 
we apply some one dimensional harmonic analysis machinaries, the weighted inequality for 
the Hilbert transform $\tilde \Hg$ and the Hardy-Littlewood maximal operator $\Mg$. 
However, as the comparison of formulae above with those in the odd dimensional case suggests, 
the analysis in even dimensions becomes much more intricate.  In Subsection 5.2 we shall 
indicate how to modify the argument in subsection 5.1 for dealing with the operators 
produced by $\l^j \log^k \l D_{jk}$.  

In Section 6, we prove that the high energy part $W_>$ is bounded in $L^p$ 
for any $1\leq p \leq \infty$. As the high energy part is insensitive to the low energy 
singularities and as the argument used for the same purpose in \cite{Y-odd} for odd dimensions 
applies, 
we shall only very briefly sketch the proof. In Section 7, we show the continuity of $W$ 
in Sobolev spaces and complete the proof of \refth(even). For $1<p<\infty$, this follows 
from the intertwining property 
$ W= (H-z)^{-j} W (H_0-z)^{j}$ and the well known mapping property 
of the resolvent. For $p=1$ and $p=\infty$, we may adopt the commutator argument 
as in \cite{Y-d3} and we omit the discussion.

We use the same notation and conventions as in [I]: 
For $u \in \Hg_{-\c}$ and $v\in \Hg_\c$ 
$\la u, v \ra = \int_{\R^n} \overline{u(x)}v(x) dx$ 
is the standard coupling of functions; $|u\ra \la v| = u \otimes v$ 
will be interchangeably used to denote the rank 1 operator 
$\f \mapsto \la v, \f \ra u $. For Banach spaces $X$ and $Y$, 
$\Bb(X,Y)$ (resp. $\Bb_\infty(X,Y)$) is the Banach space of bounded 
(resp. compact) operators from $X$ to $Y$, 
$\Bb(X)=\Bb(X,X)$ (resp. $\Bb_\infty(X)=\Bb_\infty(X,X)$).   
The identity operator is denoted by $1$. 
The norm of $L^p$-spaces, $1\leq p \leq \infty$, is denoted by $\|u\|_p=\|u\|_{L^p}$. 
We write $\Sg(\R^m)$ for the space of rapidly decreasing functions. 
The Fourier transform is defined by 
\[
\hat u(\xi)= \Fg u(\xi)= \frac{1}{(2\pi)^{m/2}}\int_{\R^m} e^{-ix\xi}u(x) dx
\]
and $\Fg^\ast u(\xi)=\Fg u(-\xi)$ is the conjugate Fourier transform. 
For functions $f$ on the line $f^{(j)}$ is the $j$-th derivative of $f$, $j=1,2, \ldots$. 
For $a\in \R$, $a_+$ or $a_-$ is an arbitary number larger or smaller than $a$ 
respectively; $[a]$ is the largest integer not larger than $a$.  
When $I$ is an open or closed interval which contains $0$, 
$C_{0\ast}^s(I)$ is the set of functions of order $C^s$ on $I$ which vanishes at 
$\l=0$ along with the derivatives upto the order $[s]$. We sometimes say that 
$u$ is of order $C_{0\ast}^s$ when $u \in C_{0\ast}^s(I)$.

\section{Preliminaries} 

\subsection{Free resolvent} 
Recall that $\Hg_\c = L^2_{\c}(\R^m, \ax^{2\c}dx)$ and $\Si$ is the  unit sphere of 
$\R^m$. As is well known the Fourier transform $\Fg$ is an isomorphism from $\Hg_\c$ 
to the Sobolev space $H^\c(\R^m)$ and the mapping 
\begin{equation}\lbeq(sob)
\tilde \Gamma \colon  H^{\gamma} (\R^m)\ni u \mapsto \l^{\frac{m-1}{2}}  u(\l \cdot ) \in 
H^{\gamma}_0((0,\infty), L^2(\Si))
\end{equation}
is bounded if $0 \leq \c <\frac{m}2$. The upper bound 
for $\c$, however, is relevant only at $\l=0$ and, for any $\ep>0$,  the map \refeq(sob) 
is bounded for any $0\leq \c$ 
if $H^{\c}_0((0,\infty), L^2(\Si))$ is replaces by $H^{\c}((\ep,\infty), L^2(\Si))$. 
It follows by the Sobolev embedding theorem that the $\Bb(\Hg_\c, L^2(\Si))$-valued function  defined by 
\[
\Ga(\l) \colon  \Hg_\c \ni u \mapsto  \l^{(m-1)/2}\hat u(\l \cdot ) \in L^2(\Si), 
\]
is of class $C^{\c-\frac12}$ over $[0,\infty)$ and vanishes at $\l=0$ along with the derivatives 
up to the order $[(\c-\frac12)_-]$ if $\frac12<\c<\frac{m}2$; and it is of class $C^{\c-\frac12}$ 
over $(\ep,\infty)$ for any $\frac12<\c$ and $\ep>0$. 
We shall use the following well known lemma on the division in Sobolev spaces. 
The lemma is a result of repeated application of Hardy's inequality when $s$ is an integer 
and from the complex interpolation theory when $s$ is not an integer. 
\begin{lemma}\lblm(division)  For any $s>0$, the operator $f(x) \mapsto x^{-s}f(x)$ is bounded from 
$H^{\c}_0(\R^+, L^2(\Si))$ to $H_{0}^{\c-s}(\R^+, L^2(\Si))$. 
\end{lemma} 
We define operator valued function $A(\l)$ for $\l \in \R$ by 
\begin{equation} 
A(\l)u(x) = \frac1{(2\pi)^m}
\int_{\Si} \int_{\R^m} e^{i\l\w (x-y)}u(y)dy d\w, \quad x \in \R^m. \lbeq(Adef)
\end{equation}
It is clear that $A(\l)$ is even with respect to $\l \in \R$ and 
$\m^{m-1}A(\m)= \Ga(\m)^\ast \Ga(\m)$.  
We shall use the following expressions for $G_0(\l)$, $\l \in \C^+$. 
\begin{eqnarray}
G_0(\l) = \frac1{2\l}\left( \int^\infty_0 \frac{\Ga(\m)^\ast \Ga(\m) }{\m-\l} d\m 
-\int^\infty_0 \frac{\Ga(\m)^\ast \Ga(\m) }{\m + \l} d\m \right) \hspace{0.2cm} \lbeq(e0-a) \\
= \int_0^\infty\frac{\m^{m-1}A(\m)}{\m^2-\l^2}d\m= 
\frac12 \int^\infty_{-\infty} \frac{\m^{m-2}{\rm sign  \m} \ A(\m)}{\m-\l} d\m . \lbeq(e0-c)
\end{eqnarray}
It can be see from the last expression,  $G_0(\l)$ become logarithmically singular 
when it is differentiated by $\l$ more than $m-3$ times.  
The following lemmas are basic to the following analysis.  We let 
$D_1$, $D_2$ and $D_3$ be the closed domains of the first quadrant of $(k,\ell)$ plane 
defined by 
\[\br{c}
D_1=\{(k,\ell): k,\ell \geq 0, \, k+\ell\leq m-1, \ell\leq k\}, \\ 
D_2=\{(k,\ell): k,\ell \geq 0, \, k\leq \frac{m-1}2, \ell\geq k\}, \\ 
D_3=\{(k,\ell): k,\ell \geq 0, \, k+\ell\geq m-1, \frac{m-1}2 \leq k \leq m-1\}. \er 
\]
They have disjoint interiors and 
$D_1\cup D_2 \cup D_3 = \{(k,\ell):  0\leq k \leq m-1, 0\leq \ell\}$. 
Define the function $\s_0(k,\ell)$ for $0\leq k \leq m-1$ and $0\leq \ell$ by 
\begin{equation}\lbeq(kle) 
\s_0(k,\ell) = \left\{\br{ll} \frac{k+\ell+1}2, & \quad (k,\ell) \in D_1 \\
\ell+\frac12, & \quad (k,\ell) \in D_2, \\ k+\ell -\frac{m-2}2, & \quad (k,\ell) \in D_3 .
\er \right.
\end{equation}
The function $\s_0(k,\ell)$ is continuous, separately increasing with resepect to $k$ and $\ell$ 
and, on lines $k+\ell=c$ with fixed $c$,  decreases with $k$. 

\begin{lemma}\lblm(Ar-0) Let $\ell\geq 0$ be an integer and let $0\leq k \leq m-1$. 
Let $\s_0= \s_0(k,\ell)$ be as above and $\s>\s_0$. 
Then, $\l^{m-1-k} A^{(\ell)} (\l)$ is a $\Bb(\Hg_\s, \Hg_{-\s})$ valued function 
of $\l\in \R$ of class $C^{\s-\s_0}$. 
\end{lemma} 
\begin{proof} Define $\r(\l)u(\w)=\hat u(\l \w)$ for $\l \in \R$ and $\w\in \Si$ 
and write $\Ga_\l=\Ga(\l)$ and $\r_\l=\r(\l)$ for shortening formulae.  
We also write $\Xg_\s \equiv \Bb(\Hg_\s, L^2(\Si))$. 
We have $(A(\l)u, v) = \la \r_\l u, \r_\l v\ra $. By differentiation,  
\[
\r^{(k)}_\l u(\w)= \frac1{(2\pi)^{\frac{m}2}}\int_{\R^m}(i\w x)^k e^{i\l\w x}u(x)dx
= \sum_{|\a|=k} C_\a \w^\a \r_\l(x^\a u)(\w).
\]
It follows by Leibniz' rule that 
\[
(A^{(\ell)}(\l)u, v) = \sum_{|\a|+|\b|=\ell}C_{\a\b} \la \w^\a \r_\l (x^\a u), \w^\b \r_\l (x^\b u) \ra.
\]
In terms of $\Ga_\l$ we may write this in the form   
\[
\l^{m-1-k} (A^{(\ell)}(\l)u, v) = 
\l^{-k} \sum_{|\a|+|\b|=\ell}C_{\a\b} \la \w^\a \Ga_\l (x^\a u), \w^\b \Ga_\l (x^\b u) \ra 
\]
It is an elementary to check that $\s_0=\s_0(k,\ell)$ is equal to  
\[
\max_{|\a|+|\b|=\ell} 
\min\{\max(a+|\a|+\tfrac12, b+|\b|+\tfrac12): 0 \leq a, b \leq \tfrac{m-1}2, a+b=k\}.
\]
It follows that, if $\s>\s_0$ then for any $\a,\b$ such that $|\a|+|\b|=\ell$ we can find 
$0 \leq a,b\leq \frac{m-1}2$ such that 
\begin{equation}\lbeq(ab)
a+b=k, \quad a<\s-|\a|-\frac12, \ \mbox{and}\ \ b<\s-|\b|-\frac12. 
\end{equation}
For these $a, b$,   
$\l^{-a} \Ga_\l \ax^{|\a|}$ and $\l^{-b} \Ga_\l \ax^{|\b|}$ are 
$\Xg_\s $-valued continuous. Indeed, if $a=\frac{m-1}2$, then, 
$\l^{-a} \Ga_\l \ax^{|\a|}= \r_\l  \ax^{|\a|}$ 
is a $\Xg_\s$-valued function of class $C^{\s-|\a|-\frac{m}{2}}$ 
by virtue of Sobolev embedding theorem because $\s-|\a|>\frac{m}2$; 
if $a<\frac{m-1}2$, then, $\Ga_\l \ax^{|\a|}$ is 
$\Xg_\s$-valued function of class $C^{\c_-}$, $\c=\min(\frac{m}2, \s-|\a|-\frac12)$ 
on $\l \in [0,\infty)$ which vanishes at $\l=0$ along with the derivatives of order up to $[\c]$, 
and  $\l^{-a} \Ga_\l  \ax^{|\a|}$ is of class $C^{\c-a}$ as a $\Xg_\s$-valued function. 
A similar proof applies to $\l^{-b} \Ga_\l \ax^{|\b|}$.  This and the identity 
\[
\l^{-k} \la \w^\a \Ga_\l (x^\a u), \w^\b \Ga_\l (x^\b u) \ra 
= \la \w^\a \l^{-a} \Ga_\l (x^\a u), \w^\b \l^{-b} \Ga_\l (x^\b u) \ra 
\]
imply that $\l \mapsto \l^{m-1-k} A^{(\ell)} (\l) $ is $\Bb(\Hg_\s, \Hg_{-\s})$ valued continuous 
on $\R$. To conclude the proof, it suffices to show that it is of class $C^{(\s-\s_0)_-}$ when 
$0\leq \s-\s_0 \leq 1$. However, if $\s>\s_0+1$, a differentiation implies that 
$\ax^{-\s}\l^{m-1-k} A^{(\ell)} (\l) $ is of class $C^1$. The lemma then 
follows by interpolation. 
\end{proof}

\begin{corollary}\lbcor(Ar-c) Let $0\leq a\leq m-1$ and  $b\geq 0$. 
Let $j\geq 0$ and $\s >\s_0(a,b+j)$. 
Then, $\l^{m-1-a} A^{(b)} (\l)$ is a $\Bb(\Hg_\s, \Hg_{-\s})$-valued 
continuous function of $\l \in \R$ of class $C^{j+(\s-\s_0)_-}$.  
\end{corollary} 
\begin{proof} It suffices to show that 
$\l \mapsto \l^{m-1-a-a'}A^{(b+b')} (\l) \in \Bb(\Hg_\s, \Hg_{-\s})$ 
are continuous if $a'+b'\leq j$ and $a+a'\leq m-1$. This follows from \reflm(Ar-0) 
since, on the segment $\{(k,\ell): k+\ell=a+b+j', 0\leq a\leq k\leq m-1\}$, 
$\s_0$ attains its maximum at $(a,b+j')$ and $\s_0(a,b+j')$ increases with $0\leq j'$. 
\end{proof}

The following \reflm(Ar) is a slight improvement of \refcor(Ar-c) for small $\s$. We omit the 
proof as it is identical with that of Lemma 2.1 of [I] for the odd dimensional case. 

\begin{lemma}\lblm(Ar) Let $\frac12<\s,\t<\frac32$ be such that $\s+\t>2$ and define 
$\r_0=\t+\s-2$. Then, as a 
$B(\Hg_\s, \Hg_{-\t})$-valued function, $\l^{m-2}A(\l)$ is of class $C^\r$ 
for any $\r<\r_0$ in $\R$ and of class $C^{\min(\s-\frac12, \t-\frac12)}$ in $\R\setminus \{0\}$. 
\end{lemma}

\begin{lemma}\lblm(Ag) 
{\rm (1)} Let $1/2<\s$. Then, 
$G_0(\l)$ is a $\Bb_\infty (\Hg_{\s}, \Hg_{-\s})$ valued function of $\l \in \Cb^+\setminus\{0\}$ 
of class $C^{(\s-\frac12)_-}$. For non-negative integers $j<\s-\frac12$, 
\begin{equation}\lbeq(Ag)
\|G^{(j)}_0(\l)\|_{\Bb(\Hg_{\s}, \Hg_{-\s})} \leq C_{j\s} |\l|^{-1}, \quad |\l|\geq 1. 
\end{equation} 
{\rm (2)} Let $\frac12< \s, \t< m-\frac{3}{2}$ satisfy $\s+\t>2$. 
Then, $G_0(\l)$ is a $\Bb_\infty(\Hg_\s, \Hg_{-\t})$-valued function of $\l \in \Cb^+$ of 
class $C^{\r_{\ast-}}$, $\r_\ast=\min(\t+\s-2, \t-1/2, \s-1/2)$.  
\end{lemma} 
\begin{proof} The first statement is well known and follows immediately from \refeq(e0-a) 
and the property of $\Ga(\l)$ stated at the beginning of this subsection.  
By virtue of \refcor(Ar-c) and \reflm(Ar), ${\rm sign \m}\ \m^{m-2}A(\m)$ is a 
$\Bb_\infty(\Hg_\s, \Hg_{-\t})$-valued function of $\m \in \R$ since $\r_\ast<m-2$. 
We apply the Privaloff theorem to the last expression  of \refeq(e0-c). The second statement follows. \end{proof} 

The $(m-2)$-th derivative of ${\rm sign \m}\ \m^{m-2}A(\m)$ 
contains Heaviside type singularity at $\m=0$ and, for any large $\s$, 
$\l \mapsto \ax^{-\s}G_0^{(m-2)}(\l)\ax^{-\s}\in \Bb(\Hg)$ is not continuous at $\l=0$. 
We now examine this singularity. Let   
\[
J_k(\l)= \frac{1}{\l^k} \left(G_0(\l)-G_0(0)-\cdots - G_0^{(k-1)}(0)\frac{\l^{k-1}}{(k-1)!}\right). 
\]

\begin{proposition}\lbprp(even-exp)  Let $m \geq 4$ be even. Then:
\ben 
\item[{\rm (1)}]  Let $k=0,1, \ldots, m-3$ and $0<\r<m-2-k$. Let $\s >\s_0(k+1,\r)$. 
Then $J_k(\l)$ is a $\Bb(\Hg_{\s}, \Hg_{-\s})$-valued function of $\l \in \R$ of class $C^{\r}$.  
\item[{\rm (2)}] For $0<\pm \l<\frac18$, $G_0(\l)$ has the following expression:    
\[
\sum_{j=0}^{\frac{m-4}{2}}\l^{2j}(-\lap)^{-j-1} + \l^{m-2}\Big(\pm \frac{i\pi}{2}A(\l) -\log |\l|\, A(\l) \Big)
+ \l^{m-2} F(\l),
\]
where $\l \mapsto F(\l)$ is even and, for $k=0, \ldots, m-1$, $\l^{m-1-k}F(\l)$ 
satisfies the same smoothness property as $\l^{m-1-k}A(\l)$ as stated in \refcor(Ar-c) 
and \reflm(Ar). 
\een 
\end{proposition} 
\noindent 
Remark that proof of \reflm(Ar-0) implies that 
for $\l \mapsto \sum_{j=0}^{\frac{m-4}{2}}\l^{2j}(-\lap)^{-j-1}$ 
is a $\Bb(\Hg_{(\frac{m-1}2)_+},\Hg_{-(\frac{m-1}2)_+})$ valued polynominal and hence is analytic. 
\begin{proof} If $k=0$, statement (1) is contained in \reflm(Ag) (2). Let $k>0$. 
Substituting 
$\sum_{j=0}^{k-1} \l^j\m^{-j-1} + (\m-\l)^{-1}\l^k \m^{-k}$ for $(\m-\l)^{-1}$ in the second equation of 
\refeq(e0-c), we have for $\l \in \C^+$ that 
\[
G_0(\l) = \sum_{j=0}^{k-1} \frac{\l^j}{2} 
\int^\infty_{-\infty} \m^{m-j-3}{\rm sign} \m \ A(\m) d\m 
+ \l^k \int^\infty_{-\infty} 
\frac{\m^{m-1}{\rm sign}\m \ A(\m)u d\m}{2\m^{k+1}(\m-\l)}. 
\]
Since $A(\m)$ is even, the integrals in the sum vanish for odd $j$ and for even $j$ 
\[
\frac{1}{2} \int^\infty_{-\infty} \m^{m-j-3}{\rm sign} \m \ A(\m) d\m 
=\int^\infty_0 \frac{\m^{m-1}A(\m)}{\m^{j+2}} d\m = (-\lap)^{-(\frac{j+2}2)} \l^j.  
\]
Thus, we have for $\l \in \C^+$    
\begin{equation}\lbeq(jdef)
J_k(\l)= \frac12 \int^\infty_{-\infty} \frac{\m^{m-2-k}{\rm sign}\m \ A(\m)}{\m-\l}d\m . 
\end{equation} 
If  $\s >\s_0(k+1,\r)$ and $\r<m-2-k$, 
$\m \mapsto \m^{m-2-k}{\rm sign}\m \ A(\m)\in \Bb(\Hg_{\s}, \Hg_{-\s})$ 
is of class $C^\r$ by virtue of \refcor(Ar-c) and (1) follows by Privaloff's 
theorem.   
\noindent 
(2) We substitute  
$\m^{m-2}=(\m^2-\l^2)(\m^{m-4}+\l^2 \m^{m-6}+ \cdots + \l^{m-4})+\l^{m-2} $ 
in the first of \refeq(e0-c). The result is: 
\begin{eqnarray}
\ds G_0(\l)= \sum_{j=0}^{\frac{m-4}{2}} \l^{2j} (-\lap)^{-j-1}
\ds + \l^{m-2} \int^\infty_{0}  \frac{\m A(\m)}{\m^2-\l^2} d\m , \quad \l \in \C^+. \hspace{0.1cm} \lbeq(e-exp)
\end{eqnarray}
Take an even function $\chi \in C^\infty_0(\R)$ such that $\chi(\m)=1$ for $|\m |\leq 1/4$ 
and $\chi(\m)=0$ for $|\m|\geq 1/2$, and split the last integral: 
\begin{equation}\lbeq(e-int)
\int^\infty_{0}  \frac{\m A(\m)}{\m^2-\l^2} d\m
=\int^\infty_{0}  \frac{\m \chi(\m) A(\m)}{\m^2-\l^2} d\m
+ \int^\infty_{0}  \frac{\m (1-\chi(\m))A(\m)}{\m^2-\l^2} d\m
\end{equation}
The second integral yields a $\Bb(\Hg)$--valued analytic function of $\l$ in a neighbourhood of 
the interval $(-\frac18, \frac18)$ and we include it in $F(\l)$. 
We denote $B(\m)=\chi(\m)A(\mu)$, write the first integral in the form 
\begin{equation}\lbeq(e0)
\frac12 \left(\int^\infty_{0}  \frac{B(\m)}{\m+ \l} d\m 
+ \int^\infty_{0}  \frac{B(\m)}{\m-\l} d\m \right)
\end{equation}
and take the boundary values at $-\frac18< \l <\frac18$:  
\begin{equation}\lbeq(e0a)
\frac12 \left(\int^1_0  \frac{B(\m)}{\m+ |\l|} d\m 
\pm i\pi B(\l) + {\rm p.v.} \int_0^{\infty} \frac{B(\m)}{\m-|\l|} d\m  \right),  
\quad 0<\pm \l <\tfrac18. 
\end{equation} 
To fix the idea we let $0<\l<\frac18$. We split the domain of integral 
of the second integral $[0,\infty)=[0,2\l) \cup[2\l,\infty)$. The integral over $[2\l,\infty)$ 
is equal to $\int_\l^1 B(\m-\l)\m^{-1}d\m$, and we add it to the first 
integral which is equal to $\int_\l^1 B(\l+\m)\m^{-1}d\m$. We write the sum in the form 
\[
\left(\int_{0}^1 -\int_{0}^\l \right) \frac{B(\m+\l)+B(\m-\l)-2B(\l)}{2\m} d\m -(\log \l) A(\l) 
\]
and add this to 
\[
{\rm p.v.} \int_0^{2\l} \frac{B(\m)}{\m-\l} d\m= 
\int_0^{\l}\frac{(B(\l+\m)-B(\l-\m))}{2\m} d\m. 
\]
We arrive at the desired expression with 
\begin{eqnarray}
F(\l)= \int^\infty_{0}  \frac{\m (1-\chi(\m))A(\m)}{\m^2-\l^2} d\m
- \int_0^{\l}\frac{(B(\l-\m)-B(\l))}{\m} d\m  \notag \\
+ \int_{0}^1\frac{B(\m+\l)+B(\m-\l)-2B(\l)}{2\m} d\m .\hspace{1cm} \lbeq(ead)
\end{eqnarray}
It is immediate to check that $F(\l)=F(-\l)$. 
We prove that $\l^{m-1-k}F(\l)$ satisfies the desired smoothness 
properties on $(-\frac18,\frac18)$. The first integral defines a $\Bb(\Hg)$-valued analytic 
as remark above and we ignore it. Let $k=m-1$ first. 
We have $\s(m-1,j)=j+\frac{m}2$. Hence, if $\s>\frac{m}2$ and $0 \leq t<\s-\frac{m}2$, 
$B(\m)$ is a $\Bb(\Hg_\s, \Hg_{-\s})$valued function of class  $C^{t}$ and  
the last two integrals satisfy the same property. Indeed, if  $0<t<1$. Then 
\[
\left\|\int_\l^{\l+h}\frac{B(\l+h-\m)-B(\l+h)}{\m} d\m\right\| \leq \int_\l^{\l+h} \m^{t-1}d\m 
\leq C h^t. 
\]
Since $\|(B(\l+h-\m)-B(\l+h))-(B(\l-\m)-B(\l))\|\leq C\min(\m^t, h^t)$, 
\begin{eqnarray*}
\left\|\int_0^{\l}((B(\l+h-\m)-B(\l+h))-(B(\l-\m)-B(\l)))\frac{d\m}{\m}\right\| \\
\leq \int_0^{h} C\m^{t-1} d\m +  Ch^t \int_h^\l \frac{d\m}{\m} \leq C h^t (1+|\log h|)h^t.  
\end{eqnarray*}
This prove that the first integral on the right of \refeq(ead) is of class $C^{t_-}$. 
When $t \geq 1$, we differentiate it and apply similar estimates.  In this way we prove  
that it is of class $C^{(\s-\frac{m}2)_-}$. The proof for the second integral is simplier. 
We next let $0\leq k \leq m-2$. Write $m-1-k=t$, 
\begin{multline} \lbeq(2-19)
\l^{t}\int_{0}^{\l}\frac{B(\l-\m)-B(\l)}{\m} d\m 
= \int_{0}^{\l}\frac{(\l-\m)^{t}B(\l-\m)-\l^{t}B(\l)}{\m} d\m \\
+ \sum_{\ell=1}^{t} \left(\br{c} t \\ \ell \er \right) \int_{0}^{\l} \m^{\ell-1}(\l-\m)^{t-\ell}B(\l-\m)d\m
\end{multline}
The argument for the case $k=0$ implies that the first integral on the right 
satisfies the same smoothness property as $\l^{m-1-k} A(\l)$ as stated in \refcor(Ar-c). 
We write the second integral in the form 
\[
\int_{0}^{\l} \m^{t-\ell}(\l-\m)^{\ell-1}B(\m)d\m
= (\ell-1) ! \int_{0}^{\l} \cdots \int_0^{s_2} \int_0^{s_1} \m^{t-\ell}B(\m)d\m ds_1 \ldots ds_{\ell-1}  
\]
Since $\s(k,j) \geq \s(k-\ell,j-\ell)$ for any $k,\ell\geq 0$, the operator valued function defined by 
the right hand side 
enjoys the desired smoothness property. To prove the same for the last integral in \refeq(ead): 
\begin{equation}\lbeq(2-16)
\int_{0}^1\frac{B(\m+\l)-B(\l)}{2\m} d\m+ \int_{0}^1\frac{B(\m-\l)-B(\l)}{2\m} d\m, 
\end{equation}
we multiply it by $\l^{t}$, $t=m-1-k$ and write the resulting function as in \refeq(2-19). 
For the first integral, as previously, it suffices to show 
\begin{equation}\lbeq(2-17)
\int_{0}^{1} (\l+\m)^{t-\ell}B(\l+\m) \m^{\ell-1}d\m
= \int_{\l}^{\l+1} \m^{t-\ell}B(\m)(\m-\l)^{\ell-1}d\m
\end{equation}
satisfies the desired property. However, the derivative of the right side is 
\[
(\l+1)^{t-\ell}B(\l +1) + (\ell-1)\int_{\l}^{\l+1} \m^{t-\ell}B(\m)(\m-\l)^{\ell-2}d\m
\]
and the first term is as smooth as $\l^{m-1}A(\l)$ as $\frac78< 1+\l$ when $-\frac18<\l<\frac18$. 
Thus, by induction, \refeq(2-17) satisfies the desired property. The same holds for the second 
integral of \refeq(2-16). We omit the details.  
\end{proof}

\subsection{Born Terms}

If we formally expand the right of $G(\l)V = (1+ G_0 (\l)V)^{-1} G_0(\l)V$ into the 
series $\sum_{n=1}^\infty (-1)^{n-1} (G_0(\l)V)^n$ 
and substitute it for $G(\l)V $ in the stationary formula \refeq(stationary), then we have $W=1-\W_1+ \W_2 - \cdots$ where  
\[
\W_n u = \frac{1}{\pi i} \int_0^\infty (G_0(\l) V)^n (G_0(\l)-G_0(-\l)) u \l d\l,  
\ n=1,2, \ldots.  
\]
The sum $I - \W_1+ \cdots + (-1)^n \W_n$ is called the $n$-th Born approximation of $W_-$ 
and individual $\W_n$ is called the $n$-Born term. The following lemma is proved 
in any dimension $m \geq 3$ (\cite{Y-d3}) and it will be used for studying both the low 
and the high energy parts of $W$.   

\begin{lemma}\lblm(wr1) Let $\s>1/m_\ast$. Then there exists a constant $C>0$ such that 
\begin{eqnarray} 
\|\W_{1} u \|_{W^{k,p}} \leq C \sum_{|\a|\leq k} \|\Fg \ax^{\s}(\pa^\a V)\|_{L^{m_\ast}(R^m)} 
\|u\|_{W^{k,p}}, \hspace{2cm} \\ 
\|\W_{n} u \|_{W^{k,p}} \leq C^n \Big(\sum_{|\a|\leq k} \|\Fg \ax^{2\s}(\pa^\a V)\|_{L^{m_\ast}(R^m)} \Big)^n 
\|u\|_{W^{k,p}} ,\quad n=2, \ldots 
\end{eqnarray}
for any $1\leq p \leq \infty$. 
\end{lemma}

\section{Threshold singularities}

The resolvent $G(\l)=(H-\l^2)^{-1}$ of $H=-\lap +V$ is a $\Bb(\Hg)$-valued meromorphic 
function of $\l\in \C^+$ with possible poles $i\k_1, \ldots, i\k_n$ on $i\R^+$ such that 
$-\k_1^2, \ldots, -\k_n^2$ are eigenvalues of $H$ and outside the poles we have    
\begin{equation}\lbeq(res-eq) 
G(\l)=(1+ G_0(\l)V)^{-1}G_0(\l), \quad \l \in \C^+. 
\end{equation}
For $\l \in \R$, $G_0(\l)V \in \Bb_\infty(\Hg_{-\c})$ for 
all $\frac12<\c<\d-\frac12$ and $1+ G_0(\l)V$, $\l\not=0$, is invertible if and only 
if $\l$ is an eigenvalue of $H$ (\cite{Ag}). 
Since positive eigenvalues are absent from $H$ (\cite{Kato-e}), \refeq(res-eq) 
is satisfied for all $\l \in \Cb^+ \setminus \{0\}$ and $G(\l)$ considered 
as a $\Bb(\Hg_\c, \Hg_{-\c})$ valued function satisfies the same regularity 
properties as $G_0(\l)$ as stated in \reflm(Ag) except 
possibly at $\l=0$. We omit the proof of the following well known lemma 
(see \cite{Ag}, \cite{Ku-0}). 

\begin{lemma}\lblm(Ag-t) Let $\frac12<\c<\d-\frac12$. Then, 
$G(\l)$ is a $\Bb_\infty (\Hg_{\c}, \Hg_{-\c})$ valued function of $\l \in \Cb^+\setminus\{0\}$ 
of class $C^{(\c-\frac12)_-}$. For $0 \leq j<\c-\frac12$, 
\begin{equation}\lbeq(Ag-t)
\|G^{(j)}(\l)\|_{\Bb(\Hg_{\c}, \Hg_{-\c})} \leq C_{j\c} |\l|^{-1}, \quad |\l|\geq 1. 
\end{equation} 
\end{lemma} 

Following \cite{JK}, with $D_0=G_0(0)$ as in \refprp(even-exp) we define:   
\begin{equation}\lbeq(Mg)
\Ng=\left\{\f \in \Hg_{-\c}: (1+ D_0 V)\f =0 \right\}. 
\end{equation} 
It is well known (\cite{JK}, \cite{Y-s}) that $\Ng$ is finite dimensional and it is independent 
of $1/2<\c<\d-1/2$; $-(Vu,u)$ defines an inner product of $\Ng$; 
and if $\{\f_1, \ldots, \f_d\}$ is an orthonormal basis of $\Ng$, $\{-V\f_1, \ldots, -V\f_d\}$ 
is the dual basis of the dual space $\Ng^\ast=\left\{\p \in \Hg_\c: (1+ VG_0(0))\p =0 \right\}$. 
It follows that the spectral projection $Q$ in $\Hg_{-\c}$ for the eigenvalue $-1$ of $G_0(0)V$ 
is given by $Q= -\sum_{j=1}^d \f_j \otimes (V\f_j)$. We set $\Qb=1-Q$. 

\begin{lemma}\lblm(e-prop) Let $D_2$ be as in \refprp(even-exp) and let $\f \in \Ng$. Then: 
\begin{equation}\lbeq(e-prop)
\mbox{$V \f \in \Hg_{\left(\d+\frac{m-4}{2}\right)_-}$; 
$|(D_2 V\f)(x)|\leq C \ax^{4-m}$ and $D_2V\f \in \Hg_{\left(\frac{m-8}2\right)_-}$. }
\end{equation} 
\end{lemma} 
\begin{proof} The lemma follows since $\f \in \Ng$ satisfy $|\f(x)|\leq C \ax^{-(m-2)}$ 
and $D_2$ has the integral kernel $C|x-y|^{4-m}$. 
\end{proof}

By virtue of \refeq(e-prop), $\Ng$ coincides with the eigenspace $\Eg$ of $H$ with eigenvalue 
$0$ if $m \geq 6$ and the following defnition is consistent with the one given in the introduction.    

\begin{definition}\lbdf(gene) We say that the operator $H$ is of generic type 
if $\Ng=\{0\}$ and that $H$ is of exceptional type if otherwise.  
\end{definition}

\subsection{Generic Case} 
When $H$ is of generic type, $G(\l)$ as a $\Bb(\Hg_\c, \Hg_{-\c})$ valued function, 
$\frac12<\c<\d-\frac12$, satisfies the same regularity properties as $G_0(\l)$ as stated 
in \reflm(Ag) on $\R$. We write $M(\l)=I+ G_0(\l)V$ in what follows. 

\begin{definition} For an integer $\r>0$ and a $C^{\r-1}$ function $f(\l)$ 
defined on an open interval $I$ containing $0$ we say $f$ is of class $C^\r_\ast$ on $I$ if 
$f \in C^\r(I \setminus \{0\})$  and it satisfies 
$\| f^{(\r)}(\l)\| \leq C \la \log \l \ra^N $ for constants $C>0$ and $N>0$, $\l \not=0$.  
\end{definition}

\begin{lemma}\lblm(basic) Let $\frac12<\c, \t <\d-\frac12$ be such that $\c+\t>2$. Let 
$\r_0= \min(\c-1/2, \d-\c-1/2)$ and $\r_{\ast}=\min(\c-1/2,\t-1/2, \t+\c-2)$.
Suppose $H$ is of generic type. Then: 
\ben
\item[{\rm (1)}] If $\r_0 \leq m-2$, $M^{-1}(\l)$ is a $\Bb(\Hg_{-\c})$ valued function of $\l$ of 
class $C^{(\r_0)_-}$. If $\r_0>m-2$, it is of class $C^{(\r_0)_-}$ for $\l \not=0$ 
and of class $C^{m-2}_\ast$ on $\R$. 
\item[{\rm (2)}] For any $\l \in \R$, $M(\l)^{-1}-1$ 
may be extended to a bounded operator from $\Hg_{-\d+\c}$ to $\Hg_{-\t}$. 
If $\r_\ast\leq m-2$, it is a $\Bb(\Hg_{-\d+\c}, \Hg_{-\t})$-valued function 
of class $C^{(\r_\ast)_-}$. If $\r_\ast > m-2$, it is of class  $C^{(\r_\ast)_-}$ 
for $\l\not=0$ and of class $C^{m-2}_\ast$ on $\R$.  
If $m=4$ and $\r_\ast > 3$, $\l (M(\l)^{-1}-1)$ is of class  $C^{(\r_\ast)_-}$  
for $\l\not=0$ and of class $C^{3}_\ast$ on $\R$. 
\een
\end{lemma} 
\begin{proof} We prove the estimates 
$\|\pa_\l^{m-2}M^{-1}(\l)\|_{\Bb(\Hg_{-\d+\c}, \Hg_{-\t})}\leq C \la \log \l\ra$ 
only, assuming $\r_\ast > m-2$, as the rest may be proved, by virtue of \reflm(Ag), by 
an almost word by word repetition of the proof of Lemma 2.7 of [I]. By using the 
identity $\pa_\l M^{-1}(\l)=-M^{-1}(\l) G_0'(\l)V M^{-1}(\l)$ we $m-2$ times formally 
differentiate $M^{-1}(\l)-I$.  This produces a linear combination 
over $j_1+ \cdots j_k = m-2$, $j_1, \ldots, j_k \geq 1$ of  
\[
M^{-1}(\l) G_0^{(j_1)}(\l)V M^{-1}(\l) \cdots M^{-1}(\l) G_0^{(j_k)}(\l) V M^{-1}(\l) .
\]
If $k\geq 2$, this is bounded in $\Bb(\Hg_{-\d+\c}, \Hg_{-\t})$ near $\l=0$ 
by the proof of Lemma 2.7 of [I]; and if $k=1$, this is bounded by $C\la \log \l\ra$ 
by virtue of \refprp(even-exp) (2) and of the estimate  
\[
\|\pa_\l^{m-2} (\l^{m-2} \log \l A(\l))\|_{\Bb(\Hg_{\c}, \Hg_{-\t})} \leq C \la \log\l \ra , 
\quad |\l|<1 
\]
obtained via \refcor(Ar-c) and \reflm(Ar-0). The desired estimate follows.  \end{proof}

\subsection{Exceptional Case}

In this subsection we assume $H$ is of exceptional type. 
Then $(1+ G_0(\l)V)^{-1}$ is singular at $\l=0$. 
When $m$ is even, the logarithmic singularities appear in addition to those 
due to the $0$ eigenspace of $H$ and the analysis 
becomes more complex than in odd dimensions. In this subsection 
we prove the following expansion formulae for $(1+ G_0(\l)V)^{-1}$. Recall 
that $P_0$ is the orthogonal projection onto the zero eigenspace of $H$.

\begin{proposition}\lbprp(sing-exp) 
{\rm (1)}  Let $m=6$ and $|V(x)|\leq C \ax^{-\d}$ with $\d>10$. Then, 
with $E(\l)$ such that $VE(\l)$ satisfies the condition $(K)_\r$ with $\r>m+1$,   
\begin{equation}\lbeq(sing-1) 
(1+ G_0(\l)V)^{-1}-1= 
\frac{P_0 V}{\l^2} + \sum_{j=0}^{\frac{m-2}2} \sum_{k=1}^{2} D_{jk} \l^j \log^k \l + E(\l). 
\end{equation}
Here $D_{jk}$ are finite rank operators of the form 
\begin{equation}
 D_{jk}=P_0 V D_{jk}^{(1)}P_0V +D_{jk}^{(2)}P_0V + P_0 V D_{jk}^{(3)}, 
\end{equation}
where $D_{jk}^{(1)}\in \Bb(\Ng), \ D_{jk}^{(2)}\in \Bb(\Ng, \Hg_{-3_+}), \  D_{jk}^{(3)} 
\in \Bb(\Hg_{-\d+3_+}, \Ng) $ 

\noindent {\rm (2)} Let $m \geq 8$ and $|V(x)|\leq C \ax^{-\d}$ with $\d>m+3$. Then, with 
$E(\l)$ such that $VE(\l)$ satisfies the condition $(K)_\r$ with $\r>m+1$,  
\begin{equation}\lbeq(sing-2) 
(1+ G_0(\l)V)^{-1}-1= 
\frac{P_0 V}{\l^2} + c_m \ph \otimes (V\ph) \l^{m-6} \log \l + E(\l). 
\end{equation}
Here $\ph = P_0 V$ with $V$ being considered as a function. If $m \geq 12$, then 
$c_m \ph \otimes (V\ph) \l^{m-6} \log \l$ may be included in $E(\l)$. 
\end{proposition}

The rest of this subsection is devoted to the proof of 
\refprp(sing-exp).   
We use the following lemma as in the odd dimensional case. 

\begin{lemma}\lblm(alg) Let $\Xg= \Xg_0 \dot + \Xg_1$ be a direct sum 
decomposition of a vector space $\Xg$. Suppose that a linear operator 
$L$ in $\Xg$ is written in the form 
\[
L= \begin{pmatrix} L_{00} & L_{01} \\ L_{10} & L_{11} \end{pmatrix}
\]
in this decomposition and that $L_{00}^{-1}$ exists. Set 
$C= L_{11}-L_{10}L_{00}^{-1}L_{01}$. 
Then, $L^{-1}$ exists if and only if $C^{-1}$ exists. 
In this case  
\begin{equation}\lbeq(L00)
L^{-1}= \begin{pmatrix} L_{00}^{-1}+ L_{00}^{-1}L_{01}C^{-1}L_{10}L_{00}^{-1}  &  
-L_{00}^{-1}L_{01}C^{-1} \\ -C^{-1}L_{10}L_{00}^{-1} &  C^{-1} 
\end{pmatrix} .
\end{equation}
\end{lemma} 

Using the spectral projections $Q$ and $\Qb=1-Q$, we decompose $\Hg_{-\c}= \Qb \Hg_{-\c}\dot + \Ng$ 
as a direct sum. With respect to this decomposition, we write 
\begin{equation}\lbeq(Mmat)
M(\l) = 
\begin{pmatrix} \Qb M(\l) \Qb & \Qb M(\l) Q \\ 
QM(\l)\Qb & QM(\l)Q \end{pmatrix}
\equiv \begin{pmatrix} L_{00}(\l) & L_{01}(\l) \\ 
L_{10}(\l) & L_{11}(\l) \end{pmatrix}, 
\end{equation}
where the right side is the definition. 
We begin by studying $L_{00}^{-1}(\l)$. Since $L_{00}(0) \in \Bb(\Qb\Hg_{-\c})$ 
is invertible by the separation of spectrum theorem for compact operators,  
$L_{00}(\l)$ is also invertible for small $|\l|<\l_0$. We omit the proof of the following 
lemma  which is similar to that of \reflm(basic). 

\begin{lemma}\lblm(basic-m) Let $\frac12<\c, \t <\d-\frac12$ and $\c+\t>2$. Let  
$\r_0= \min(\c-1/2, \d-\c-1/2)$ and $\r_{\ast}=\min(\c-1/2,\t-1/2, \t+\c-2)$. Then: 
\ben
\item[{\rm (1)}] If $\r_0\leq m-2$, 
$L^{-1}_{00}(\l)$ is a $\Bb(\Qb\Hg_{-\c})$ valued function of $\l\in (-\l_0,\l_0)$  
of class $C^{(\r_0)_-}$. If $\r_0>m-2$, it is of class $C^{(\r_0)_-}$ for $\l\not=0$ 
and of class $C^{m-2}_\ast$ on $(-\l_0,\l_0)$.   
\item[{\rm (2)}] For any $\l \in \R$, $L_{00}^{-1}(\l)-\Qb$ may be extended to a bounded operator 
from $\Qb\Hg_{-\d+\c}$ to $\Qb\Hg_{-\t}$. If $\r_\ast \leq m-2$, it is of class 
$C^{(\r_\ast)_-}$ as a $\Bb(\Qb\Hg_{-\d+\c}, \Qb\Hg_{-\t})$-valued function. 
If $\r_{\ast}>m-2$, then it is of class $C^{(\r_\ast)_-}$ for $\l\not=0$ and 
of class $C^{m-2}_\ast$ on $(-\l_0,\l_0)$. 
\een
\end{lemma} 

Removing the singular part $-\l^{m-2}\log\l\, A(\l)$ from $G_0(\l)$, 
we define 
\[
G_{0reg}(\l)=G_0(\l)+\l^{m-2}\log\l \, A(\l), \quad 
N(\l) = \Qb(1+ G_{0reg}(\l)V)\Qb. 
\]
If $\c$ and $\r_0$ is as in \reflm(basic-m), \refprp(even-exp) implies that
$N(\l)$ is a $\Bb(\Qb\Hg_{-\c})$ valued function of $\l \in \R$   
of class $C^{(\r_0)_-}$ and $N(\l)$ is invertible in $\Qb\Hg_{-\c}$ for 
$\l\in (-\l_0, \l_0)$ if $\l_0>0$ is chosen small enough. We write 
\begin{equation}\lbeq(def-l)
\tilde L(\l) = L_{00}^{-1}(\l) -\Qb, \quad X(\l) = N^{-1}(\l), \quad \tilde X(\l)=X(\l) - \Qb. 
\end{equation}
We omit the proof of the following lemma which is also similar to that of 
\reflm(basic).  
\begin{lemma}\lblm(basic-a) Let $\c,\t$ and $\r_0, \r_\ast$ be as in \reflm(basic-m). Then: 
\ben
\item [{\rm (1)}] 
$X(\l)$ is a $\Bb(\Qb\Hg_{-\c})$ functions of $\l \in (-\l_0,\l_0)$ of class $C^{(\r_0)_-}$.  
\item[{\rm (2)}] 
For $\l \in (-\l_0, \l_0)$, $\tilde X(\l)$ extends to a bounded operator 
from $\Qb\Hg_{-\d+\c}$ to $\Qb\Hg_{-\t}$ and it is $\Bb(\Qb\Hg_{-\d+\c},\Qb \Hg_{-\t})$-valued function 
of class $C^{(\r_\ast)_-}$.  
\een
\end{lemma}

We define $Y(\l)=L_{00}^{-1}(\l) - X(\l)$. In what follows we shall often use the arguments 
similar to the ones which will be used in (i) to (iv) of the proof of the following 
corollary. We use the following elementary lemma which follows by a direct estimate:  
\begin{lemma}\lblm(aux) Suppose $f(x)$ is of class $C^s_{0\ast}(\R)$, $0<s\leq 1$, then 
$\log x\, f(x)$ is of class $C^{s_-}_{0\ast}(\R)$. 
\end{lemma} 
 
\begin{corollary}\lbcor(basic) Let  $\c, \t$ and $\r_\ast$ be as in \reflm(basic-m) and  
$j, k=0,1, \ldots$. Then,  $Y_{jk}(\l)\equiv \l^j (\log  \l)^k Y(\l)$, $\l \not=0$, may 
be extended to a bounded operator 
from $\Qb\Hg_{-\d+\c}$ to $\Qb\Hg_{-\t}$. Define $Y_{jk}(0)=0$. 
If $\r_\ast \leq m-2$, then $Y_{jk}(\l)$ is of class $C^{(\r_\ast)_-}$ as a 
$\Bb(\Qb\Hg_{-\d+\c}, \Qb\Hg_{-\t})$-valued function of $\l \in (-\l_0,\l_0)$.  
If $\r_{\ast}>m-2$, then it is of class $C^{(\r_\ast)_-}$ for $\l\not=0$ and 
of class $C^{m-2}_\ast$ on $(-\l_0,\l_0)$.  
\end{corollary} 
\begin{proof}
We insert $L_{00}^{-1}(\l)=\Qb + \tilde L(\l)$ and 
$X(\l)=\Qb + \tilde X(\l)$ into 
\begin{equation}\lbeq(co)
Y(\l) =L_{00}^{-1}(\l)(\l^{m-2}\log\l\, \Qb A(\l)V\Qb) X(\l) 
\end{equation} 
This produces four terms and the smoothness property of each operator outside $\l=0$ 
is easy to check. We we treat them near $\l=0$.  

\noindent 
{\rm (i)} $\l^{m-2}\log\l\, \Qb A(\l)V\Qb$ enjoys the desired property by virtue of \reflm(Ar); 

\noindent {\rm (ii)} To see the same for $\tilde L(\l)(\l^{m-2}\log\l\, \Qb A(\l)V\Qb)$, 
we compute the $l$-th derivative via Leibniz' formula. If $l=a +b <\r_\ast$, 
we choose $\k$ such that $b+\frac12<\k<\d-a-\frac12$, $\t+ (\d-\k)>2$ and $\k+\c>2$. 
This is possible since $\d-a-\frac12 > b+\frac12$, 
$b+\frac12<\d+\t-2$ and $2-\c<\d-a-\frac12$. Then, 
as a $\Bb(\Qb\Hg)$-valued function,  
\[
\ax^{-\t} \tilde L^{(a)}(\l)\ax^{\k} \cdot \ax^{-\k}(\l^{m-2}\log\l\, \Qb A(\l)V\Qb)^{(b)} \ax^{\d-\c}
\]
is continuous with respect to $\l \in (-\l_0,\l_0) \setminus \{0\}$. If $b<m-2$, 
this is continuous also at $\l=0$, and is bounded by $\la \log\l \ra$ if $b=m-2$.  

\noindent 
{\rm (iii)} The argument which is entirely similar to the one used in (ii) proves that 
$\Qb(\l^{m-2}\log\l\,  A(\l))V\Qb \tilde X(\l)$ satisfies the corollary. 

\noindent 
{\rm (iv)} In view of the result in (ii), we estimate for $l=a +b < \r_\ast$ 
\[
\ax^{-\t} \{\tilde L(\l) \l^{m-2}\log\l\, \Qb A(\l)V\Qb\}^{(a)} \ax^{\k} \cdot 
\ax^{-\k} \tilde X^{(b)}\ax^{\d-\c}
\]
by choosing $\k$ such that $b+\frac12<\k<\d-a-\frac12$, $\t+\d-\k>2$ and $\k+ \c>2$. 
Such $\k$ exists by the same reason as in (ii). Then, as a $\Bb(\Qb\Hg)$-valued 
function, this is continuous with respect to $\l \in (-\l_0,\l_0) \setminus \{0\}$. If 
$a<m-2$ it is continuous also at $\l=0$, and is bounded by $\la \log\l \ra$ if $a=m-2$. 
Hence $\tilde L(\l)(\l^{m-2}\log\l\, \Qb A(\l)V\Qb) \tilde X(\l)$ has the desired property. 
\end{proof}  
Since the logarithmic  singularity appears in the form $\l^{m-2}\log \l \ A(\l)$ in $G_0(\l)$ as 
in \refprp(even-exp) and $\l^{m-2}\log \l$ is less singular in higher dimensions, the proof of 
the proposition becomes 
easier as the spatial dimension $m$ increases. Thus, we study the case $m=6$ first and then 
discuss the case $m \geq 8$ only briefly. 
{\it In what follows we shall indiscrimately write $E_0(\l)$ for $\Bb(\Ng)$ valued functions
which satisfy the following property for some $N\geq 0$: }
\begin{equation}\lbeq(e0op)
\br{c}
\mbox{$E_0(\l)$ is of class $C^{\frac{m+2}2}((-\l_0, \l_0)\setminus\{0\})$ 
and $C^{\frac{m-2}2}(-\l_0, \l_0)$ and} \\
\|E_0^{(\frac{m}2)} (\l)\| \leq C \la \log \l \ra^N, \quad 
\|\l\, E_0^{(\frac{m+2}2)} (\l)\| \leq C \la \log \l \ra^N. 
\er 
\end{equation}
Note that the function of class $C^{\frac{m+2}2}_\ast$ on $(-\l_0, \l_0)$ clearly satisfies the 
condition \refeq(e0op). We often omit the variable $\l$ of operator valued functions.
Note that $m-2 \geq \frac{m+2}2$ when $m \geq 6$ with strict inequality when $m>6$. 

\subsubsection{Proof of \refprp(sing-exp) for $m=6$} 
In view of \reflm(alg), we first study $C(\l)=L_{11}-L_{10}L_{00}^{-1}L_{01}$. 
We have
\[
G_0(\l)=D_0 + \l^2 D_2 - \l^{4}(\log \l ) A(\l) + \l^{4} F(\l)
\]
by virtue of \refprp(even-exp). Since $(1+D_0 V)Q=Q(1+VD_0)=0$, 
\begin{equation}\lbeq(L11) 
\br{l}
L_{11}(\l)= \l^2 Q(D_2 -  \l^2 \log \l A(\l)+ \l^2 F(\l))VQ. \\ 
L_{01}(\l) = \l^2 \Qb(D_2 - \l^2 \log \l A(\l)+ \l^2 F(\l))V Q .   \\
L_{10}(\l) = \l^2 Q(D_2 - \l^2 \log \l A(\l) + \l^2 F(\l))V \Qb .  \er
\end{equation}
We know that $QD_2VQ$ is invertible in $\Ng$ and $(QD_2VQ)^{-1}=P_0 V$,
$P_0 VQ=P_0V$ and $VQP_0 = VP_0$. 
Then, $C(\l)$ may be written in the form 
\begin{align} 
C(\l) = \l^2(Q D_2 VQ)( 1  - P_0 V E_2^\ast(\l)), \hspace{2cm} \lbeq(cl) \\ 
E^\ast_2(\l)= \l^2 F_{00} + \l^2 \log \l  F_{01} 
+\l^4 F_{20} +  \l^4 \log \l F_{21} + \l^2 E_{0}(\l), \notag
\end{align}
where $F_{00}(\l), F_{01}(\l), F_{20}(\l)$ and $F_{21}(\l)$ are defined by  
\begin{equation} \lbeq(5-10)
\br{c}
F_{00}(\l)= -Q(F(\l)-D_2 V\Qb  L^{-1}_{00}\Qb  D_2)VQ, \quad F_{01}= QA(\l)VQ , \\
F_{20}(\l)= Q (F(\l) V \Qb L^{-1}_{00}\Qb D_2 + D_2 V \Qb L_{00}^{-1}\Qb  F(\l) )VQ, \\
F_{21}(\l)= -Q (A(\l) V \Qb  L^{-1}_{00}\Qb D_2 + D_2 V\Qb  L_{00}^{-1}\Qb A(\l) )VQ, 
\er
\end{equation} 
and 
$E_0(\l)=\l^4 Q(\log \l A(\l) - F(\l))V\Qb L^{-1}_{00}(\l) \Qb(\log \l A(\l)-F(\l))VQ$ 
is of class $C_\ast^4$ thanks to \refeq(e-prop) (recall $\d>10$). We write $\tilde F_{jk}(\l)$ for the operator 
obtained from $F_{jk}(\l)$ of \refeq(5-10) by replacing $L_{00}^{-1}(\l)$ by $X(\l)=N^{-1}(\l)$. 

\begin{lemma}\lblm(ju) As $\Bb(\Ng)$-valued functions of $\l \in (-\l_0,\l_0)$, 
\ben 
\item[{\rm (1)}] $F_{00}(\l)$, $F_{20}(\l)$ and $F_{21}(\l)$ 
are of class $C^{4}_\ast$;  
\item[{\rm (2)}]  $\tilde F_{00}(\l)$, $F_{01}(\l)$, $\tilde F_{20}(\l)$ and $\tilde F_{21}(\l)$   
are of class $C^{(\d- 4)_-}$; 
\item[{\rm (3)}]  $ QA(\l)V\Qb \tilde L(\l) \Qb F(\l) V Q$ is of class $C_\ast^{4}$;
\item[{\rm (4)}] $QA(\l)V\Qb \tilde X(\l) \Qb F(\l) V Q$ is of class $C^{(\d- 4)_-}$. 
\een
The same holds for the operators which are obtained by replacing one or both of $A(\l)$ and $F(\l)$ 
by the other operator.  
\end{lemma} 
\begin{proof}  We prove statements (2). The proof for others is similar. 

\noindent  
{\rm (i)} \refprp(even-exp) and properties \refeq(e-prop) of $\f \in \Ng$ imply 
that $QF(\l)VQ$ is of class $C^{(\d-2)_-}$; and  
the operators $Q F(\l) V \Qb  D_2 VQ$ and $Q D_2 V \Qb F(\l)VQ$ are of class $C^{(\d-4)_-}$. 
The same holds when $F(\l)$ is replaced by $A(\l)$. 

\noindent 
{\rm (ii)}  By virtue of \reflm(basic-a) and \refeq(e-prop), 
$ QD_2 V X(\l)D_2 VQ$ is of class $C^{(\d-\frac32)_-}$.

\noindent 
{\rm (iii)} $QF(\l) V \tilde X(\l)D_2 VQ$ is of class $C^{(\d-\frac{7}{2})_-}$. 
To see this we differentiate it by $\l$ by using Leibniz' formula. Then, by virtue 
of \reflm(basic-a) and \refeq(e-prop), for some $\ep>0$, 
$\ax^{-\d-1+\ep} F^{(k_1)}(\l) V \tilde X^{(k_2)}(\l)\ax^{1+\ep}$ 
is $\Bb(\Hg)$-valued continuous as long as 
\[
k_1+3+ k_2 +\tfrac12<\d, \ k_1+3<\d+1, \ k_2+\tfrac12 <\d -1 
\] 
and the latter inequalities trivially hold if $k_1+k_2 <\d-\frac72$. 
Similar argument implies that the same holds for $Q D_2 V \Qb \tilde X(\l)\Qb F(\l)VQ$ 
and for the operators obtained by replacing $F(\l)$ by $A(\l)$.

Combining (i), (ii) and (iii) we obtain statement (2).   
\end{proof} 

\begin{lemma} There exist $\l_0>0$ such that for $\l \in (-\l_0, \l_0)\setminus \{0\}$ 
$C(\l)$ is invertible in $\Ng$ and $C(\l)^{-1}$ may be written in the form 
\begin{equation} \lbeq(cc-1)
\l^{-2}P_0 V  
+ P_0V \Big(\log\l D_{10} + \sum_{1\leq k\leq j\leq 2}\l^j (\log\l)^k D_{jk} +  E_0(\l)\Big)P_0V
\end{equation}
with $D_{jk}\in \Bb(\Ng)$ and a $\Bb(\Ng)$-valued $C^4_\ast$ function $E_0(\l)$. 
\end{lemma}
\begin{proof} By virtue of \reflm(ju), $\|P_0 V E_2^\ast(\l)\| \to 0$ as $\l \to 0$. 
It follows that $C(\l)$ is invertible for small $\l \not=0$ and 
\[
C^{-1}(\l)= \l^{-2} \sum_{n=0}^\infty (P_0V E^\ast_2(\l))^n P_0 V
\]
Here $\l^{-2}\sum_{n=3}^\infty (P_0V E^\ast_2(\l))^n P_0 V$ 
is of the form $P_0 V E_0(\l) P_0 V$ with a $C^4_\ast$ function 
$E_0(\l)$ again by \reflm(ju). We also put all terms into $E_0(\l)$ which is 
produced by $\l^{-2}\sum_{n=1}^2(P_0V E^\ast_2(\l))^n P_0 V$ and 
which do not contain any $\log \l$ factors or contain factors  
$\l^j$ with $j \geq 3$. In this way we obtain  
\begin{equation}\lbeq(cinv-1)\br{l}
C^{-1}(\l) = \l^{-2}P_0 V + \log \l P_0V  F_{01} P_0V  \\ 
\quad\qquad + \l^2\log\l\, P_0V (F_{21}+ F_{00}P_0VF_{01}+ F_{01}P_0VF_{00})P_0V  \\
\qquad\quad + \l^2\log^2\l(P_0VF_{01})^2P_0V + E_0(\l) . 
\er 
\end{equation} 
The equation \refeq(cinv-1) remains valid if $F_{00}, F_{01}$ and $F_{21}$ are 
replaced by $\tilde F_{00}, \tilde F_{01}$ and $\tilde F_{21}$ respectively 
because the difference is of class $C^4_\ast$  by virtue of \refcor(basic).   
We then expand various operators in the resulting equation in powers of $\l$ 
by using Taylor's formula. More specifically, we expand 

\noindent 
{\rm (a)}  
$P_0V F_{01}(\l) P_0V= P_0V A(\l)VP_0V$ upto the order $\l^2$ with the remainder $\l^3 R_{1}(\l)$,  
$R_1(\l)$ being a $\Bb(\Ng)$-valued function of class 
$C^{\r}$ for any $\r< \d-5$ by virtue of the property \refeq(e-prop) of eigenfunctions 
and of \reflm(Ar-0). It follows, since $\d-5>4$, that $\l^3 \log\l\, R_{1}(\l)$ satisfies 
property \refeq(e0op) and it can be absorbed into $E_0(\l)$; 

\noindent 
{\rm (b)} If we replace 
$P_0V (F_{21}(\l)+ F_{00}(\l)P_0VF_{01}(\l)+ F_{01}(\l)P_0VF_{00}(\l))P_0V$ 
by the constant operator obtained by setting $\l=0$, the difference 
is $\l$ times a $\Bb(\Ng)$-valued function $R_{2}(\l)$ of class $C^\r$ for any $\r<\d-5$ by 
virtue of \reflm(ju), and  $\l^3\log\l\, R_{2}(\l)$ may be put into $E_0(\l)$. 

\noindent 
{\rm (c)}  If we replace $A(\l)$ by $A(0)$ in $(P_0VF_{01}(\l))^2P_0V=(P_0VA(\l) VP_0)^2V$, 
then the difference is $\l$ times a $\Bb(\Ng)$-valued function 
$R_{3}(\l)$ of class $C^4$ and $\l^3\log^2\l\,  R_{3}(\l)$ may be absorbed 
into $E_0(\l)$. 

Equation \refeq(cinv-1) and (a), (b) and (c) imply the lemma. 
\end{proof}

We denote $C_r(\l)=C^{-1}-\l^{-2}P_0 V$, the second member of \refeq(cc-1). 

\begin{lemma}\lblm(c-step1) There exist $\l_0>0$ such that for $\l \in (-\l_0, \l_0)\setminus\{0\}$  
\begin{equation}\lbeq(cc-2)
L^{-1}_{00}(\l)L_{01}(\l) C_r(\l) = \l^{2} \log \l\, D_{21}^{(1)}P_0V 
+ R_{01}(\l)P_0V .  
\end{equation}
Here $D_{21}^{(1)}\in \Bb(\Ng,\Hg_{(-1)_-})$ and $R_{01}(\l)$ is such that 
as $\Bb(\Ng, \Hg)$ valued functions, $\ax^{-(\s+1)_+}R_{01}(\l)$ is 
of class $C^\s$ for $0\leq \s \leq 2$; for $\s=3$ and $4$, $\ax^{-(\s+\frac12)_+}R_{01}(\l)$ is 
of class $C^\s$ for $\l\not=0$ and for some $N>0$ 
\begin{equation}
\|\ax^{-(\frac72)_+}R_{01}^{(3)}(\l)\|_{\Bb(\Ng, \Hg)} + 
\|\ax^{-(\frac{9}2)_+}\l R_{01}^{(4)}(\l)\|_{\Bb(\Ng, \Hg)}\leq C \la\log \l\ra^N.   
\end{equation}
\end{lemma}
\begin{proof} Since $L_{01}=\Qb(1+ G_0 V)Q$ is a $\Bb(\Ng, \Hg_{-(\c+\frac12)_+})$-valued function of $\l$ 
of class $C^{\c}$ for $\c<4$ and of class $C^4_\ast$ if $\c\geq 4$,  \reflm(basic-m) implies 
that 
\[
L_{00}^{-1}L_{01}(\l) E_0(\l)P_0V= (\Qb L_{01}(\l) E_0(\l) + \tilde L(\l)L_{01}(\l) E_0(\l))P_0 V
\]
satisfies the same property. We have only to consider 
$\l^j (\log\l)^k L_{00}^{-1}L_{01}D_{jk}$. (In this proof we ignore the factor $P_0 V$ 
for shortening formulas). As a 
$\Bb(\Ng, \Hg_{-(\c+\frac12)_+})$-valued function 
$\l^{4+j}(\log\l)^k \Qb(\log \l A(\l)+ F(\l))V Q$ is of class $C^{\c}$ if $\c<4$ 
and of class $C^4_\ast$ if $\c\geq 4$ by virtue of \refcor(Ar-c), \reflm(Ar) and 
\refprp(even-exp). It follows  
by writing again $L_{00}^{-1}=\Qb + \tilde L(\l)$ and applying \reflm(basic-m) that  
\[
L^{-1}_{00}(\l)(\l^{m-2+j}(\log\l)^k \Qb(\log \l A(\l)+ F(\l))V Q)D_{jk} 
\]
shares this property. Writing $L_{00}^{-1}=\Qb+Y+ \tilde X$, we are left with 
\begin{align} 
\l^{2+j}(\log\l)^k L^{-1}_{00}(\l)\Qb D_2VQD_{jk}= \l^{2+j}(\log\l)^k \Qb D_2VQD_{jk} \notag \\
+ \l^{2+j}(\log\l)^k Y(\l)\Qb D_2VQD_{jk} + \l^{2+j}(\log\l)^k \tilde X(\l)\Qb D_2VQD_{jk} 
\lbeq(5-20)
\end{align}
Recalling that $\Qb D_2V \in \Bb(\Ng, \Hg_{-1-\ep})$ for any $\ep>0$,  
we put the first term on the right to $\l^{2} \log \l\, D_{21}^{(1)}$ if $j=0$ and 
into $R_{01}(\l)$ otherwise. Note that if $j=0$, we have only $k=1$ term, see \refeq(cc-1).  
\refcor(basic) with $\c=\d-1-\ep$ and $\t=\s+1+\ep$ implies that 
the second term may also be put into $R_{01}(\l)$. To deal with the last term, 
we write $\tilde X(\l)=\tilde X(0)+ (\tilde X_1(\l)-\tilde X(0))$. Remarking that 
$\tilde X(0)\Qb D_2VQ \in \Bb(\Ng, \Hg_{-\frac12-\ep})$ for any $\ep>0$ we put    
$\l^{2+j}(\log\l)^k \tilde X(0)\Qb D_2VQD_{jk}$ into $\l^{2} \log \l\, D_{21}^{(1)}$ if $j=0$ 
and into $R_{01}(\l)$ otherwise. Finally,  
\[
\l^{2+j}(\log\l)^k(\tilde X(\l)-\tilde X(0))\Qb D_2VQD_{jk} 
\]
may also be put into $R_{01}(\l)$. This can be seen by diffentiating 
via Leibniz' rule and by applying \reflm(basic-a). This completes the proof. 
\end{proof}

Recall $\s_0(3,\s)= \frac{\s}2 + 2$ for $\s \leq 3$. 
In the following two lemmas, we set 
\begin{equation}\lbeq(s-def)
 \c(\s)= \left\{\br{ll} \max(\s_0(3,\s), 3), \quad & \mbox{if $\s <2$}, \\
\s+1, &\mbox{if $2 \leq \s \leq 4$}. \er \right.
\end{equation}
We remark that the condition $\d>10=m+4$ when $m=6$ is originated friom the fact that 
$\c(0)=3$ and that, for $VR_{02}(\l)P_0V$ and $VP_0 V R_{30}(\l)$ to satisfy the property 
$(K)_\r$ with $\r>m+1$, we need $\d-3>m+1$. 

\begin{lemma}\lblm(c-step2)  With $D_{2,1}^{(2)}\in \Bb(\Ng, \Hg_{(-3)_-})$ 
and $R_{02}(\l)$, we have 
\begin{equation}\lbeq(cc-3)
L^{-1}_{00}(\l)L_{01}(\l) C^{-1}(\l)= \l^2 \log\l D_{2, 1}^{(2)}P_0V + R_{02}(\l)P_0V. 
\end{equation}
where, as a $\Bb(\Ng, \Hg)$ valued functions, $\ax^{-\c(\s)_+}R_{02}(\l)$ is 
of class $C^\s$ on $(-\l_0,\l_0)$ for $0\leq \s \leq 2$, for $\l\not=0$  for $\s=3, 4$ 
and 
\begin{equation}
\|\ax^{-4_+}R_{02}^{(3)}(\l)\|_{\Bb(\Ng, \Hg)} + 
\|\ax^{-5_+}\l R_{02}^{(4)}(\l)\|_{\Bb(\Ng, \Hg)}\leq C \la\log \l\ra. 
\end{equation} 
\end{lemma}
\begin{proof} In view of \reflm(c-step1) it suffices to prove the lemma with 
$\l^{-2}P_0 V$ in place of $C^{-1}(\l)$. We multiply the following by $L_{00}^{-1}(\l) $ from the left: 
\[
L_{01}(\l)\l^{-2} P_0 V = \Qb D_2VP_0 V - \l^2\log\l\, \Qb A(\l) VP_0V +\l^2\Qb F(\l)VP_0 V. 
\]

\noindent 
{\rm (i)}  We may include 
$L^{-1}_{00}(\l)\Qb D_2VP_0 V=(\Qb + \tilde L(\l))\Qb D_2VP_0 V $ 
into $R_{02}(\l)$ by virtue of \reflm(basic-m).    

\noindent 
(ii) In $L^{-1}_{00}(\l) \Qb \l^2\log\l\, A(\l) VP_0 V $ 
we substitute $Y(\l) + \tilde X(\l)+\Qb$ for $L_{00}^{-1}$. 
We may put 
$Y(\l)\Qb \l^2\log\l\, A(\l) VP_0 V = \l^2\log\l\, Y(\l) \cdot \Qb  A(\l) VP_0 V$ 
into $R_{02}(\l)$ by virtue of \refcor(basic). We write in the form 
\begin{equation} \br{l}
\l^2 \log \l \tilde X(\l) \Qb A(\l) V = \l^2 \log \l \tilde X(0) \Qb A(0)V \\
+ \l^2 \log \l (\tilde X(\l)- \tilde X(0)) \Qb A(0)V + \l^2 \log \l \tilde X(\l) (A(\l)-A(0))V.
\er
\end{equation} 
Then, $\tilde X(0) \Qb A(0)V \in \Bb(\Ng, \Hg_{(-\frac12)_-})$ and we put the first term 
on the right into $\l^2\log\l\, D_{21}^{(2)}$; it is easy to check by using \reflm(basic-a) (2) 
that the last two terms satisfy the properties of $R_{02}(\l)$. We write 
$\Qb \l^2 \log\l\, A(\l)V P_0 V$ as 
\[
\Qb \l^2 \log\l\, A(0)V P_0 V + \Qb \l^2 \log\l\, (A(\l)-A(0))V P_0 V.
\]
Since  $\Qb A(0)V P_0 V \in \Bb(\Ng, \Hg_{-3_+})$, we put the first term into 
$\l^2 \log\l D_{21}^{(2)}$. It can be checked that the second term satisfies the 
properties of $R_{02}$ by differentiating it by $\l$ and by applying \reflm(Ar-0) 
and \refcor(Ar-c). 

\noindent
(iii) Since $\l^2 F(\l)V$ is also is of class $C^\s$ as a $\Bb(\Ng, \Hg_{-\c(\s)_+})$-valued 
function by virtue of \refprp(even-exp), we may put  
$L^{-1}_{00}(\l) \Qb \l^2 F(\l)VP_0 V$ into $R_{02}(\l)$. 
This completes the proof. 
\end{proof} 

We omit the proof of the following lemma which goes entirely in parallel with that of 
the previous \reflm(c-step2). 
\begin{lemma}\lblm(c-step3) There exists an operator $D_{12}^{(3)}\in \Bb(\Hg_{(-\d+3)_+}, \Ng)$ 
and an operator valued function $R_{03}(\l)$ which satisfies the property below such that  
\begin{equation}\lbeq(cc-4)
C^{-1} L_{10}(\l) L_{00}(\l)= \l^{2} \log\l\, P_0 V D_{k}^{(3)} + P_0 V R_{30}(\l).  
\end{equation}
Here, as a $\Bb(\Hg, \Ng)$ valued function, $R_{03}(\l)\ax^{(\d-\c(\s))_-}$ is 
of class $C^\s$  on $(-\l_0,\l_0)$ if $0\leq \s \leq 2$, for $\l\not=0$ if $\s=3, 4$, 
and with some $N>0$ 
\[
\|R_{03}^{(3)}(\l)\ax^{(\d-\c(3))_-}\|_{\Bb(\Hg, \Ng)} + 
\| R_{03}^{(4)}(\l)\ax^{(\d-\c(4))_-}\l\|_{\Bb(\Hg, \Ng)}\leq C \la \log \l \ra^N .
\]
\end{lemma}

Since $VP_0VE_0(\l)P_0V$, $VR_{01}(\l)P_0V$, $VR_{02}(\l)P_0V$ and 
$VP_0 V R_{03}(\l)$ satisfy property $(K)_\r$ with $\r>m+1$, 
the following lemma completes the proof of \refprp(sing-exp) for $m=6$. 
\begin{lemma}\lblm(c-step4) The operator valued function 
$VL_{00}^{-1}L_{01}C^{-1}L_{10}L_{00}^{-1}$  satisfies 
the condition $(K)_\r$ with $\r>m+1$. 
\end{lemma} 
\begin{proof} We substitute $\l^{-2}P_0 V + C_r(\l)$ for $C(\l)$  
and write $L_{01}\l^{-2} P_0 VL_{10}$ in the form  
\[
\Qb (\l D_2 -\l^3 \log \l A(\l)+ \l^3 F(\l))VP_0 V (\l D_2 -\l^3 \log \l A(\l)+ \l^3 F(\l))\Qb.
\]
It follows by virtue of \refcor(Ar-c) that 
this is a $\Bb(\Hg_{-\d+\s_0(2,j)_+}, \Hg_{-\s_0(2,j)_+})$-valued function 
of class $C^j$ on $(-\l_0,\l_0)$ for $0\leq j \leq 2$, on  $(-\l_0,\l_0) \setminus\{0\}$ for $j=3$ 
and $j=4$,  and the thrid and fourth derivatives are bounded by $C\la\log\l \ra$ and 
$C|\l|^{-1}\la\log\l \ra$ in respective norms. It follows, by writing 
$L_{00}^{-1}(\l) = \Qb + (L_{00}^{-1}-\Qb)$ 
as usual, that $VL_{00}^{-1}L_{01}(\l^{-2} P_0 V)L_{10}L_{00}^{-1}$  satisfies 
the condition $(K)_\r$ with $\r>m+1$. It is then obvious that so does 
$VL_{00}^{-1}L_{01}C_r(\l)L_{10}L_{00}^{-1}$. The lemma follows. 
\end{proof} 

\subsubsection{The case $m\geq 8$ is even} 
Let now $m \geq 8$. Define $F_0(\l), F_1(\l)$ and $F_2(\l)$ by 
\[\br{c}
F_{0}(\l)= D_0 + \l^2 D_2+ \cdots + \l^{m-4}D_{m-4}+ \l^{m-2}F(\l)\\ 
F_2(\l) = D_2+ \cdots + \l^{m-6}D_{m-4}+ \l^{m-4}F(\l), \\
F_4(\l) = D_4+ \cdots + \l^{m-8}D_{m-4}+ \l^{m-6}F(\l) \er
\]
so that $G_0(\l) = F_0(\l)- \l^{m-2}\log\l\, A(\l)$, 
$F_0= D_0 +\l^2 F_2(\l)$ and $F_2(\l)= D_2 + \l^2 F_4(\l)$.  
Since $(1+ D_0 V ) Q=0$, we then have   
\begin{equation}\lbeq(m8)
L_{11}(\l)= \l^2 Q(D_2 + \l^2 F_4(\l) - \l^{m-4}\log \l A(\l)) VQ;  
\end{equation}
and $L_{10}(\l)$ and $L_{01}(\l)$ are obtained from \refeq(m8) by replacing one of $Q$ by $\Qb$ 
as in \refeq(L11). Recall that $(Q D_2V Q)^{-1} = P_0 V$, $VQP_0=VP_0$ and $P_0VQ=P_0V$. 

\begin{lemma}\lblm(3-2-2) Let $\ph=P_0 V$ where $V$ is considered as a function. Then 
there exists $\l_0$ such that     
\begin{equation}\lbeq(hc-1)
C^{-1}(\l) = \l^{-2}P_0V + c_m \l^{m-6}\log \l\, \ph \otimes (V \ph)  + P_0V E_0(\l)P_0 V, 
\end{equation}
where $c_m= (2\pi)^{-\frac{m}2}(m!!)^{-1}$ and 
$E_0(\l)$ is a $\Bb(\Ng)$-valued function of $\l \in (-\l_0,\l_0)$ which satisfies the 
property \refeq(e0op). 
\end{lemma}
\begin{proof} In this proof the smoothness of operator valued functions will be referered to 
as $\Bb(\Ng)$ valued functions. By virtue of \refprp(even-exp), $E_{01}(\l) \equiv QF_2(\l)VQ$  
is of class $C^{\frac{m+2}2}$. Likewise \reflm(basic-m), \refprp(even-exp), property \refeq(e-prop) 
of $\f \in \Ng$ and that $2(m-4) >\frac{m+2}2$ imply that 
\[\br{c}
E_{02}(\l)\equiv \l^{2(m-4)}(\log \l)^2  QA(\l)V\Qb L_{00}^{-1}(\l)\Qb A(\l)VQ, \\
E_{03}(\l)\equiv  Q  F_1 V \Qb L_{00}^{-1}(\l)\Qb F_1 V Q, \\
F_{20}(\l)= Q(F_1 V\Qb L_{00}^{-1}\Qb A + AV\Qb L_{00}^{-1}\Qb F_1)VQ \\
\er
\]
are all of class $C^{\frac{m+2}2}$. With these definitions, we may write 
\begin{equation}\lbeq(800) 
L_{10}(\l)L_{00}(\l)^{-1}L_{01}(\l)= \l^4 (E_{02}+ E_{03} - \l^{m-4}\log \l F_{20}(\l)), 
\end{equation}
Thus, defining  
\[\br{c} 
\tilde E_0(\l)= P_0 V (E_{01}(\l)- E_{02}(\l)- E_{03}(\l)), \\ 
\tilde F_{10}(\l)= P_0VA(\l)VQ, \quad \tilde F_{20}(\l)= P_0VF_{20}(\l), 
\er
\]
subtracting \refeq(800) from \refeq(m8) and factoring out $\l^2 QD_2 VQ$, we obtain 
\[
C(\l)= \l^2 QD_2 VQ(1 - \l^{m-4}\log\l\, \tilde F_{10}(\l)+ 
\l^{m-2}\log \l\, \tilde F_{20}(\l) + \l^2 \tilde E_0(\l)) .
\]
It follows that $C(\l)$ is invertible in $\Ng$ for $0<|\l| < \l_0$ for small enough $\l_0$  and 
\[
C^{-1}(\l) = 
\l^{-2} \sum_{n=0}^\infty (\l^{m-4}\log\l\, \tilde  F_{10}(\l)-\l^{m-2}\log \l  \tilde  F_{20}(\l)-\l^2  \tilde 
E_0(\l))^n P_0 V .
\]
It is easy to see by counting the powers of $\l$ in front of powers of $\log \l$ that 
the series over $2 \leq n <\infty $ produces a function of class $C^{\frac{m+2}2}$. 
Thus, writing $E_0(\l)$ for $C^{\frac{m+2}2}$ functions indiscriminately, we have 
\[
C^{-1}(\l)= \l^{-2} P_0 V + \l^{m-6}\log \l\,  \tilde F_{10}(\l) P_0 V 
- \l^{m-4}\log \l\,  \tilde F_{20}(\l)P_0V + E_{0}(\l) .
\]
Since $F_{20}(\l)$ is of class $C^{\frac{m+2}2}$ as mentioned above and $m-4 \geq \frac{m}2$ 
if $m \geq 8$, $ \l^{m-4}\log \l\, \tilde F_{20}(\l)P_0V$ satisfies the property \refeq(e0op). 
If we expand $A(\l)= A(0)+ \l A'(0) + A_2(\l)$ with $A_2(\l)=A(\l)-A(0)-\l \, A'(0)$ in 
\[
\l^{m-6}\log \l\, \tilde F_{10}(\l) P_0 V =\l^{m-6}\log \l\,P_0 V A(\l)V P_0 V, 
\]
then, $\l^{m-6}\log \l\,P_0 V A_2(\l) V P_0 V $ satisfies the property \refeq(e0op).  
Since $A(0)= c_m 1 \otimes 1$ and $A'(0)=0$, the lemma follows.  
\end{proof}

\reflm(3-2-2) and the following lemma complete the proof of \refprp(sing-exp) for $m \geq 8$. 
We use the following short hand notation.  
\[\br{c}
R_l(\l)= C^{-1}(\l) L_{10}(\l)L_{00}^{-1}(\l), \quad 
R_r(\l)=L_{00}^{-1}(\l) L_{01}(\l) C^{-1}(\l) \\
R_c(\l)=L_{00}^{-1}(\l) L_{01}(\l)C^{-1}(\l)L_{10}(\l) L_{00}^{-1}(\l)
\er
\]
\begin{lemma} For sufficiently small $\l_0>0$ the following properties are satisfied:   

\noindent 
{\rm (1)} For $\s\leq \frac{m-2}2$, $R_l(\l)$ is a $\Xg_\s \equiv \Bb(\Hg_{(\s+2-\d)_+}, \Ng)$-valued 
function of class $C^{\s}$ on $(-\l_0,\l_0)$; it is of class $C^{\s}$ for $\l\not=0$ for 
$\frac{m}2 \leq \s \leq \frac{m+2}2$  and 
\[
\|R_l^{(\frac{m}2)}(\l)\|_{\Xg_{\frac{m}2}} + 
|\l|\|R_l^{(\frac{m+2}2)}(\l)\|_{\Xg_{\frac{m+2}2}} \leq C\la \log \l \ra. 
\]

\noindent 
{\rm (2)} For $\s\leq \frac{m-2}2$, $R_r(\l)$ is a $\Yg_\s\equiv \Bb(\Ng, \Hg_{-(\s+2)_+})$-valued function 
of class $C^{\s}$; it is of class $C^{\s}$ for $\l\not=0$ 
for $\frac{m}2\leq \s \leq  \frac{m+2}2$ and 
\[
\|R_r^{(\frac{m}2)}(\l)\|_{\Yg_{\frac{m}2}} + 
|\l|\|R_r^{(\frac{m+2}2)}(\l)\|_{\Yg_{\frac{m+2}2}} \leq C\la \log \l \ra.
\]

\noindent 
{\rm (3)} For $\s\leq \frac{m-2}2$, $R_c(\l)$ is a $\Zg_\s\equiv \Bb(\Hg_{(\s+2-\d)_+}, \Hg_{-(\s+2)_+})$-valued function 
of class $C^{\s}$. Moreover it is of class $C^{\s}$ for 
$\l\not=0$ for $\frac{m}2\leq \s \leq \frac{m+2}2$  
and 
\[
\|R_c^{(\frac{m}2)}(\l)\|_{\Zg_{\frac{m}2}} + 
|\l|\|R_c^{(\frac{m+2}2)}(\l)\|_{\Zg_{\frac{m+2}2}}  \leq C\la\log \l\ra.
\]
\end{lemma}
\begin{proof} We have $\d-\frac{m}2>\frac{m+2}2$ and 
\[
P_0 V L_{01}(\l)V\Qb L_{00}^{-1}(\l)= \l^2 P_0 V J_2(\l)V\Qb(\Qb+ \tilde L(\l)). 
\]
\refprp(even-exp) and \refcor(Ar-c) imply that $T(\l)=P_0 V J_2(\l)V\Qb$ satisfies 
the property of $R_l(\l)$ of the lemma 
(recall that $J_2(\l)$ contains $\l^{m-4}\log\l\, A(\l)$ and $m-4\geq \frac{m}{2}$). 
Since 
\[
\l^2 C^{-1}(\l)= P_0 V + c_m \l^{m-4}\log \l \, \ph \otimes V\ph + \l^2 P_0 V E_0(\l) P_0V 
\]
satisfies property \refeq(e0op), statement (1) follows. We likewise see that 
\[
\tilde T(\l)= (\Qb + \tilde L(\l))\Qb J_2(\l) VQ 
\]
satisfies the property of $R_r (\l)$ of the lemma. Then statement (2) follows 
since $\l^2 C^{-1}(\l)$ satisfies the property \refeq(e0op). 
Statement (3) is obvious since $C_c(\l)= \tilde T(\l) C_l(\l)$, $C_l(\l)$ satisfies (1) 
and $\tilde T(\l)$ satisfies the property of $R_r (\l)$. 
\end{proof}

\section{Low energy estimate I, generic case} 

In the following two sections, we study the low energy part $W_<$ of the wave operator $W_-$. 
We take and fix $\l_0>0$ arbitrarily if $H$ is generic type, otherwise small enough so 
that \refprp(sing-exp) is satisfied. We take cut-off functions $\Phi$ and $\Psi$ as in the 
introduction and define $W_<$ as in \refeq(low):  
\[
W_< = \Phi(H)\Phi(H_0) -\int^\infty_0 \Phi(H)G(\l)V(G_0(\l)-G_0(-\l))\Phi(H_0)\l \frac{d\l}{\pi i}. 
\]
In this section we study $W_<$ in the case that $H$ is of generic type and 
prove the following proposition. We assume that $V$ satisfies the condition 
\begin{equation}\lbeq(cond-t)
\mbox{ $\Fg(\ax^{2\s}V) \in L^{m_\ast}(\R^m)$ and $|V(x)|\leq C \ax^{-\d}$ 
for some $\d>m+2$}. 
\end{equation}

\begin{proposition}\lbprp(low-r) Let $m \geq 6$ be even and let $V$ satisfy \refeq(cond-t). 
Suppose that $H$ is of generic type. 
Then $W_<$ is bounded in $L^p(\R^m)$ for all $1\leq p \leq \infty$. 
\end{proposition}
The integral kernels $\Phi_0(x,y)$ and $\Phi(x,y)$ of $\Phi(H_0)$ and 
$\Phi(H)$ respectively are continuous and bounded by $C_N \la x-y \ra^{-N}$ for any $N$ 
(\cite{Y-d4}) and a fortiori $\Phi(H)$ and $\Phi(H_0)$ are bounded in $L^p$ for all 
$1\leq p \leq \infty$. Hence, we have only to discuss the operator defined by 
the integral 
\begin{equation}\lbeq(low-1op)
\int_0 ^\infty \Phi(H)G(\l)V(G_0(\l)-G_0(-\l))\l \Phi(H_0) d\l. 
\end{equation}
By iterating the resolvent equation, 
we have, with  $L(\l)=(1+G_0(\l)V)^{-1} -1$ as in \reflm(basic), 
\[
G(\l)= G_0(\l) -G_0(\l)VG_0(\l) - G_0(\l)VL(\l) G_0(\l).  
\]
We substitute this for $G(\l)$ in \refeq(low-1op).  The first two terms produce the Born 
approximation $\Phi(H)(\W_1-\W_2)\Phi(H_0)$, which is bounded in $L^p(\R^m)$ for all $1\leq p \leq \infty$ 
by virtue of \reflm(wr1). The last term produces  
\begin{equation}\lbeq(low-2op)
\int_0 ^\infty \Phi(H)G_0(\l)VL(\l)G_0(\l)V(G_0(\l)-G_0(-\l))\l \Phi(H_0) \tilde \Phi(\l)d\l. 
\end{equation}
where we have introduced another cut off function $\tilde \Phi(\l)\in C_0^\infty(\R)$ which satisfies  
\[
\tilde \Phi(\l)\Phi(\l^2)=\Phi(\l^2), \ \mbox{and $\tilde \Phi(\l)=0$ for $|\l |\geq \l^2_0$}.
\]
We prove that \refeq(low-2op) is bounded in $L^p(\R^m)$ for all $1\leq p \leq \infty$ 
in the following slightly more general setting for a later purpose. 
Note that, by virtue of  \reflm(basic), $VL(\l)G_0(\l)V$ satisfies property the $(K)_\r$ 
with $\r=\d-1-\ep$ for any $\ep>0$ and, by choosing $\ep>0$ small enough, 
$\r$ can be taken larger than $m-1$ as $\d>m+2$. 

\begin{proposition}\lbprp(2) Let $m \geq 6$. 
Suppose $K(\l)$ satisfies property $(K)_{\r}$ for some $\r>m+1$. 
Let $\Phi, \tilde \Phi \in C_0^\infty(\R)$ be as above and $\W$ be defined by 
\begin{equation}\lbeq(Omdef)
\W = \int_0^\infty \Phi(H)G_0 (\l)K(\l) (G_0(\l)-G_0(-\l))\Phi(H_0)\l \tilde\Phi(\l)d\l. 
\end{equation}
Then, $\W$ is an integral operator with admissible integral kernel.
\end{proposition} 

We prove \refprp(2) by using a series of lemma. We first remark that 
\refeq(Omdef) may be considered as Riemann integral of 
$\Bb(\Hg_\c, \Hg_{-\c})$ valued continuos function and that $\W$ may be extended 
to a bounded operator in $\Hg$. Indeed, since the multiplication by $\ax^{-\c}$, 
$\c>1$ is $H_0$-smooth in the sense of Kato (\cite{KaS1}), we have  
\begin{eqnarray*}
|\la \W f, g \ra |\leq \sup_{\l \in \R}\|\tilde \Phi(\l) \ax^{\c}K(\l)\ax^{\c}\|_{\Bb(\Hg)} 
\|\ax^{-\c}G_0(\l)\Phi(H_0) f\|_{L^2(\R;\Hg, \l d\l)} \\
\times \|\ax^{-\c}G_0(\l)\Phi(H) g\|_{L^2(\R;\Hg, \l d\l)}\leq C \|f\|\|g\|. 
\end{eqnarray*}
Define $\W(x,y)=\W_+(x,y)-\W_-(x,y)$, where 
\begin{equation}\lbeq(ker-1)
\W_\pm(x,y)=\int^\infty_{0} \la K(\l)G_0(\pm\l)\Phi_0(\cdot, y), G_0(-\l)\Phi(\cdot, x) \ra \l \tilde\Phi(\l) d\l. 
\end{equation}
\begin{lemma}  The function $\W(x,y)$ is continuous and $\W$ is an integral operator 
with the integral kernel $\W(x,y)$.
\end{lemma} 
\begin{proof} For $\c>1$, $x\mapsto\Phi(\cdot, x)$ and $y \mapsto \Phi_0(\cdot,y)$ are 
$\Hg_{\c}$-valued continuous, and $\W_\pm(x,y)$ are continuous functions of $(x,y)$. 
For $f,g \in C_0^\infty(\R^m)$, 
$\Phi(H_0)f (\cdot) = \int \Phi(\cdot, y) f(y) dx$ and 
$\Phi(H)g (\cdot) = \int \Phi(\cdot, x) g(x) dx$ 
converge as Riemann integrals in $\Hg_\c$. It follows by Fubini's theorem that  
\[
\la \W f, g \ra
= \sum_{\pm}\pm \int \la K(\l)G_{0}(\pm\l)\Phi(H_0)f, G_{0}(-\l)\Phi(H)g \ra \tilde \Phi(\l) d\l 
\]
is equal to ${\ds \int \W(x,y)f(y)\overline{g(x)}dydx}$. The lemma follows. \end{proof}

Introducing the notation  
\begin{eqnarray} 
& G_{0l}(\l, \cdot,y )= e^{-i\l|y|}G_0(\l)\Phi_0(\cdot, y), \   
G_{0r}(\l, \cdot, x)= e^{-i\l|x|}G_0(\l)\Phi(\cdot, x) \lbeq(tG), \\
& F_{\pm }(\l,x,y)= \la K(\l)G_{0l}(\pm \l, \cdot,y), G_{0r}(\l,\cdot,x)\ra \tilde \Phi(\l) , 
\lbeq(Fr) 
\end{eqnarray} 
we write \refeq(ker-1) in the form 
\begin{equation}\lbeq(Wr)
\W_\pm (x,y) = \int^\infty_{0} e^{i\l(|x|\pm |y|)} F_{\pm} (\l,x,y)\l d\l .
\end{equation}

\begin{lemma}\lblm(r-e) Let $\c>\frac12$ and $\b\geq 0$ be an integer and let $x, y \in \R^m$.  
Then: 

\noindent
{\rm (1)}  As $\Hg$ valued functions of $\l$, $\la \cdot \ra^{-\b-\c}G_{0l}(\l,\cdot, y)$ and 
$\la \cdot \ra^{-\b-\c}G_{0r}(\l,\cdot, x)$ are of class $C^\b(\R)$ for 
$0 \leq \b \leq \frac{m+2}2$ if $m\geq 8$; they are of class $C^\b(\R)$ for 
$\b \leq \frac{m}2$ and of class $C^{\b}_\ast(\R)$ for $\b=\frac{m+2}2$ if $m=6$.  

\noindent 
{\rm (2)}  For $0 \leq \b \leq \frac{m+2}2$ and $\ep>0$, we have  
\begin{equation}\lbeq(low-1)
\|\la \cdot \ra^{-\b-\ep-\frac{m}2}G_{0l}^{(\b)}(\l,\cdot, y)\| \leq C 
\l^{\min\left(0, \frac{m-3}2-\b\right)}\ay^{-\frac{m-1}2}, \quad 0<|\l| <\l_0. 
\end{equation}

\noindent 
{\rm (3)} For $0 \leq \b \leq \frac{m-2}2$, we have at $\l=0$: 
\begin{equation}\lbeq(low-1l)
|G_{0l}^{(\b)}(0, z, y)| \leq C\sum_{\b_1+\b_2=\b} \frac{\az^{\b_1}}{\la z- y \ra^{m-2-\b_2}}. 
\end{equation}

\noindent 
{\rm (4)} With obvious modifications $G_{0r}(\l,z,x)$ satisfies \refeq(low-1) and \refeq(low-1l).  

\end{lemma} 
\begin{proof} Statement (1) follows from the LAP, viz. from \reflm(Ag) and 
\refprp(even-exp). 
The following proof will be a bit more than necessary for the lemma for a later purpose. 
By Leibniz' rule $G^{(\b)}_{0l}(\l, z,y)$ is a linear combination of 
\begin{eqnarray}
& \ds K_{\b_1\b_2}(\l,z,y)= \int_{\R^m} \frac{(i\p(w,z,y))^{\b_1} e^{i\l\p(w,z,y)}}{|z-w|^{m-2-\b_2}} 
H_{\b_2}(\l|w-z|)\Phi_0(w,y)dw, \notag \\
& \ds H_{\b}(s) = \int_0^\infty e^{-t}t^{\frac{m-3}2}\Big(s+\frac{it}2\Big)^{\frac{m-3}2-\b}dt  \lbeq(rkd) 
\end{eqnarray}
over the indices $\b_1, \b_2$ such that $\b_1+\b_2=\b$. Here 
$\p(w,z,y)\equiv |w-z|-|y|$ satisfies $|\p(w,z,y)|\leq |w-y|+|z|$. 
It follows when $\b \leq \frac{m-2}2$ that 
\[
|K_{\b_1\b_2}(0,z,y)|\leq C_{\b_1\b_2 N} \int_{\R^m} \frac{\az^{\b_1} \la w-y \ra^{-N}}{|z-w|^{m-2-\b_2}} dw
\leq \frac{C_{\b_1\b_2}\az^{\b_1}}{|z-y|^{m-2-\b_2}}  
\]
for any $N$ and we obtain \refeq(low-1l). 

If $\b_2 \leq \frac{m-3}2$, we have  
$\left|s+\frac{it}2\right|^{\frac{m-3}2-\b_2}\leq C (s^{\frac{m-3}2-\b_2} +|t|^{\frac{m-3}2-\b_2})$ 
and $| H_{\b_2}(s)| \leq C (s +1)^{\frac{m-3}2-\b_2} $. 
It follows that 
\begin{align} 
|K_{\b_1\b_2}(\l,z,y)|\leq & 
C \int_{\R^m} \frac{|z|^{\b_1}+ |w-y|^{\b_1}}{|z-w|^{m-2-\b_2}} 
(\l|z-w|+1)^{\frac{m-3}2-\b_2}|\Phi_0(w,y)|dw \notag  \\
\leq & C \az^{\b_1} \la z-y \ra^{-\frac{m-1}2} \ \ \mbox{for all} \ \ |\l| \leq \l_0. 
\lbeq(Kab)
\end{align} 
Hence, if $\b \leq \frac{m-3}2$,  $\la \cdot \ra^{-\b-\c}G_{0l}(\l,\cdot, y)$ is 
$\Hg$ valued function of class $C^\b$ and the estimate \refeq(low-1) follows.  

If $ \frac{m-2}2\leq \b_2 \leq \frac{m+2}2$, 
$|s+\frac{it}2|^{\frac{m-3}2-\b_2}\leq C \min\{s^{\frac{m-3}2-\b_2}$,  $|t|^{\frac{m-3}2-\b_2}\}$  
and 
\[
\begin{split}
|H_{\b_2}(s)| \leq C\left(s^{\frac{m-3}2-\b_2}\int_0^s e^{-t} t^{\frac{m-3}2}dt 
+ \int_s^\infty e^{-t}t^{m-3-\b_2}dt \right) \\
\leq C \left\{\br{l} 
\min(s^{\frac{m-3}2-\b_2}, |\log (1/s)|), \ \mbox{if $\b_2=m-2$}, \\
\min(s^{\frac{m-3}2-\b_2}, 1), \ \mbox{if otherwise.} \er \right.
\end{split}
\]
The smoothness property of $\la \cdot \ra^{-\b-\c}G_{0l}(\l,\cdot, y)$ and 
$\la \cdot \ra^{-\b-\c}G_{0r}(\l,\cdot, x)$ for $ \frac{m-2}2\leq \b \leq \frac{m+2}2$ 
follows from this estimate and Lebesgue's dominated convegence theorem.  
We have $\b_2=m-2$ if and only if $(m,\b_2)=(6, 4)$. It follows that, if $(m,\b_2)=(6, 4)$,   
\begin{equation}\lbeq(low-2a) 
|K_{\b_1\b_2}(\l,z,y) |\leq C \az^{\b_1}\min\left(
\frac{\l^{\frac{m-3}2-\b_2}}{\la z- y\ra^{\frac{m-1}2}}, 
\la \log\l \ra + \log \la z-y \ra  \right) 
\end{equation} 
and if otherwise 
\begin{equation}\lbeq(low-2b) 
|K_{\b_1\b_2}(\l,z,y) |\leq C \az^{\b_1}\min\left(
\frac{\l^{\frac{m-3}2-\b_2}}{\la z- y\ra^{\frac{m-1}2}}, 
\frac{1}{\la z- y\ra^{m-2-\b_2}} \right) .
\end{equation}
Estimate \refeq(low-1) follows from the first estimates of 
\refeq(low-2a) and \refeq(low-2b). 
(We shall use the second estimates shortly.) The proof for $G_{0r}(\l,\cdot, x)$ is similar 
and we omit it.  
\end{proof} 

By virtue of \reflm(r-e), $F_{\pm }(\l,x,y)$ is of class $C^{\frac{m+2}2}_\ast$ on 
$\R$ with respect to $\l$ for every fixed $x,y \in \R$ and it satisfies 
$|\W(x,y)|\leq C \ax^{-\frac{m-1}2} \ay^{-\frac{m-1}2}$. 
It is then easy to check that 
\begin{equation} 
\sup_{x \in \R^m} \int_{||x|-|y||<1} |\W(x,y)|dy +
\sup_{y \in \R^m} \int_{||x|-|y||<1} |\W(x,y)|dx < \infty. 
\end{equation} 
Thus, we hereafter consider $\W(x,y)$ only on the domain $||x|- |y|| >1$. 
We apply integration by parts $k=(m+2)/2$ times to  
\begin{equation}\lbeq(Wrd)
\W_{\pm}(x,y) = \frac{1}{(i(|x|\pm |y|))^k} 
\int^\infty_{0} \Big(\frac{\pa}{\pa \l}\Big)^k e^{i\l(|x|\pm |y|)} \cdot F_\pm (\l,x,y)\l d\l. 
\end{equation}
The result is that $\W(x,y)$ is the sum of 
\begin{eqnarray}
I_1(x,y)= \sum_{\pm} \frac{\pm i^{\frac{m+2}2} }{(|x|\pm |y|)^{\frac{m+2}2}} \int^\infty_{0} 
e^{i\l(|x|\pm |y|)}F_\pm^{({\frac{m+2}2})} (\l,x,y)\l d\l , \lbeq(int-term1) \\
I_2(x,y)=\sum_{\pm}\frac{\pm  (m+2)i^{\frac{m+2}2}}{2(|x|\pm |y|)^{\frac{m+2}2}} \int^\infty_{0} 
e^{i\l(|x|\pm |y|)} F_\pm^{({\frac{m}2})} (\l,x,y)d\l , \lbeq(int-term2)
\end{eqnarray}
and the boundary terms: 
\begin{equation}\lbeq(fab)
B(x,y)=\sum_{j=0}^{\frac{m-2}2} i^j (j+1) \left(\frac{ F_+^{(j)} (0,x,y)}{(|x|+|y|)^{j+2}}
-\frac{ F_-^{(j)} (0,x,y)}{(|x|-|y|)^{j+2}}\right).
\end{equation}

\begin{lemma}\lblm(bdry) The function $B(x,y)$ of \refeq(fab) is an admissible integral kernel.  
\end{lemma} 
\begin{proof}
Derivatives $F_\pm^{(j)} (0,x,y)$ are linear combinations over $\a+\b+\c=j$ of 
\begin{equation}\lbeq(fab-1)
(\pm 1)^\b \la K^{(\a)} (0)G^{(\b)}_{0l}(0, \cdot,y), G_{0r}^{(\c)}(0,\cdot,x)\ra 
\end{equation}
with coefficients $ (-1)^{\c} j! /\a! \b!\c!$. In \refeq(fab-1), we have for arbitrarily small  $\ep>0$
\begin{equation}\lbeq(b-b)
\br{l}
\| \az^{1+j-m-\ep-\b} G^{(\b)}_{0l}(0, z,y)\| \leq  C 
\left\{\br{l}  \ay^{j+2-m} \ \mbox{if} \ \b=j, \\ \ay^{j+1-m} \ \mbox{if otherwise}. \er \right.\\ 
\|\az^{1+j-m-\ep-\c} G_{0r}^{(\c)}(0, z,x)\| \leq C 
\left\{\br{l}  \ax^{j+2-m} \ \mbox{if} \ \c=j, \\ \ax^{j+1-m} \ \mbox{if otherwise}. \er \right. 
\er
\end{equation}
This can be seen as follows. By virtue of \refeq(low-1l) we have  
\[
\left| \frac{G^{(\b)}_{0l}(0, z,y)}{\az^{m-j-1+\b+\ep}}\right| \leq 
\sum_{\b_2=0}^{\b} \frac{C}{\az^{m-\b_1-1+\b_2+\ep}\la z-y \ra^{m-2-\b_2}}
\]
and the like for $G_{0r}^{(\c)}(0,\cdot,x)$. Since 
$(m-j-1+\b_2+\ep)+ (m-2-\b_2) > m $, we have either $m-j-1+\b_2 +\ep > \frac{m}2$ or 
$m-2-\b_2> \frac{m}2$. Hence  
\begin{equation}\lbeq(b-sec)
\|\az^{j+1-m-\b_2-\ep}\la z-y \ra^{2+\b_2-m} \| 
\leq C \left\{\br{l} \ay^{2+j-m}, \quad \mbox{if}\ \b=j \\
\ay^{1+j-m},  \quad \mbox{if otherwise}. \er \right. 
\end{equation} 
Since $\r>m+1$, we have for $0<\ep \leq 1$ that 
\[
\max(m-1-(j-\b)+\ep, m-1-(j-\c)+\ep)< \r-\a 
\]
and $\la \cdot \ra^{m-1-(j-\c)+\ep} K^{(\a)} (0) \la \cdot \ra^{m-1-(j-\b)+\ep} \in \Bb(\Hg)$ 
by property $(K)_\r$. Thus, \refeq(fab-1) is bounded in modulus by a constant times 
\begin{equation} 
Y_{\a\b\c}(x,y)= \left\{\br{l} \ax^{1+j-m}\ay^{1+j-m} \ \ \mbox{if} \ \ \b \not=0, j, \\
\ax^{1+j-m}\ay^{2+j-m} \ \ \mbox{if} \ \ \b = j, \\ 
\ax^{2+j-m}\ay^{1+j-m} \ \ \mbox{if} \ \ \b = 0 . \er \right. 
\end{equation} 
It follows that the $j$-th summand of \refeq(fab) is bounded by  
\begin{equation}\lbeq(e-s1)
C\sum_{\a+\b+\c=j}  \left|\frac1{(|x|+|y|)^{j+2}}- \frac{(-1)^\b }{(|x|-|y|)^{j+2}}\right| Y_{\a\b\c}(x,y)
\end{equation}
and it is an easy exercise to prove that this is an admissible integral kernel. 
(Indeed, summands with $\b\not=0, j$ are admissible by virtue of Lemma 3.6 of [I];  those  
with $\b=0$ or $\b=j$ are the same as (3.21) of [I] and the argument in [I] following 
(3.21) applies also for $\b \leq \frac{m-2}2$ or $\c \leq \frac{m-2}2$ if $m \geq 4$. )  
This completes the proof.   
\end{proof} 

\begin{lemma}\lblm(4-1) The integral kernel $I_2(x,y)$ defined by \refeq(int-term2) is admissible. 
\end{lemma}
\begin{proof} By Leibniz' rule $F_\pm^{(\frac{m}2)} (\l,x,y)$ is a linear combination  of 
\begin{equation}\lbeq(Lei)
X_{\xi,\pm }(\l,x,y)= (\pm 1)^{\b} 
\la K^{(\a)} (\l)G^{(\b)}_{0l}(\pm\l, \cdot,y), G_{0r}^{(\c)}(-\l,\cdot,x)\ra \tilde \Phi^{(\eta)}(\l)
\end{equation}
with $\pm$ independent coefficients $(-1)^\c (\frac{m}2)!/\a!\b!\c!\eta!$ over multi-indices 
$\xi=(\a,\b,\c,\eta)$ of length $|\xi|=\frac{m}2$ . Thus, if we define 
\begin{equation}\lbeq(4-1)
\W_{\xi,\pm }^{(1)}(x,y) = \int_0^\infty e^{i\l(|x|\pm |y|)} X_{\xi,\pm}(\l,x,y) d\l, 
\end{equation} 
then $I_2(x,y)$ is a linear combination over $\xi=(\a,\b,\c,\eta)$ with $|\xi|=\frac{m}2$ of 
\begin{equation}\lbeq(4-2)
I_{2,\xi}(x,y)=\left( \frac{\W_{\xi,+}^{(1)}(x,y)}{(|x|+ |y)^{\frac{m+2}2}} -
\frac{\W_{\xi,-}^{(1)}(x,y)}{(|x| -|y)^{\frac{m+2}2}}  \right).  
\end{equation} 
We estimate $I_{2,\xi}(x,y)$ for various cases of $\xi$ separately.  

\noindent {\bf (1) The case $\xi\not=(0,\frac{m}2, 0,0), (0,0, \frac{m}2,0)$.} \ 
In view of property $(K)_\r$, we estimate 
\begin{eqnarray}
|X_{\xi\pm}(\l,x,y)| \leq C \|\ax^{\r-\a}K^{(\a)}(\l)\ax^{\r-\a}\|_{\Bb(\Hg)} \hspace{3cm}\notag \\
\times \| \la \cdot \ra^{-(\r-\a)}G^{(\b)}_{0l}(\pm\l, \cdot,y)\|  
\| \la \cdot \ra^{-(\r-\a)}G^{(\c)}_{0l}(\pm\l, \cdot,x)\|.  \lbeq(star)
\end{eqnarray}
Since $\r>m+1$ and $\a+\b+\c \leq \frac{m}2$, we have for $0<\ep<1$
\begin{equation}\lbeq(room)
\max(\b+\tfrac{m}2+\ep, \c+ \tfrac{m}2+\ep) < \r-\a. 
\end{equation}
Hence, by virtue of \refeq(low-1), we have that 
\begin{equation}\lbeq(x)
|X_{\xi\pm}(\l,x,y)| 
\leq C\left\{\br{l}
\ax^{-\frac{m-1}2} \ay^{-\frac{m-1}2}, \quad \mbox{if both}\ \b, \c \leq \frac{m-3}2, \\  
\l^{-\frac12} \ax^{-\frac{m-1}2} \ay^{-\frac{m-1}2}, \quad \mbox{if one of }\ \b, \c =\frac{m-2}2, 
\er \right.
\end{equation}
where we have to modify the first line on the right by multiplying by 
$\la \log \l\ra^N $ when $\xi=(\frac{m}2,0,0,0)$. Thus after integrating 
with respect to $\l$ we obtain for $||x|-|y||>1$ that 
\begin{equation}\lbeq(4-3)
\left|\frac{\W_{\xi\pm }^{(1)}(x,y)}{\la |x|\pm |y|\ra^{\frac{m+2}2}} \right| \leq 
\frac{C}{\la |x|\pm |y| \ra^{\frac{m+2}2}\ax^{\frac{m-1}2} \ay^{\frac{m-1}2}}.  
\end{equation}
It follows that $I_{2,\xi}$ are admissible for these $\xi$'s. 

\noindent 
{\bf (2) The case $\xi=(0,\frac{m}2,0,0)$.}  Recall the definition \refeq(rkd) of 
$K_{\b_1\b_2}(\l,z,y)$. We substitute 
$\sum_{\b_1+\b_2=\frac{m}2} C_{\b_1\b_2} K_{\b_1\b_2}(\l, z, y)$ 
for $G_{0l}^{(\frac{m}2)}(\l,z,y)$ in 
\[
X_{\xi,\pm }(\l,x,y)= (\pm 1)^{\frac{m}{2}} 
\la K (\l)G^{(\frac{m}2)}_{0l}(\pm\l, \cdot,y), G_{0r}(-\l,\cdot,x)\ra \tilde \Phi(\l). 
\]
If $(\b_1,\b_2)\not=(0,\frac{m}2)$, we have 
$\b_2 \leq \frac{m-2}2$ and the first estimate of \refeq(low-2b) implies 
\begin{equation}\lbeq(kab)
\|\la \cdot \ra^{-(\b+\frac{m}2+\ep)} K_{\b_1\b_2}(\l,\cdot, y)\| \leq 
C \l^{-\frac12} \ay^{-\frac{m-1}2}.
\end{equation} 
Since $\b+\frac{m}2+\ep < \r$ for $0<\ep\leq 1$, it follows via the argument 
similar to the one used for \refeq(star) that, for $(\b_1,\b_2)\not=(0,\frac{m}2)$,  
\[
|\la K(\l) K_{\b_1\b_2}(\pm \l, \cdot, y),  G_{0r}(-\l, \cdot, x)\ra|
\leq C  \l^{-\frac12} \ax^{-\frac{m-1}2} \ay^{-\frac{m-1}2}. 
\]
This implies that all members under the summation sign of 
\[
\sum_{\b_1+\b_2=\frac{m}2} \frac{(\pm 1)^{\frac{m}2}C_{\b_1\b_2}}{(|x|\pm |y|)^{\frac{m+2}2}}  
\int_0^\infty e^{i\l(|x|\pm |y|)} 
\la K(\l) K_{\b_1\b_2}(\pm \l, \cdot, y),  G_{0r}(-\l, \cdot, x)\ra d\l, 
\]
are admissible except those with $(\b_1,\b_2)=(0,\frac{m}2)$. 
We are thus left with 
\[
I_{2r}=\sum_{\pm} \frac{\pm (\pm 1)^{\frac{m}2}}{(|x|\pm |y|)^{\frac{m+2}2}}  
\int_0^\infty e^{i\l(|x|\pm |y|)} 
\la K(\l) K_{0\frac{m}2}(\pm \l, \cdot, y),  G_{0r}(-\l, \cdot, x)\ra d\l.
\]
For proving that $I_{2r}(x,y)$ is admissible, we restore the factors $e^{i\l(|x|\pm |y|)}$ 
to the original position. Thus defining $\tilde G_{0r}(\l,\cdot,x)$ and $G_{\frac{m}2}(\l, z,y)$
by 
\[
\tilde G_{0r}(\l,\cdot,x)= G_0(\l) \Phi(\cdot, x)= e^{-i\l|x|}  G_{0r}(\l,\cdot,x)
\]
and $G_{\frac{m}2}(\l, z,y)= e^{i\l|y|}K_{0\frac{m}2}(\l,z,y)$ respectively, we rewrite, 
ignoring the unimportant constant, $I_{2r}$ in the form 
\begin{equation}\lbeq(4-6)
\sum_{\pm}\frac{(\pm 1)^{\frac{m+2}2}}{(|x|\pm |y|)^{\frac{m+2}2}} \int^\infty_{0} 
\la K(\l) G_{\frac{m}2}(\pm\l,\cdot,y), \tilde G_{0r}(-\l,\cdot,x) \ra \tilde \Phi(\l) d\l.
\end{equation} 
More explicitly $G_{\frac{m}2}(\l, z,y)$ is given by  
\begin{align}
\int_{\R^m} \frac{e^{i\l|z-w|}}{|z-w|^{\frac{m}2-2}}
\left(\int_0^\infty e^{-t}t^{\frac{m-3}2}\Big(\l|z-w|+\frac{it}2\Big)^{-\frac32}dt \right) 
\Phi_0(w,y)dw.  \lbeq(rkd-1) 
\end{align}

\begin{lemma} \lblm(p-pa) \ben \item[{\rm (1)}] There exists a constant $C>0$ such that 
\begin{eqnarray}
|\tilde  G_{0r}(\l,z,x) - \tilde  G_{0r}(0,z,x)| 
\leq C |\l| \la z- x \ra^{-\frac{m-1}2}, \lbeq(g0-t) \\
|G_{\frac{m}2}(\l,z,y)| \leq C \min\left(\l^{-\frac32}\la z- y \ra^{-\frac{m-1}2}, 
\la z- y \ra^{-\frac{m-4}2}\right). \lbeq(res-d)
\end{eqnarray} 
\item[{\rm (2)}] For a fixed $(z,y)$, $G_{\frac{m}2}(\l,z,y)$ is continuous with respect to $\l \in \R$; 
for a fixed $y$ and, for $\c>\frac32$, $G_{\frac{m}2}(\l,\cdot,y)$ is $\Hg_{-\c}$ valued  
integrable on $\R$ and is continuous for $\l \not=0$. 
\item[
{\rm (3)}] The integrand of \refeq(rkd-1) is integrable with respect to $(t,\l, w)$. 
\een
\end{lemma} 
\begin{proof} (1) Write the convolution kernel of $G_0(\l)-G_0(0)$ 
in the form 
\[ 
\frac{C_m e^{i\l|x|}}{|x|^{m-2}}
\int^\infty_0 e^{-t}t^{\frac{m-3}2}\left\{\Big(\frac{t}2-i\l|x|\Big)^{\frac{m-3}2}-\Big(\frac{t}2\Big)^{\frac{m-3}2}
\right\}dt+ \frac{C_m'(e^{i\l|x|}-1)}{|x|^{m-2}}
\]
and estimate it by $C |\l| (|x|^{3-m}\ax^{\frac{m-5}2} + |x|^{3-m})$ for $|\l|\leq \l_0$. This yields 
\refeq(g0-t) since $\frac{m-1}2\leq m-3$ for $m\geq 6$. Estimate \refeq(res-d) is contained in \refeq(low-2b). 

\noindent (2) The continuity of $\l \mapsto G_{\frac{m}2}(\l,z,y)$ is obvious by 
Lebesgue's dominated convergence theorem. Then the second statement follow 
from the estimate \refeq(res-d) which also implies 
$|G_{\frac{m}2}(\l,z,y)|\leq C \l^{-\frac12}\la z- y \ra^{-\frac{m-3}2}$ by interpolation. 

\noindent (3) Integrating with respect to $\l$ first, we have 
\begin{align*}
\int_{\R^m} \int_0^\infty  \int_{\R} \frac{e^{-t}t^{\frac{m-3}2}}
{|z-w|^{\frac{m}2-2}(|\l||z-w|+t)^{\frac32}} |\Phi_0(w,y)|dw dt d\l \\
= C \int_{\R^m}\frac{|\Phi_0(w,y)|}{|z-w|^{\frac{m-2}2}} dw \cdot 
\int_0^\infty e^{-t}t^{\frac{m-6}2}dt < \infty. 
\end{align*}
\end{proof} 

We first show that it is sufficient to show that \refeq(4-6) is admissible if 
$K(\l)$,  
$\tilde  G_{0r}(-\l,\cdot,x)$ and $\tilde \Phi(\l)$  are replaced by $K(0)$, $\tilde  G_{0r}(0,\cdot,x)$ 
and the constant function $1$ respectively. 

\noindent 
(i) Let $I_{2r}^{(1)}(x,y)$ be defined by \refeq(4-6) with $K(0)$ in place of 
$K(\l)$. Via Taylor's formula, $K(\l)-K(0)= \l \int_0^1 K'(\th\l)d\th$ and 
property $(K)_\r$ implies 
that 
\[
\left\| \ax^{\r-1} \Big(\int_0^1 K'(\th\l)d\th\Big) \ax^{\r-1} \right\|_{\Bb(\Hg)} \leq C .
\]
Hence, using \refeq(low-1) for $\tilde G_{0r}(-\l,\cdot,x)$ and \refeq(res-d), we obtain 
\begin{multline*}
|\la (K(\l)-K(0)) G_{\frac{m}2}(\pm\l,\cdot,y), \tilde G_{0r}(-\l,\cdot,x) \ra| \\
\leq C \|\la \cdot \ra^{-\r+1} \l G_{\frac{m}2}(\l,\cdot,y)\|
\|\la \cdot \ra^{-\r+1} \tilde G_{0r}(\l,\cdot,x)\|
\leq C\l^{-\frac12} \ax^{-\frac{m-1}2}\ay^{-\frac{m-1}2}. 
\end{multline*}
It follows after integration with respect to $\l$ that  
\begin{equation}\lbeq(resi-1)
|I_{2r}(x,y)-I_{2r}^{(1)}(x,y)| \leq 
\frac{C}{\ax^{\frac{m-1}2} \ay^{\frac{m-1}2} \la |x|-|y| \ra^{\frac{m+2}2}}
\end{equation}
and $I_{2r}(x,y)-I_{2r}^{(1)}(x,y)$ is an admissible kernel. 

\noindent 
(ii)  We then let $I_{2r}^{(2)}(x,y)$ be defined by \refeq(4-6) with $K(0)$ and 
$\tilde  G_{0r}(0,\cdot,x)$ in places of $K(\l)$ and $\tilde  G_{0r}(\l,\cdot,x)$ 
respectively.   Then, trading the factor $\l$ of \refeq(g0-t) 
for estimating $G_{\frac{m}2}(\l,z,y)$ as above, we obtain  
\begin{equation}\lbeq(resi-2)
|I_{2r}^{(1)}(x,y)-I_{2r}^{(2)}(x,y)| \leq 
\frac{C}{\ax^{\frac{m-1}2} \ay^{\frac{m-1}2} \la |x|-|y| \ra^{\frac{m+2}2}}
\end{equation}
and $I_{2r}^{(1)}(x,y)-I_{2r}^{(2)}(x,y)$ is also admissible. 

\noindent 
(iii) In view of arguments in (i) and (ii), to see that we may further replace 
$\tilde \Phi(\l)$ by $\tilde \Phi(0)=1$ it is sufficient to notice that 
$|G_{\frac{m}2}(\pm\l, z,y)|\leq C\la \l \ra^{-\frac32}\la z- y\ra^{-\frac{m-1}2}$ 
on the support of $1-\Phi(\l)$, which is obvious from \refeq(res-d). 

\vspace{0.2cm}
\noindent 
Thus the problem is reduced to proving that 
\begin{equation}\lbeq(7)
\tilde I_2(x,y)= \sum_{\pm}\frac{(\pm 1)^{\frac{m+2}2}}{(|x|\pm |y|)^{\frac{m+2}2}} \int^\infty_{0} 
\la K(0) G_{\frac{m}2}(\pm\l,\cdot,y), \tilde G_{0r}(0,\cdot,x) \ra d\l
\end{equation} 
is an admissible kernel. Since $G_{\frac{m}2}(\l,\cdot,y)$ satisfies the continuity 
property of \reflm(p-pa), $\ax^{\c}K(0)\ax^{\c} \in \Bg(\Hg)$ and 
$\tilde G_{0r}(0,\cdot,x) \in \Hg_{-\c}$ for some $\c>\frac32$, we may perform 
the integration in \refeq(7) before taking the inner product and write the 
integral in the form   
\[
\Big\la K(0) \int^\infty_{0}G_{{\frac{m}2}}(\pm\l,\cdot,y)d\l,\tilde G_{0r}(\cdot,x)\Big\ra. 
\] 
Here the integral on the right is the Riemann integral of an $\Hg_{-\c}$ 
valued function, however, \reflm(p-pa) (2) implies that we may replace it  
by the standard Riemann integral of the scalar continuous function 
$G_{\frac{m}2}(\pm\l,z,y)$. 
Then, by virtue of \reflm(p-pa) (3), we may integrate \refeq(rkd-1) with respect 
to $\l$ first via Fubini's theorem. For $a >0 $ and $ t>0$ we have by residue theorem 
that 
\[
\int_{0}^\infty e^{i\l a} \Big(\l a+\frac{it}2\Big)^{-\frac32}d\l 
= -\int_{0}^\infty e^{-i\l a} \Big(-\l a+\frac{it}2\Big)^{-\frac32}d\l 
\]
and both sides are bounded in modulus by $Cat^{-\frac12}$. It follows that    
\begin{align}
\int^\infty_{0}G_{\frac{m}2}(\l, z,y)d\l = - \int_0^{\infty}G_{\frac{m}2}(-\l,z,y)d\l
\equiv J(z,y),  \lbeq(Jd) \\
\mbox{and} \quad |J(z,y)| \leq \int_{\R^m} \frac{C |\Phi_0(w,y)|}{|z-w|^{\frac{m-2}2}}dw 
\leq \frac{C}{\la z-y \ra^{\frac{m-2}2}}. \lbeq(Jest)
\end{align}
Thus, we have $|\la K(0)J(\cdot,y),G_{00}(\cdot,x)\ra| \leq C\ax^{-(m-2)} \ay^{-\frac{m-2}2}$ 
and 
\begin{equation}\lbeq(est-i1)
|\tilde I_2(x,y)|\leq C \left|\frac{1}{(|x|+|y|)^{\frac{m+2}2}} -\frac{1}{(|y|-|x|)^{\frac{m+2}2}}\right| 
 \ax^{-(m-2)} \ay^{-\frac{m-2}2} . 
\end{equation}
The right side is the same as the summand in \refeq(e-s1) with $j=\b=\frac{m-2}2$ and $\a=\c=0$ 
and, hence, $\tilde I_2(x,y)$ is admissible.  

\noindent
{\bf (3) The case $\xi=(0,0,\frac{m}2, 0)$ }. 
Define $\tilde G_{\frac{m}2}(\l, z,x)$ by \refeq(rkd-1) with $\Phi(w,x)$ in place of $\Phi_0(w,y)$ 
and   
\[
\tilde J(z,x) = \int^\infty_{0} \tilde G_{\frac{m}2}(-\l,z,x)d\l .
\]
Proceeding virtually in the same way as in the case $\xi=(0,\frac{m}2, 0,0)$, 
we see that it suffices to show that 
\[
I_3(x,y)=\left( \frac1{(|x|+|y|)^{\frac{m+2}2}} -\frac1{(|x|-|y|)^{\frac{m+2}2}}\right)  
\la K(0)  \tilde G_{0l}(0,\cdot,y ) , \tilde J(\cdot,x) \ra 
\]
is admissible. It is obvious from the argument which lead to \refeq(Jest) that  
$|\tilde J(z,x)| \leq C \la z- x \ra^{-\frac{m-2}2}$ and have 
\[
|\la K(0)  \tilde G_{0l}(0,\cdot,y ) , \tilde J(\cdot,x) \ra|\leq C\ax^{-\frac{m-2}2} \ay^{-(m-2)}.
\] 
Thus, $I_3(x,y)$ is bounded by the right of \refeq(est-i1) with $x$ and $y$ interchanged 
and is therefore admissible. This completes the proof of \reflm(4-1). 
\end{proof}

\begin{lemma}\lblm(4-22) The integral kernel $I_4(x,y)$ defined by the integral \refeq(int-term1):  
\[
I_4(x,y)=\sum_\pm \frac{\pm i^{\frac{m+2}2}}{(|x|\pm |y|)^{\frac{m+2}2}} 
\int^\infty_{0} e^{i\l(|x|\pm |y|)} F_\pm^{(\frac{m+2}2)}(\l,x,y) \l d\l 
\] 
is admissible. 
\end{lemma}
\begin{proof} We proceed as in the proof of \reflm(4-1). Let as in \refeq(Lei): 
\[
X_{\xi,\pm }(\l,x,y)= (\pm 1)^{\b} 
\la K^{(\a)} (\l)G^{(\b)}_{0l}(\pm\l, \cdot,y), G_{0r}^{(\c)}(-\l,\cdot,x)\ra \tilde \Phi^{(\eta)}(\l)
\]
for $\xi=(\a,\b,\c,\eta)$ and define  
\begin{equation}\lbeq(4-11)
\W_{\xi\pm }^{(2)}(x,y) = \int_0^\infty e^{i\l(|x|\pm |y|)} X_{\xi,\pm}(\l,x,y)\l d\l. 
\end{equation} 
By Leibniz' formula we have  
\begin{equation}\lbeq(4-21)
I_4(x,y)=  \sum_{|\xi|=\frac{m+2}2}C_\xi 
\left( \frac{\W_{\xi,+}^{(2)}(x,y)}{(|x|+ |y)^{\frac{m+2}2}} 
-\frac{\W_{\xi-}^{(2)}(x,y)}{(|x| -|y)^{\frac{m+2}2}}  \right). 
\end{equation} 
Let first $\xi\not=(0,\frac{m+2}2, 0,0), (0,0, \frac{m+2}2, 0)$. 
Since $\a+ \max(\b+\tfrac{m}2, \c+ \tfrac{m}2) \leq m+1$ and $\r>m+1$, there exists 
$\ep>0$ such that $\max(\b+\tfrac{m}2, \c+ \tfrac{m}2)+\ep<\r$.  
By virtue of \refeq(low-1) and the property $(K)_\r$, we have with this $\ep>0$ that   
\begin{eqnarray}
|\l| |X_{\xi,\pm}(\l,x,y)| \leq |\l| \|\ax^{\r-\a}K^{(\a)}(\l) \ax^{\r-\a}\|
\|\ax^{-(\r-\a)}G_{0l}^{(\b)}(\l,\cdot, y)\|_{\Hg}  \notag \\ 
\times \|\ax^{-(\r-\a)}G_{0r}^{(\c)} (\l,\cdot, x)\|_{\Hg} 
\leq C |\l|^{-\frac12}\la \log \l \ra^{N} \ay^{-(\frac{m-1}2)} \ax^{-(\frac{m-1}2)}. \hspace{0.5cm} \lbeq(cores)
\end{eqnarray}
This implies that for the summands in \refeq(4-21) with these $\xi$ we have   
\begin{equation}\lbeq(4-9) 
\left|\frac{\W_{\xi,\pm}^{(2)}(x,y)}{(|x|\pm |y)^{\frac{m+2}{2}}}\right|\leq 
\frac{C}{\la |x|\pm |y| \ra^{\frac{m+2}2}} \cdot \frac{1}{\ay^{\frac{m-1}2} \ax^{\frac{m-1}2}} 
\end{equation}
and these are therefore admissible. We are left with those either with $\xi=(0,\frac{m+2}2, 0,0)$ or 
$\xi=(0, 0, \frac{m+2}2,0)$ and we shall deal with the former case  
only as the other case may be treated similarly. So let 
$\xi=(0, 0, \frac{m+2}2,0)$ in what follows. We subsitute 
$\sum_{\b_1+\b_2=\frac{m+2}2} C_{\b_1\b_2} K_{\b_1\b_2}(\l, z, y)$ 
for $G_{0l}^{(\frac{m+2}2)}(\l,z,y)$ in \refeq(Lei) and plug this into \refeq(4-11). 
This produces several functions indexed by $\b_1$ and $\b_2$ in the obvious manner 
and, by virtue of \refeq(low-2b) and estimates corresponding to \refeq(cores), they 
are all admissible except the one with index $(\b_1,\b_2)= (0,\frac{m+2}2)$ which is written 
in the form as follows as in \refeq(4-6) after restoring the exponents $e^{i\l(|x|\pm |y|)}$ to the original position: 
\begin{equation}
\sum_{\pm}\frac{\pm (\pm 1)^{\frac{m+2}2}}{(|x|\pm |y|)^{\frac{m+2}2}} \int^\infty_{0} 
\la K(\l) G_{\frac{m+2}2}(\pm\l,\cdot,y), \tilde G_{0r}(-\l,\cdot,x) \ra \Phi(\l)\l d\l. \lbeq(4-10)
\end{equation} 
The same argument as in the proof of \reflm(4-1) shows that  
it suffices to show that \refeq(4-10) is admissible 
after replacing $\tilde G_{0r}(\l,z,x)$ by $G_{0r}(0,z,x)$, $K(\l)$ by $K(0)$ and $\Phi(\l)$ 
by the constant function $1$. In this case the residue theorem implies, for $a >0 $ and $ t>0$, that 
\[
\int_{0}^\infty e^{i\l a} \Big(\l a+\frac{it}2\Big)^{-\frac52}\l d\l 
= \int_{0}^\infty e^{-i\l a} \Big(-\l+\frac{it}2\Big)^{-\frac52} \l d\l 
\]
and the both sides are bounded in modulus by $C a^{-2} t^{-\frac12}$. Thus, we have 
\begin{equation}
\int_0^\infty  G_{\frac{m+2}2}(\l,\cdot,y) \l d\l = \int_0^\infty  G_{\frac{m+2}2}(-\l,\cdot,y) \l d\l 
\end{equation} 
and both sides are bounded in modulus by $C \la z- y \ra^{-\frac{m-2}2}$. 
It follows that \refeq(4-10) with this change is bounded in modulus by 
\[
C\left|\frac{1}{(|y|-|x|)^{\frac{m+2}2}} - \frac{1}{(|y|+|x|)^{\frac{m+2}2}}\right| 
\frac{1}{\ay^{\frac{m-2}2} \ax^{m-2}} 
\]
and is admissible. This completes the proof of \reflm(4-22) and therefore that of \refprp(2).  
\end{proof}

\section{Low energy estimate II, Exceptional case} 
In this section we discuss the low energy part $W_<$ in the case 
when $H$ is of exceptional type,  assuming 
\begin{equation}\lbeq(assm)
\mbox{$|V(x)|\leq C \ax^{-\d}$ with $\d > m+3$ if $m\geq 8$ and $\d>10$ if $m=6$. }
\end{equation}
so that the results of \refprp(sing-exp) apply. 
We substitute \refeq(sing-1) when $m=6$ or \refeq(sing-2) when 
$m=8$ for $L(\l)=(1+ G_-0(\l)V)^{-1}-I$ in formula \refeq(low-2op). 
As $VE_0(\l)$ satisfies property $(K)_\r$ with $\r>m+1$, \refprp(2) implies that  
$E_0(\l)$ produces an operator with admissible integral kernel. Thus, we have only to discuss  
operators produced by singular parts $\l^{-2}P_0V $ and 
$\sum_{ab} \l^a (\log \l)^b \, D_{ab}$ (note that we have changed indices $j,k$ to $a,b$). 
In Subsection 5.1 we prove that the operator produced by $\l^{-2}P_0V $, 
\begin{equation}\lbeq(sing-main)
W_{s,m} = \int_0^{\infty} G_0(\l) VP_0 V ( G_0(\l) - G_0(-\l)) \tilde \Phi (\l) \l^{-1} d\l, 
\end{equation}
is bounded in $L^p$ for $\frac{m}{m-2}<p < \frac{m}2$ 
and in Subsection 5.2 we indicate how the argument in Subsection 5.1 can be modified to prove 
the same for $\Phi(H)W_{s,ab}\Phi(H_0)$ where 
\begin{equation}\lbeq(logs)
W_{s,ab} = \int_0^{\infty}G_0(\l) V D_{ab}( G_0(\l) - G_0(-\l))\tilde \Phi (\l) \l^{a+1} (\log \l)^b \, d\l.
\end{equation}

\subsection{Estimate for $W_{s,m}$} 
In this subsection, we prove the following proposition.  We shall often write 
$\n=(m-3)/2$. 

\begin{proposition}\lbprp(sing)
Let $V$ satisfy \refeq(assm). 
Then, for any $\frac{m}{m-2} < p < \frac{m}{2} $, there exists a constant $C_p$ such
that 
\begin{equation}
\| W_{s,m} u \|_p \leq C_p \| u \|_p, \quad u \in C^{\infty}_0 (\R^m ).
\end{equation}
\end{proposition}

We first state two lemmas which will be used in what follows for proving the proposition. 
The first one can be found in
\cite{Stein}.
\begin{lemma} \lblm(weightlm)
The function $|r|^a$ on $\R$ is a one dimensional $(A)_p$ weight if and only if $ -1 < a < p-1$.
The Hilbert transform $\tilde{ \mathcal H}$ and the Hardy-Littlewood maximal operator $\mathcal M$
are bounded operators in $L^p( \R , w(r)dr )$ for $(A)_p$ weights $w(r)$.
\end{lemma}
For a function $f$ on $\R^m$, $M(r, u)$ is the spherical average of $f$:
\begin{equation*}
M(r,u) = \frac{1}{| \Sigma |} \int_{\Sigma} u(r\omega ) d\omega , \qquad r\in \R.  
\end{equation*}

\begin{lemma}
Let $m\geq 3$. Let $\psi \in L^1 (\R^m )$ and 
$u \in {\mathcal S}(\R^m)$. Then
\bea
F_{\psi, u}(\l)= \langle \psi , (G_0(\l)- G_0(-\l) )u \rangle \hspace{5cm} \notag\\ 
= C_m\int_0^{\infty} e^{-t} t^{\frac{m-3}2} 
\lf(\int_{\R} e^{-i \l r } (t + 2i \l r )^{\frac{m-3}2}r M(r, \psib \ast \check{u} )dr\ri) dt 
\lbeq(usformula)
\eea
where $\check{u}(x) = u(-x)$ and 
$C_m= -|\Sigma|/(4\pi)^{\frac{m-2}2 } \Gamma (\frac{m-2}2) =-2^{\frac{m-4}2}/(m-3)!$. 
\end{lemma}  
\begin{proof} Recall \refeq(co-ker). By Fubini's theorem and by using polar coordinates, 
\begin{multline*}
\langle \psi , G_0(\l) u \rangle = \int_{\R^m} G_0(\l, y) (\psi \ast \check{u})(y)dy \\
=-C_m 
\int_0^{\infty} e^{-t} t^{\nu - \half} \lf(
\int_{0}^{\infty} e^{i \l r } (t - 2i \l r )^{\nu -\half}r M(r, \psib \ast \check{u} ) dr \ri) dt.
\end{multline*}
Since $M(r)=M(-r)$, It follows that 
$-\langle \psi , G_0(-\l) u \rangle $ is given by 
\begin{multline*}
C_m \int_0^{\infty}  e^{-t} t^{\nu - \half} \lf(
\int_{0}^{\infty} e^{-i \l r } (t +2i \l r )^{\nu -\half}r M(r, \psib \ast \check{u} )dr\ri) dt \\
=-C_m \int_0^{\infty} e^{-t} t^{\nu - \half} \lf(
\int_{-\infty}^{0} e^{i \l r } (t - 2i \l r )^{\nu -\half}r M(r, \psib \ast \check{u} )dr\ri) dt. 
\end{multline*}
Adding the two equations and changing $r \to -r$, we obtain the lemma. 
\end{proof}

For fixed $f,g \in L^1(\R^m)$, we define the operator $Z=Z(f \otimes g)$ by 
\begin{equation}\lbeq(zj)
Z(f \otimes g)u = \int_0^{\infty} G_0(\l) (f \otimes g)  
\lf( G_0(\l) - G_0(-\l) \ri) \tilde\Phi (\l) \l^{-1} d\l
\end{equation}
If we write $P_0 = \sum_{j=1}^d \f_j \otimes \f_j$ in terms of   
an orthonormal basis of $P_0 \Hg$, we have $W_{s,m} = \sum_{j=1}^d Z((V\f_j) \otimes (V\f_j))$.  

\begin{lemma}\lblm(zlm) With suitable constants $C_{jk}$ we have 
\begin{equation}\lbeq(z)
Zu(x) = \sum_{j,k=0}^{\frac{m-4}{2}} C_{jk}
\int_{\R^m} \frac{f(y) K_{jk}u(|x-y|)}{|x-y|^{m-2}}dy 
\end{equation}
where with $M(r, \overline{g} \ast \check{u} ) = M(r)$, 
$K_{jk}u(|x-y|)$, $0\leq j,k \leq \frac{m-4}{2}$, are defined by 
\begin{multline}
K_{jk}u(\rho) = \rho^j \int_0^{\infty} \, e^{i \l \rho} \l^{j+k-1} \tilde\Phi(\l)
\Big\{ \int_0^{\infty}\int_0^{\infty} \; e^{-(t+s)}  \times  \\ 
t^{2\nu -\frac32 -k} s^{2\nu -\frac32 -j} (s-2i\l \rho)^{\frac12} 
\Big( \int_{\R}e^{-i\l r} (t+2i\l r )^{\frac12} r^{k+1} M(r)dr \Big) dt ds \Big\} d\l .
\lbeq(kjkop)
\end{multline}
\end{lemma}
\begin{proof} We remark that, for $u\in C^{\infty}_0(\R^m)$,  \refeq(kjkop) is well defined for all 
$j,k$  because $M(r)$ is smooth, $\langle r \rangle^{m-1} M^{(\ell)}(r)$ is integrable 
and $\int_{\R}  r M(r)dr =0$ because $M(r) $ is even. 
By virtue of \refeq(usformula), we have  
\[
Zu(x) = \int_{0}^\infty G_0(\l)f(x) \cdot  F_{g, u}(\l)d\l 
\]
We substitute \refeq(usformula) for  $F_{g, u}(\l)$ and the expression 
\[
\frac{1}{(4\pi)^{\frac{m-3}2}\Ga(\frac{m+2}2)}\int_{\R^m} \frac{e^{i\l|x-y|}f(y)}{|x-y|^{m-2}} 
\left(\int_0^\infty e^{-s}s^{\frac{m-3}2}(s-2i\l|x-y|)^{\frac{m-3}2}ds\ri)dy
\]
for $G_0(\l)f(x)$. 
We then change the order of integrations with respect to $d\l$ and $dy$  
and, using the binomial formula, write 
\[
(t+2i\l r)^{\frac{m-3}2}= \sum_{j=0}^{\frac{m-4}2} \binom{\frac{m-4}2}{j} 
t^{\frac{m-4}2-j} (2i\l r)^j (t+2i\l r)^{\frac12} 
\] 
and similarly for $(s-2i\l \r)^{\frac{m-3}2}$. The lemma follows. 
\end{proof} 

In what follows we fix $f, g$ which satisfy for some $\ep>0$ and $C>0$  
\begin{equation}\lbeq(fg)
|f(x)|\leq C \ax^{-m-\ep}, \quad |g(x)|\leq C \ax^{-m-\ep}
\end{equation} 
and define operator $W_{jk}$ for $0\leq j,k \leq \frac{m-4}{2}$ by 
\begin{equation}\lbeq(wjk-def)
W_{jk}u(x) =\int_{\R^m} \frac{f(y) K_{jk}u(|x-y|)}{|x-y|^{m-2}}dy 
\end{equation}
so that $Z = \sum C_{jk}W_{jk}$. We use the following lemma. 

\begin{lemma}\lblm(mest) Let $M(r)=M(r, \gb\ast u)$. Then, for $\frac{m}{m-2} < p< \frac{m}{2}$, 
we have 
\[
\int_0^\infty \la r \ra |M(r)|dr \leq C \|u\|_p.
\]
\end{lemma} 
\begin{proof} Let $q=\frac{p}{p-1}$ be the conjugate exponent of $p$. We have 
\begin{multline*}
\int_{\R} \la r \ra |M(r)| dr \leq 
C \int_{\R^m} \frac{\ax |(\overline{g} \ast \check{u})(x)|}{|x|^{m-1} }dx \\
\leq C \int_{|x|<1 }\frac{\| \gb \ast \check{u}\|_{\infty} }{|x|^{m-1}}dx  
 +
C  \lf( \int_{|x|>1 }\frac{\|\gb \ast \check{u}\|_{p}}{|x|^{q(m-2)}}dx \ri)^{\frac{1}{q}}
\leq C (\|\gb\|_q+ \|\gb \|_{1})\| u \|_{p}
\end{multline*}
since $q(m-2)>m$ for $\frac{m}{m-2} < p< \frac{m}{2}$. 
\end{proof}

It is clear that $V\f_j $, $j=1, \ldots, d$, satisfy the condition \refeq(fg) and 
\refprp(sing) follows from the following proposition. 

\begin{proposition}\lbprp(wjk) Let $f, g$ satisfy \refeq(fg).  Then, 
$W_{jk}$, $0 \leq j,k \leq \tfrac{m-4}2$, are bounded in 
$L^p(\R^m)$ for $\frac{m}{m-2} < p< \frac{m}{2}$.   
\end{proposition}
We prove \refprp(wjk) for various cases of $j,k$ separately. 
By interpolation, we have only to show \refprp(wjk) for $p=\frac{m}{m-2-\ep} $ 
and $p=\frac{m}{2+\ep}$ with arbitrary small $\ep>0$. We denote the Hilbert transform 
by $\tilde \Hg$ and ${\mathcal H} = (1 + \tilde{\mathcal H} )/2$. 
By lemma \reflm(weightlm) $|r|^{m-1-p\theta}$ is a one dimensional $(A)_p$ 
weight if and only if $0< \frac{m}{p} -\theta <1$,
viz. 
\begin{equation}
\begin{array}{cl}
m-3-\ep < \theta < m-2-\ep & \text{if }\, p=m/(m-2-\ep), \\
1+\ep < \theta < 2+\ep & \text{if } \, p=m/(2+\ep).
\end{array}
\lbeq(theta)
\end{equation}

\noindent 
{\bf (1) The case $j,k\geq 1$.} If $1\leq j,k \leq \frac{m-4}2$ the integrand of 
\refeq(kjkop) is integrable with respect to $dtdsdrd\l$ and we are free to change the order 
of integration. Thus, we may write  
\begin{align}
K_{j,k}u(\rho) = \int_{\R} M(r) T_{jk}(\r,r) dr, \hspace{3cm}\lbeq(kjkop-1) \\
T_{jk}(\r,r)= \int_0^{\infty}\int_0^{\infty} e^{-(t+s)}  t^{2\nu -\frac32 -k} s^{2\nu -\frac32 -j} 
J_{jk}(s,t,\r,r)  dt\, ds, \lbeq(tjk) \\ 
J_{jk} = \rho^j  r^{k+1}  
\int_0^{\infty} e^{i \l(\r- r)} \l^{j+k-1} \tilde\Phi(\l) (s-2i\l \rho)^{\frac12}  (t+2i\l r )^{\frac12}d\l. 
\lbeq(jjk) 
\end{align}

\begin{lemma}\lblm(t2est) Let $j,k\geq 1$. Then, with a constant $C=C_{jk}$, we have 
\begin{equation}
\lf| T_{jk}(\rho,r) \ri| \leq 
C \lf| \frac{\la \rho \ra^{j+1/2}
 r^{k+1} \langle r \rangle^{1/2}  }{ \langle r - \rho \rangle^{j+k}} \ri|
\lbeq(t2est)
\end{equation} 
Estimate \refeq(t2est) holds when $\tilde\Phi$ is replaced by any smooth function with 
compact support and $t^{2\n-\frac32-k}$ and/or $s^ {2\n-\frac32-j}$ by $t^a$ and/or $s^b$ 
with $a,b \geq 0$.  
\end{lemma} 
\begin{proof} Since \refeq(t2est) is obvious for $|\r-r|\leq 1$, we prove it 
only for $|\r-r|\geq 1$. By integrating by parts $j+k$ times with respect to $\l$, we have 
\begin{multline*}
J_{jk}(s,t,\r,r)= \frac{i^{j+k}\sqrt{st}\rho^j  r^{k+1}}{(\r - r)^{j+k}} (j+k-1)! + 
\\
-\frac{i^{j+k}\rho^j  r^{k+1}}{(\r - r)^{j+k}} \int_0^{\infty}  e^{i \l (\r-r)}
\lf\{ \l^{j+k-1} \tilde\Phi(\l) (s-2i\l \rho)^{1/2}  (t+2i\l r )^{1/2} \ri\}^{(j+k)}d\l. 
\lbeq(intbypart)
\end{multline*}
We insert this into \refeq(tjk). The boundary term produces 
\begin{equation}
T_{jk,b}(\r,r) = C \frac{\rho^j r^{k+1} }{(r - \rho )^{j+k}}, 
\lbeq(t1est)
\end{equation}
and this  clearly satisfies \refeq(t2est). We compute the derivative via Leibniz' rule: 
\begin{multline*} 
\lf( \frac{d}{d\l} \ri)^{j+k} \lf\{ \l^{j+k-1} \tilde\Phi(\l) (s-2i\l \rho)^{1/2}  (t+2i\l r )^{1/2} \ri\} \\
= \sum_{a+b+c=j+k} C_{abc}\Psi_{a}(\l) 
(s-2i\l \rho)^{1/2-b} (2i\rho)^b   (t+2i\l r )^{1/2-c} (-2ir)^{c} 
\end{multline*}
where $\Psi_{a}(\l)= \{\l^{j+k-1} \tilde\Phi(\l)\}^{(a)}$. Denoting summands on the right 
by $E_{abc}=E_{abc}(\l,s,t,\r,r)$, we define  
\begin{equation} 
J_{abc}(s,t,\r,r) \equiv 
\frac{i^{j+k}\r^j r^{k+1}}{(\r - r)^{j+k}} \int_0^{\infty}  e^{i \l (\r-r)}E_{abc}(\l,s,t,\r,r)d\l, 
\end{equation}
and $T_{abc}(\r,r)$ by the right of \refeq(tjk) with $J_{abc}$ replacing $J_{jk}$. By using obvious estimates 
$|(s-2i\l \rho)^{-1}(2i\l\rho)|\leq 1 $ and $|(t+2i\l r )^{-1}(-2i\l r)|\leq 1$, we obtain 
\[
|E_{abc}|\leq C |\Psi_{a}(\l)| \times 
\left\{\br{ll} (s^\frac12 +|\l \rho|^\frac12) (t^\frac12 + |\l r|^\frac12 ), & \mbox{if}\ b=c=0; \\
(\r r)^\half |\l|^{1-(b+c)}, & \mbox{if}\ b, c \not=0; \\ 
 r ^{\frac12}(s^\frac12+ |\l \rho|^{\frac12}) |\l|^{\frac12-c} ,  & \mbox{if}\ b=0, c \not=0; \\ 
 r ^{\frac12}(t^\frac12+ |\l \rho|^{\frac12}) |\l|^{\frac12-b} ,  & \mbox{if}\ b\not=0, c =0. 
\er \right.  
\]
Note that $\l^{1-(b+c)}\Psi_{a}(\l) $, $\l^{\frac12-b} \Psi_{a}(\l)$ and 
$\l^{\frac12-c} \Psi_{a}(\l)$ are integrable functions with compact supports in respective cases. 
It immediately follows that 
\begin{equation}
|T_{abc}(\r,r)| \leq C \frac{\rho^j \la \r \ra ^\half r^{k+1} \la r\ra^\half}{\la \r-r\ra^{j+k}} 
\end{equation}
and, by summing up, we obtain  \refeq(t2est). 
The only property of $\tilde\Phi$ which is used in the argument above is that it is smooth and compactly 
supported; all estimate above go through if $e^{-t}t^{2\nu -\frac32 -k}$ or $e^{-s}s^{2\nu -\frac32 -j}$ are 
replaced by $e^{-t}t^{a}$ or $e^{-s}s^{b}$, $a,b \geq 0$, and the last statement of the 
lemma follows.  
\end{proof}

\begin{lemma}\lblm(wjkp)  Let $j,k \geq 1$ and let $T_{jk}(\r,r)$ satisfy \refeq(t2est). 
Let $W_{jk}$ be defined by \refeq(wjk-def) with $K_{jk}$ given by \refeq(kjkop-1). 
Then $W_{jk}$ satisfies \refprp(wjk). 
\end{lemma}
\begin{proof} By splitting the domain of integration, we estimate
\begin{eqnarray}
|W_{jk}u(x)| & \leq &  \lf( \int_{|x-y|<1} + \int_{|x-y|\geq 1} \ri) 
\frac{|f(y) K_{jk}(|x-y|) |}{|x-y|^{m-2}} dy   \nonumber \\ 
 & \equiv & I_1(x) + I_2(x) \lbeq(i1i2)
\end{eqnarray} 
By Young's inequality, \refeq(t2est) and \reflm(mest)  
\begin{align*}
\| I_1\|_p & \leq C\| f \|_p \left(\int_{|x|\leq 1} \frac{|K_{jk}(x)|}{|x|^{m-2}} dx\right) 
\leq C  \sup_{|\r|<1}| K_{jk}(\r)| \\
& \leq C  \sup_{0\leq \rho \leq 1} 
 \int_{\R} \frac{|r|^{k+1} \la r \ra^{\frac12}}{ \langle r-\rho  \rangle^{j+k} } |M(r)|dr   
\leq C \int_{\R} |r \, M(r)| dr \leq C \|u\|_p. 
\end{align*}
Let $p=\frac{m}{2+\ep}$, $0<\ep<1$ and choose $\theta=2$, see \refeq(theta). 
By Young's inequality we have by using polar coordinates 
that 
\begin{align} 
\| I_2\|_p & \leq  \| f \|_1
\lf( \int_1^{\infty}  \rho^{m-1} 
\lf| \frac1{\r^{m-2}}\int_{\R} T_{jk}(\rho, r ) M(r)dr \ri|^p 
d\rho \, \ri)^{\frac{1}{p}}  \notag \\
& \leq C  \int_1^{\infty} \rho^{m-1-p\th} 
\lf( \frac{\rho^{j + \half} }{ \rho^{m-4} } \int_{\R} dr \, 
\frac{ |r|^{k+1} \langle r \rangle^{\frac12}  }{  \langle r - \rho \rangle^{j+k}} |M(r)| \ri)^{p} d\r.  
\lbeq(1-7)
\end{align}
Since $|r|^{k-1} \langle r \rangle^{\frac12} \leq C \langle r - \rho \rangle^{k-\half} \rho^{k-\half}$ 
(recall $k \geq1$) and $m-4\geq j+k$, 
\[
\frac{\r^{j+\frac12} |r|^{k+1} \langle r \rangle^{\frac12}  }{\r^{m-4}  \langle r - \rho \rangle^{j+k}}
\leq C \frac{\r^{j+k} |r|^{2}  }{\r^{m-4}  \la r - \rho \ra^{j+\frac12}}
\leq C \frac{|r|^{2} }{\langle r - \rho \rangle^{j+\frac12}}, \quad \r >1. 
\]
Hence,  the right of \refeq(1-7) is bounded by 
\[
C\int_1^{\infty} \rho^{m-1-p\th} \Mg(|r|^2 M)(\r)^{p} d\r 
\leq C \lf( \int_0^{\infty}r^{m-1} | M(r)|^{p} dr \ri)^{\frac1{p}}  
\]
by virtue of the weighted inequality for the maximal functions. 
By H\"older's inequality  the right side is bounded by 
$C \| \gb \ast \check{u}\|_p \leq  C \|u \|_{p}$.  

\noindent 
When $p= \frac{m}{m-2-\ep}$, $0<\ep<1$, we choose $\theta = m-3$. Again by using 
Young's inequality and \refeq(t2est) 
\begin{equation}
 \| I_2\|_{p}^{p} \leq C
 \int_1^{\infty} \rho^{m-1-p\th}
\lf( \rho^{j-\frac12}  \int_{\R} 
\frac{ |r|^{k+1} \la r \ra^{\frac12}  }{  \la r - \rho \ra^{j+k}} |M(r)|dr \ri)^{p} d\r 
\end{equation}
Since $ \rho^{j - \half} \leq \la r \ra^{j - \half} \langle r - \rho \rangle^{j-\half} $  
and $m-1-p\th$ is an $(A)_p$ weight, the right hand side is 
further estimated by 
\begin{multline*}
C\int_1^{\infty} \rho^{m-1-p\th}
\lf( \int_{\R} 
\frac{ |r|^{k+1} \la r \ra^{j}  }{  \la r - \rho \ra^{k+\frac12}} |M(r)| dr \, \ri)^{p}d\rho \\
\leq C\int_1^{\infty} \rho^{m-1-p\th}\Mg(|r|^{k+1} \la r \ra^{j} M(r))(\r)^{p}d\rho \\
\leq C\int_0^{\infty} r^{m-1-p\th}(|r|^{k+1} \la r \ra^{j} M(r))^{p}dr 
\end{multline*}
Since $k+j+1 \leq m-3=\th$, the last integral is bounded by a constant times  
\begin{multline*}
\int_0^{1} r^{m-1-p(\th-k-1)} M(r)^{p}dr + 
\int_1^\infty  r^{m-1} M(r)^{p}dr \\
\leq C\int_{|x|<1}\frac{|(\gb \ast \check{u})(x)|^p}{|x|^{p(\th-k-1)}}dx 
+ \|\gb  \ast \check{u}\|^p  
\leq C (\|g\|_q+ \|g\|_{1})^p \| u \|_{p}^p, 
\end{multline*}
where $q$ is the conjugate exponent of $p$. This completes the proof. 
\end{proof}

\noindent 
{\bf (2) The case $j=0$, $k\geq 1 $}. We now prove \refprp(wjk) for $j=0$ and 
$1\leq k \leq \n-1= \frac{m-4}2$ by induction on $k$, using also the already proven 
result for the case $j,k \geq 1$.  For this and for the purpose for the next case (3) we define as follows. 
\begin{definition} {\rm (1)} We say $J_{0k}(s,t,\r,r)$ is {\it $\ell$-admissible} if  
operators $W_{0k, ln}$, $0\leq l,n \leq \n-\ell$, defined by \refeq(wjk-def) 
with $K_{0k, ln}(\r)=\int T_{0k,ln}(\r,r)M(r)dr$ in place of $K_{jk}(\r)$ are bounded in $L^p(\R^m)$ for 
$\frac{m}{m-2}<p<\frac{m}2$ where 
\begin{equation}\lbeq(tjk1)
T_{0k,ln}(\r,r)= \int_0^{\infty}\int_0^{\infty} e^{-(t+s)}  t^{2\nu -\frac32 -k+n} s^{2\nu -\frac32 -l} 
J_{0k}(s,t,\r,r)  dt\, ds.
\end{equation}
{\rm (2)} We say $J_{j0}(s,t,\r,r)$ is {\it $\ell$-admissible} if 
operators $W_{j0, ln}$, $0\leq l,n \leq \n-\ell$, defined by \refeq(wjk-def)  
with $K_{j0, ln}(\r)=\int T_{j0,ln}(\r,r)M(r)dr$ in place of $K_{jk}(\r)$ are bounded in $L^p(\R^m)$ for 
$\frac{m}{m-2}<p<\frac{m}2$ where 
\begin{equation}\lbeq(tjk-1)
T_{j0,ln}(\r,r)= \int_0^{\infty}\int_0^{\infty} e^{-(t+s)}  t^{2\nu -\frac32 -n} s^{2\nu -\frac32 -j+l} 
J_{j0}(s,t,\r,r)  dt\, ds.
\end{equation}
\end{definition}
\noindent 
Compare \refeq(tjk1) or \refeq(tjk-1) with \refeq(tjk). Note that the exponents 
$2\nu -\frac32 -k+n, 2\nu -\frac32 -l$ are not smaller than $1$ for all relevant $l,n$.  
It suffices to prove the following lemma. 

\begin{lemma} \lblm(induction) 
Let $J_{0k}(s,t,\r,r)$, $1 \leq k\leq \n-1$,  be defined by \refeq(jjk) 
with $\tilde\Phi \in C_0^\infty(\R)$ 
which is a constant near $\l=0$. Then, $J_{0k}(s,t,\r,r)$ are $k$-admissible. 
\end{lemma} 
\begin{proof}  We prove the lemma by induction on $k$. We begin with the case $k=1$.

\begin{lemma}\lblm(prev-1) For all $0\leq l,n \leq \n-1$, $T_{01, ln}(\r,r)$ satisfies the estimate 
\begin{equation}\lbeq(t01)
|T_{01, ln}(\r,r)|\leq C \frac{|r|^2 (\la \r\ra + \la r \ra)}{\la r-\r \ra^2}
\end{equation} 
\end{lemma}
\begin{proof}
This is obvious when $|\r-r|\leq 1$ and we assume $|\r-r|>1$ in what follows. 
Integrating by parts twice with respect to $\l$, we have 
\begin{multline}
J_{01}(s,t, \rho,r) = \frac{ir^{2}}{(\r - r)} \sqrt{st} +  
\frac{ir^{2}}{(\r - r)^{2}} (\r(t/s)^\frac12 - r (s/t)^\frac12) \\  
+\frac{r^{2}}{(\r - r)^{2}}
\int_0^{\infty} \, e^{i \l (\r-r)}\Big(\frac{\pa}{\pa \l}\Big)^2 
\Big(\tilde\Phi(\l) (s-2i\l\rho)^{1/2}  (t+2i\l r )^{1/2} \Big) d\l 
\lbeq(intbypart2)
\end{multline}
We substitute this for $J_{01}$ in \refeq(tjk1). Then the functions produced by the boudary terms 
are bounded by 
\[
C\Big(\frac{r^2}{\la \r-r \ra} + \frac{r^2(|\r|+|r|+1)}{\la \r-r \ra^2}\Big)\leq 
C \frac{|r|^2 (\la \r\ra + \la r \ra)}{\la \r-r \ra^2} .
\]
Denoting by $'$ the deivative with respect to the variable $\l$, we compute:   
\begin{multline*}
\Big(\tilde\Phi(\l) (s-2i\l\rho)^{1/2}  (t+2i\l r )^{1/2} \Big)''= 
\tilde\Phi''(\l) (s-2i\l\rho)^{1/2}  (t+2i\l r )^{1/2} \\
+2 \tilde\Phi'(\l) \Big((s-2i\l\rho)^{1/2}  (t+2i\l r )^{1/2} \Big)'
+\tilde\Phi(\l)\Big( (s-2i\l\rho)^{1/2}  (t+2i\l r )^{1/2} \Big)''.
\end{multline*}
Since $\tilde\Phi'(\l)=0$ near $\l=0$ and  
$|(s-2i\l\rho)^{-\frac12} 2\r|  \leq C (\r/s)^\frac12$ for $|\l| \geq C >0$, this is bounded 
in modulus by a constant times 
\begin{align*}
& |\tilde\Phi''(\l)|(s^\frac12+|\l\rho|^\frac12)(t^\frac12+ |\l r|^\frac12)+ 
|\tilde\Phi'(\l)|(\r/s)^\frac12(t^\frac12+|\l r|^\frac12) \\ 
& + |\tilde\Phi'(\l)|(r/t)^\frac12 (s^\frac12+ |\l\r|^\frac12)+ 
|\tilde\Phi(\l)||t\r-sr|^2 (s+|\l r|)^{-\frac32}  (t+|\l\r|)^{-\frac32} .
\end{align*}
Hence integrating by $dtds$ first and using also elementary estimates 
\begin{align}
& \lf| \int_0^{\infty} \frac{e^{-t}t^a dt}{(t+ |\l r |)^{b}} \ri| \leq C \langle \l r \rangle^{-b}
\qquad 0< b <a+1, \lbeq(int) \\
& \lf| \int_0^{\infty}  \frac{|\tilde\Phi(\l)|}
{\langle \l r \rangle^{a}\langle \l \rho \rangle^{b} } \ri| \leq C 
\frac{1}{\langle  r \rangle+\langle  \rho \rangle }, \quad a,b>0, a+b>1, 
\end{align}
we obtain estimate \refeq(t01). 
\end{proof}

\begin{lemma}\lblm(1adm) Let $J_{01}(s,t,\r,r)$ be defined by \refeq(jjk) 
with $\tilde\Phi \in C_0^\infty(\R)$ 
which is a constant near $\l=0$. Then, $J_{01}(s,t,\r,r)$ is $1$-admissible. 
\end{lemma} 
\begin{proof} We have  
${\ds \frac{|r|^2 (\la \r\ra + \la r \ra)}{\la r-\r \ra^2}
\leq \frac{2|r|^2 \la \r\ra}{\la r-\r \ra^2} + \frac{|r|^2 }{\la r-\r \ra}} $. 
The first term on the right satisfies \refeq(t2est) with $j=k=1$ and, 
by virtue of \reflm(wjkp), it suffices to show 
that $W_{01}$ is bounded in $L^p(\R^m)$  for $\frac{m}{m-2}<p<\frac{m}2$ if  
\begin{equation}\lbeq(aux-1)
|T_{01}(\r,r)|\leq C \frac{|r|^2 }{\la r-\r \ra}\leq C \lf( |r|+ \frac{|r||\r|}{\la\r-r \ra}\ri) .
\end{equation}
We estimate $|W_{01}u(x)|\leq I_1(x)+ I_2(x)$ as in \refeq(i1i2). For $I_1(x)$ we use 
the first of \refeq(aux-1) and proceed as in the proof of \reflm(wjkp). We have 
\[
\|I_1\|_p \leq C \|f\|_p \sup_{|\r|\leq 1} \int_{\R}\frac{|r|^2|M(r)|dr}{\la \r-r\ra} 
\leq C \|f\|_p \int_{\R}|r||M(r)|dr  \leq C \|u\|_p.  
\]
For $I_2(x)$ we use Young's inequality and the second of \refeq(aux-1) to obtain: 
\begin{multline}\lbeq(I2)
\|I_2\|_p 
\leq \|f\|_1 \lf( \int_1^{\infty}  \rho^{m-1} 
\lf| \frac1{\r^{m-2}}\int_{\R}\, |r| |M(r)| dr \ri|^p 
d\rho \, \ri)^{\frac{1}{p}} \\
+ \|f\|_1 \lf( \int_1^{\infty}  \rho^{m-1} 
\lf| \frac1{\r^{m-3}}\int_{\R}\frac{|r| |M(r)|}{\la \r -r \ra} dr \ri|^p 
d\rho \, \ri)^{\frac{1}{p}} 
\end{multline}
The first term is bounded by ${\ds C \int_{\R}\, |r| |M(r)| dr \leq   C \|u\|_p }$ 
since $p(m-2)>m$ for $\frac{m}{m-2}<p< \frac{m}{2}$.  For estimating the second, take $\ep>0$ 
arbitraily small and fix $p \in (\frac{m}{m-2-\ep}, \frac{m}{2+\ep})$. Take 
$0<\ep'<\ep$ and choose $\frac{m}{p}-1<\th<\frac{m}{p}$ sufficiently close to $\frac{m}p-1$ 
so that $m-1-p\th$ is an $(A)_p$ weight and so that $1+\ep'<\th\leq m-3-\ep'$. Then, using   
${\ds \la \r-r \ra^{-1} \leq C_{\ep'}\la \r \ra^{\ep'}\la  r \ra^{\ep'}\la \r- r \ra^{-(1+\ep')}}$,  
we estimate the second integral by a constant times 
\begin{multline*}
\lf( \int_1^{\infty}  \rho^{m-1} 
\lf( \frac1{\r^{m-3-\ep'}}\int_{\R}\frac{|r| \la r \ra^{\ep'} |M(r)|}{\la \r -r \ra^{1+\ep'}} dr \ri)^p 
d\rho \, \ri)^{\frac{1}{p}} \\
\leq C\lf( \int_1^{\infty}  \rho^{m-1-p\th} 
\lf( \int_{\R}\frac{|r| \la r \ra^{\ep'} |M(r)|}{\la \r -r \ra^{1+\ep'}} dr \ri)^p d\rho \, \ri)^{\frac{1}{p}} \\
\leq C \lf( \int_{\R} r^{m-1-p(\th-1)}\la r \ra^{p\ep'} |M(r)|^p dr \ri)^{\frac{1}{p}} 
\leq  C\lf( \int_{\R^m} \frac{\la x\ra^{p{\ep}'} |\gb \ast u(x)|}{|x|^{p(\th-1)}}dx \ri)^{\frac1{p}}.
\end{multline*} 
Since $p\ep'< p(\th-1)<m$, the right hand side is bounded by $C\|u\|_p$. 
This proves that $\|I_2\|_p \leq C \|u\|_p$.  This completes the proof.  
\end{proof}

\noindent 
{\bf Completetion of the proof of \reflm(induction).}
The lemma is satisfied when $k=1$ by virtue of \reflm(prev-1). We let $k \geq 2$ and 
assume that the lemma is already proved for smaller values of $k$. We write 
$r^{k+1}= r^k \r - r^k (\r-r)$ in the definition \refeq(jjk) for $J_{0k}(s,t,\r,r)$ 
and apply integration by part to the integral containing $r^k (\r-r)$. We obtain 
\begin{multline}
J_{0k}(s,t,\rho,r) = J_{1(k-1)}(s,t,\rho,r) \\
- i r^{k}\int_0^{\infty}\,  e^{i \l (\r-r)} \Big(\frac{\pa}{\pa \l}\Big) 
\Big(\l^{k-1} \tilde\Phi(\l) (s-2i\l \rho)^{\frac12}(t+2i\l r )^{\frac12}\Big) d\l .
\lbeq(0kj) 
\end{multline}
Thanks to results in case (1), $J_{1(k-1)}(s,t,\rho,r)$ is $(k-1)$-admissible and it may be ignored.  
We insert the following for the derivative in the integrand:   
\begin{multline*}
(k-1) \l^{k-2} \tilde\Phi (s-2i\l \rho)^{\frac12}(t+2i\l r )^{\frac12}
+\l^{k-1} \tilde\Phi' (s-2i\l \rho)^{\frac12}(t+2i\l r )^{\frac12} \\
-2i\r \l^{k-1} \tilde\Phi\Big(\frac{\pa}{\pa s}\Big) (s-2i\l \rho)^{\frac12}(t+2i\l r )^{\frac12}
+ \l^{k-1} \tilde\Phi (s-2i\l \rho)^{\frac12} ir (t+2i\l r )^{-\frac12}
\end{multline*}
The first term produces 
$(k-1) J_{0(k-1)}(s,t,\r,r)$, which is $(k-1)$-admissible by induction hypothesis;  
the second does $J_{0(k-1)}(s,t,\r,r)$ with $\l \tilde\Phi'(\l)$ replacing 
$\tilde\Phi$, which is also $(k-1)$-admissible since $\l\tilde\Phi'(\l)=0$ near $\l=0$. 
Define  
\[
J_{0k(3)} \equiv 
2r^{k}\r \int_0^{\infty}\,  e^{i \l (r-\r)} 
\l^{k-1} \tilde\Phi(\l) \Big(\frac{\pa}{\pa s}\Big) (s-2i\l \rho)^{\frac12} \cdot (t+2i\l r )^{\frac12} d\l 
\]
and substitute this for $J_{jk}(s,t,\r,r)$ in \refeq(tjk1).  
This yields after integration by parts with respect to the $s$-integral  
\[
-2 T_{1(k-1), ln}(\r, r) + 2 (2\n-\tfrac32-l) T_{1(k-1), (l+1)n}(\r, r) 
\]
and the result of case (1) implies $J_{0k(3)}(s,t,\rho,r)$ is $k$-admissible. 
We rewrite the last term $\l^{k-1} \tilde\Phi (s-2i\l \rho)^{\frac12} ir (t+2i\l r )^{-\frac12}$ in the form 
\[
\tfrac{1}2\l^{k-2} \tilde\Phi(\l) (s-2i\l \rho)^{\frac12}(t+2i\l r )^{\frac12}
-\l^{k-2} \tilde\Phi(\l) (s-2i\l \rho)^{\frac12} t \Big(\frac{\pa}{\pa t}\Big) (t+2i\l r )^{\frac12}
\]
The first term again  produces $\frac12 J_{0(k-1)}(s,t,\r,r)$. Define  
\[
J_{0k(4)}(s,t,\rho,r)\equiv 
r^{k}\r \int_0^{\infty}\,  e^{i \l (r-\rho)} 
\l^{k-2} \tilde\Phi(\l)(s-2i\l \rho)^{\frac12} \cdot \Big(\frac{\pa}{\pa t}\Big)(t+2i\l r )^{\frac12} d\l
\]
and substitute $J_{0k(4)}(s,t,\rho,r)$ for $J_{jk}(s,t,\r,r)$ in \refeq(tjk1).   
This yields, after integration by parts with trespect to the $t$-integral, 
\[
- T_{0(k-1),l(n+1)}(\r,r)+ (2\n -\tfrac12-k+n) T_{0(k-1), ln}(\r,r) 
\]
It follows by induction hypothesis that $J_{0k(4)}(s,t,\rho,r)$ is also $k$-admissible. 
This completes the proof. 
\end{proof}

\noindent
{\bf (3) The case $j \geq 1$ and $k=0$}.  We next prove \refprp(wjk) for 
$j \geq 1$ and $k=0$. It suffices to prove the following lemma. 

\begin{lemma}\lblm(jj0) Let $J_{j0}(s,t,\r,r)$, $j=1, \ldots, \n-1$, be defined by \refeq(jjk) 
with $\tilde\Phi \in C_0^\infty(\R)$ 
which is a constant near $\l=0$. Then, $J_{j0}(s,t,\r,r)$  are $j$-admissible. 
\end{lemma} 
\begin{proof} We prove the lemma by induction on $j$. Thus, we let $j=1$ first. Comparing 
definitions of $J_{01}$ and $J_{10}$ and \refeq(t01), it is obvious that 
\[
|T_{10, (ln)}(\r,r)|\leq C \frac{|r|\r (\la \r \ra + \la r \ra)}{\la \r-r \ra^2} 
\]
Since $\la \r \ra + \la r \ra\leq 4(\la \r-r\ra +|r|)$ and 
$|r|^2 \r\la \r-r \ra^{-2}$ satisfies \refeq(t2est) with $j=k=1$, 
it suffices to show that $W_{10}$ has the desired property when 
\[
|T_{10}(\r,r)|\leq C \frac{|r|\r}{\la \r-r \ra}.
\]
However, the right hand side is the same as the second term on the right 
of \refeq(aux-1) and the proof of \reflm(1adm) shows that \reflm(jj0) 
is satisfied when $j=1$.  We then let $j \geq 2$ and assume that 
the lemma is already proved for smaller values of $j$. We write 
$\r^j r = \r^{j-1}r^2 + \r^{j-1}r (\r-r)$ in the definition \refeq(jjk) of 
$J_{j0}(s,t,\r,r)$. The first term $\r^{j-1}r^2$ 
produces $J_{(j-1)1}(t,s,\r,r)$ and we may ignore it by virtue of results in case (1).  
We need study the operators corresponding to $W_{jk}$ produced by functions 
\[
\r^{j-1}r \int_0^{\infty}\,  e^{i \l (\r-r)} \Big(\frac{\pa}{\pa \l}\Big) 
\Big(\l^{k-1} \tilde\Phi(\l) (s-2i\l \rho)^{\frac12}(t+2i\l r )^{\frac12}\Big) d\l .
\]
However, after this point the argument completely in paralell with that of 
\reflm(induction) after \refeq(0kj) and we omit the repetitious 
details. 
\end{proof}

\noindent 
{\bf (4) The case $j=k=0$.}  \ Finally we prove \refprp(wjk) for $j=k=0$. 
Recall the definition \refeq(kjkop) and \refeq(tjk). In \refeq(kjkop) 
we substitute  
\[
(s-2i\l\r)^{\frac12}((t+2i\l r )^{\frac12}-t^\frac12) + 
((s-2i\l \rho)^{\frac12}-s^\frac12)t^{\frac12}+ s^{\frac12} t^{\frac12}. 
\]
for $(s-2i\l \rho)^{\frac12} (t+2i\l r )^{\frac12}$ 
and denote by $K_j$ the operator produced by $j$-th summand, $j=1,2,3$, 
so that $K_{00}=K_{1}+ K_{2}+ K_{3}$. Define 
$W_{j}$ by \refeq(wjk-def) with $K_j$ in place of $K_{jk}$ 
so that $W_{00}=W_1+ W_2+ W_3$. For $j=1$ and $j=2$, we may change the order of integrations 
and write $K_{j}u(\r)$ in the following form: 
\[
K_{j}u(\r)= \int_{\R} T_{j}(\r,r)  r  M(r)dr, \quad j=1,2 
\]
where $T_1(\r,r)$ and $T_2(\r,r)$ are given by constants times 
\begin{multline*}
T_1(\r,r)= r\int_0^{\infty}\int_0^{\infty} \; e^{-(t+s)}t^{2\nu -\frac32} s^{2\nu -\frac32} \times \\
\times \left(\int_0^\infty e^{i \l(\r- r)} \tilde\Phi(\l)
\Big\{\frac{2ir(s-2i\l \r )^{\frac12}}{(t+2i\l r)^{\frac12} +t^\frac12} \Big\}
 d\l \right) ds dt, \\
T_2(\r,r)= r\int_0^{\infty} \; e^{-s} s^{2\nu -\frac32} 
\left(\int_0^\infty e^{i \l(\r- r)} \tilde\Phi(\l)\Big\{\frac{2i \r}{(s-2i\l \rho)^{\frac12} +s^\frac12} \Big\}
d\l \right) ds. 
\end{multline*}
\begin{lemma} We have estimate 
\begin{align}
& |T_1(\r,r)|\leq C 
\left(|r|+ \frac{|r|\r}{\la r-\r\ra}+ \frac{|r| \la \r \ra}{\la r-\r \ra^2}+
\frac{|r|^2 \la \r \ra^\frac12}{\la r-\r \ra^2}\right), \lbeq(012) \\ 
& |T_2(\r,r)|\leq C \frac{|r| \la \r \ra}{\la \r-r \ra}.   \lbeq(012-a)
\end{align}
\end{lemma} 
\begin{proof} The estimates are trivial for $|r-\r|\leq 1$ and we suppose $|\r-r|>1$.  
We first prove \refeq(012-a). Integrating by parts, we estimate the inner intergral 
by the boundary contribution $\r|\r-r|^{-1}s^{-\frac12}$ plus  
\begin{multline*}
\lf|\frac{1}{\r-r}\Big(\int^\infty_0 
\Big(\frac{2i\r e^{i\l(r-\r)} \tilde\Phi'(\l)}{(s-2i\l \rho)^{\frac12}+s^\frac12} 
+ \frac{2i\r  e^{i\l(r-\r)} \tilde\Phi(\l) (i\r)}
{((s-2i\l \rho)^{\frac12}+s^{\frac12})^2 (s-2i\l \rho)^{\frac12}}\Big)d\l \Big)\ri| \\ 
\leq \frac{C\r}{\la r-\r\ra}\left(\frac1{\sqrt{s}} + 
\int^\infty_0 \frac{\r|\tilde\Phi(\l)|}{(|s|+|\l\r|)^{\frac32}}d\l \ri). 
\end{multline*} 
The desired estimate follows since 
\begin{equation}\lbeq(aux)
\int_0^\infty \left( \int_0^{\infty} \; e^{-s} s^{2\nu -\frac32}\frac{|\tilde\Phi(\l)|ds}{(|s|+|\l\r|)^{\frac32}} \ri)d\l 
\leq \int_0^\infty \frac{Cd\l}{\la \l\r\ra^{\frac32}} \leq \frac{C}{\r}. 
\end{equation}
For proving \refeq(012) for $T_1(\r,r)$ we apply integration by parts twice to the inner 
integral. The result is 
\begin{multline}
\frac{2r}{\r-r} \sqrt{\frac{s}{2t}} 
- \frac{r}{(\r-r)^2}\left(\frac{\r}{\sqrt{ts}}+ \frac{\sqrt{s}r}{t^\frac23}\ri) 
\\
-\frac{ir}{(\r-r)^2} \int_0^\infty \, e^{i\l (\r-r)} 
\Big (\frac{\pa}{\pa \l}\Big)^2
\Big( \tilde\Phi(\l)\frac{ (s-2i\l \r)^{\frac12}}{(t+2i\l r )^\frac12+t^\frac12} \Big)d\l. \lbeq(0t2) 
\end{multline}
We estimate the second derivative by a constant times 
\begin{multline*}
\frac{|\tilde\Phi''(\l)|(s+|\l \r|)^\frac12}{\sqrt{t}} +
\frac{\r|\tilde\Phi'(\l)|}{\sqrt{st}} + 
\frac{|\tilde\Phi'(\l)|r(s+|\l \r|)^\frac12}{(t+|\l r|)^\frac32} \\
 + \frac{|\tilde\Phi(\l)|\r^2}{\sqrt{t}(s+|\l \r|)^\frac32}
+ \frac{|\tilde\Phi(\l)|r\r}{\sqrt{s}(t+|\l r|)^\frac32}
+ \frac{|\tilde\Phi(\l)|r^2(s+|\l \r|)^\frac12}{(t+|\l r|)^\frac52} 
\end{multline*}
Desired estimate follows after integration via estimates similar to \refeq(aux).  
\end{proof} 

\begin{lemma} For $\frac{m}{m-2}<p<\frac{m}2$, $W_{1}$ and $W_2$ are bounded in 
$L^p(\R^m)$ 
\end{lemma} 
\begin{proof} $T_1(\r,r)$ is bounded by the right of \refeq(t2est) 
with $j=k=1$; $T_2(\r,r)$ by the right of \refeq(aux-1). 
The lemma follows from \reflm(wjkp) and \reflm(1adm). 
\end{proof}

Finally we deal with $W_3$. Recall that $K_3u(\r)$ is defined by \refeq(kjkop) 
with $(st)^\frac12$ replacing $(s-2i\l \rho)^{\frac12}(t+2i\l r )^{\frac12}$.  
Then, the inner most integral becomes $t$ independent and  we may integrate 
out the $(t,s)$ integral. Result is   
\begin{equation}
K_3u(\rho) = C\int_0^{\infty} \tilde\Phi(\l) e^{-i\l\r} \l^{-1} 
\lf(\int_{\R}\, e^{i\l r} rM(r) dr \ri)d\l
\end{equation}
with a suitable constant $C$. Since $M(r)$ is even, we may write 
\[
\frac1{\l} \int_{\R}\, e^{i\l r} rM(r) dr  = 
 \int_{\R}\, \frac{(e^{i\l r}-1)}{\l} rM(r) dr  = 
i \int_{\R}\, rM(r)\left( \int_0^r e^{i\l v}dv\right) dr.
\]
Thus, if we define $F(v)$ by 
\[
F(v) = \pm v \int_v^{\pm \infty } rM(r) dr, \quad  \mbox{for} \ \pm v >0
\]
and change the order of integration, we have 
\begin{equation}
K_3u(\rho) = C \int_0^{\infty}\tilde\Phi(\l) e^{-i\l \r}\lf( \int_{\R} e^{i\l v } F(v)dv \ri)d\l 
= C \lf[ {\mathcal H} (\check{\tilde\Phi}\ast F) \ri](\rho). 
\lbeq(k003)
\end{equation}
We estimate $|W_3u(x)|\leq I_1(x)+ I_2(x)$ as in \refeq(i1i2). Recall 
that for $\frac{m}{m-2}<p<\frac{m}2$ and $\frac{m}{p}-1 < \th <\frac{m}p$ we have  
$m-2-\th>0$. Let $p=\frac{m}{2+\ep}$ and $\th=2$ first for an arbitrarily small $\ep>0$. Then, 
applying \reflm(weightlm) twice, once for $\Hg$ and once for $\Mg$, we obtain  
\begin{align}
& \|I_2\|^p_p \leq 
C \|f\|_1^p \int_1^{\infty} \r^{m-1} \left(\frac{|K_3(\r)|}{\r^{m-2}}\right)^p d\r \notag \\
& \leq C  \int_0^{\infty} \r^{m-1-p\th }|{\Hg} (\check{\Phi}\ast F)(\r)|^p  d\r 
\leq C  \int_0^{\infty} \r^{m-1-p\th }|(\check{\tilde\Phi}\ast F)(\r)|^p  d\r \notag \\
& \leq C  \int_0^{\infty} \r^{m-1-p\th }|\Mg(F)(\r)|^p  d\r  
\leq C  \int_0^{\infty} \r^{m-1-2p }|F(\r)|^p  d\r . \lbeq(p-e1) 
\end{align}
We then apply Hardy's then H\"older's inequalities and estimate the right by 
\begin{equation} \lbeq(p-e2)
C \int^{\infty}_0  r^{m-1}|M(r)|^{p} dr  \leq  \int_{\R^m} | (\gb \ast \check{u})(x)|^{p} 
\leq C  \|u\|_{p}^p. 
\end{equation}
Let $q=\frac{m}{m-2-\ep}$ be the dual exponent of $p=\frac{m}{2+\ep}$. 
By H\"older's inequality  
\[
|I_1(x)|\leq 
C\lf( \int_{|y|<1 } \lf| \frac{K_3u( |y|) }{|y|^2} \ri|^pdy \ri)^{\frac1{p}}
\lf( \int_{|x-y|<1 } \lf| \frac{ |f(y)|}{|x-y|^{m-4} } \ri|^q dy \ri)^{\frac1{q} }. 
\]
The second factor on the right is an $L^p$ function of $x\in\R^m$ 
since $|f(x)|\leq C \ax^{-m-\ep}$. Then estimates 
\refeq(p-e1) and \refeq(p-e2) implies 
\[
\|I_1\|_p^p 
\leq C \int_{|y|<1 } \lf| \frac{K_3u(|y|) }{|y|^2} \ri|^pdy 
\leq C \int_0^1 \r^{m-1-2p} |{\Hg} (\check{\tilde\Phi}\ast F)(\r)|^p d\r \leq C\|u\|_p^p. 
\]
 
Let $p=\frac{m}{m-2-\ep}$ and $\th=m-3$ next. Then,  using  \reflm(weightlm) twice as 
in \refeq(p-e1), we obtain 
\begin{align} 
\|I_2\|_p & \leq 
C \|f\|_1 \lf( \int_1^{\infty} \r^{m-1} \left(\frac{|K_3u(\r)|}{\r^{m-2}}\right)^p d\r \ri)^{\frac1{p}} 
\notag \\
& \leq C \lf( \int_0^{\infty} \r^{m-1-p\th }|F(\r)|^p  d\r \ri)^{\frac1{p}}. \lbeq(p-e3)
\end{align}
Then, Hardy's inequality implies that the right side is bounded by 
\begin{equation}\lbeq(p-e4)
C \lf( \int_0^{\infty} \r^{m-1-p(\th-2)}|M(\r)|^p  d\r \ri)^{\frac1{p}}
\leq C \lf( \int_{\R^m} \frac{|(\gb \ast u)(x)|^p}{|x|^{p(\th-2)}}  dx \ri)^{\frac1{p}}. 
\end{equation}
Since $p(\th-2)<m$, the right side is bounded by $C\|u\|_p$. 
For $I_1(x)$ we proceed as previously. By H\"older's inequality  
\[
|I_1(x)|\leq 
\lf( \int_{|y|<1 } \lf| \frac{K_3 u( |y|) }{|y|^{m-3}} \ri|^pdy \ri)^{\frac1{p}}
\lf( \int_{|x-y|<1 } \lf| \frac{f(y)}{|x-y|} \ri|^q dy \ri)^{\frac1{q} }
\]
The second factor on the right is an $L^p$ function of $x\in\R^m$ as previously. 
Then \refeq(p-e3) and \refeq(p-e4) imply that the right hand side of 
\[
\|I_1\|_p 
\leq C \lf(\int_{|y|<1 } \lf| \frac{K_3(|y|) }{|y|^{m-3}} \ri|^pdy \ri)^{\frac1{p}}
\leq C \lf(\int_0^1 \r^{m-1-p\th} |{\Hg} (\check{\tilde\Phi}\ast F)(\r)|^p d\r \ri)^{\frac1{p}}
\]
is bounded by $C\|u\|_p$. This proves $W_3$ is bounded in $L^p$ for $\frac{m}{m-2}<p<\frac{m}2$ 
and completes the proof of \refprp(sing).

\subsection{Estimate for $W_{s,ab}$}
In this subsection, we indicate how the discussion in the previous subsection 
may be modified for proving the following proposition. 
\begin{proposition}\lbprp(sing-2)
Let $V$ satisfies \refeq(assm) and let $W_{ab}$ be defined by \refeq(logs).  
Then, for any $\frac{m}{m-2} < p < \frac{m}{2} $, 
\begin{equation}\lbeq(sing-52)
\|\Phi(H) W_{s,ab} \Phi(H_0) u \|_p \leq C_p \| u \|_p, \quad 
u \in C^{\infty}_0 (\R^m ).
\end{equation}
\end{proposition}
\begin{proof} By virtue of \refprp(sing-exp), $VD_{ab}$ are finite rank operators 
from $\Hg_{-(\d-3)_-}$ to $\Hg_{(\d-3)_-}$. Hence, they are finite linear combinations 
of rank one operators $f \otimes g$ with $f, g \in \Hg_{m+\ep}$ for some $\ep>0$, and  
it suffices to prove \refeq(sing-52) for $\Phi(H)\tilde Z \Phi(H_0)$, where $\tilde Z$ 
is the operator defined by 
\begin{equation}\lbeq(zj-ab)
\tilde Z = \int_0^{\infty} G_0(\l) (f \otimes g)  
\lf( G_0(\l) - G_0(-\l) \ri) \tilde\Phi (\l) \l^{a+1} \log^b \l  d\l 
\end{equation}
with such $f, g$, which is the same as \refeq(zj) if $\l^{-1}$ replaces $ \l^{a+1} \log^b \l$. 
Notice that 
$\tilde Z \Phi(H_0)u $ is given by the same formula \refeq(zj-ab) with $\Phi(H_0)g$ 
in place of $g$ and $|\Phi(H_0) g(x)|\leq C \ax^{-m-\ep}$. Thus, we may 
(and do) assume that $g$ satisfies the condition \refeq(fg), $|g(x)|\leq C \ax^{-m-\ep}$, 
and ignore $\Phi(H_0)$. After this, we proceed as in the previous section:  
Define $K_{jk}^{ab} u(\r)$ by the right side of \refeq(kjkop) 
with $\l^{j+k+1+a}(\log\l)^b $ in place of $\l^{j+k-1}$ and 
\begin{equation} 
W_{jk}^{ab}u(x)= \int_{\R^m} \frac{f(y) K_{jk}^{ab} u(|x-y|)}{|x-y|^{m-2}}dy  \lbeq(wab)
\end{equation}
so that $\tilde Z$ is a linear combination of $K_{jk}^{ab}$:   
\begin{equation}
\tilde Z u(x) = \sum_{j,k=0}^{\frac{m-4}{2}} C_{jk}W_{jk}^{ab} u(x),  \lbeq(zab)
\end{equation}
see \reflm(zlm). 
We then follow the argument in the proof of \refprp(wjk) for the case $j,k \geq 1$. 
The function $\l^{j+k+1+a}(\log\l)^b $ is certainly less singular than $\l^{j+k-1}$ at $\l=0$ 
and the proof of \reflm(t2est) implies, as previously, 
\[
 K_{jk}^{ab} u(\r) = \int_{\R} M(\gb \ast u, r) T_{jk}^{ab}(r) dr 
\]
with $ T_{jk}^{ab}(r) $ which satisfies estimate \refeq(t2est):
\[
\lf| T_{jk}^{ab}(\rho,r) \ri| \leq 
C \lf| \frac{\la \rho \ra^{j+1/2} r^{k+1} \langle r \rangle^{1/2}  }{ \langle r - \rho \rangle^{j+k}} \ri|. 
\]
We then want to apply the argument in the proof of \reflm(wjkp). 
Here it is important to observe that we may pretend 
that $f$ satisfies \refeq(fg) as well: $|f(x)|\leq C \ax^{-m-\ep}$. 
Indeed, we have 
\begin{align*}
|\Phi(H)W_{jk}^{ab}u(x)| & = 
\left|\int_{\R^m} \left(\int_{\R^m} \Phi(x,z)f(z-y)dz \right) \frac{K_{jk}^{ab} u(|y|)}{|y|^{m-2}}dy\right| \\  
& \leq C \int_{\R^m} \frac{\ay^{-(m+\ep)} |K_{jk}^{ab} u(|x-y|)|}{|x-y|^{m-2}}dy 
\end{align*}
since $|\Phi(x,z)|\leq C_N \la x-z \ra^{-N}$. Then, the proof of \reflm(wjkp) applies without any 
change and we obtain $\|\Phi(H)W_{jk}^{ab}u\|_p \leq C \|u\|_p$ for any $\frac{m}{m-2}<p<\frac{m}2$. 
\end{proof}

\section{High energy estimate}
In this section we prove the following proposition. 
Recall that $m_\ast=\frac{m-1}{m-2}$. 
\begin{proposition}\lbprp(1) Let $V$ satisfy \refeq(cond-1) and, in addition, 
$|V(x)|\leq C \ax^{-\d}$ for some $\d>m+2$. 
Let $\Psi(\l) \in C^\infty(\R)$ be such that $\Psi (\l)=0$ for $|\l|<\l_0$ for some $\l_0$. 
Then $W_>$ is bounded in $L^p(\R^m)$ for all $1\leq p \leq \infty$ 
\end{proposition}

Since the proof is entirely similar to the corresponding one in [I], we shall only sketch it very briefly 
pointing out what modifications are necessary for even dimensions.   
Iterating the resolvent equation,  we have 
$G(\l)V= \sum_1^{2n} (-1)^{j-1}(G_0(\l)V)^{j} + G_0(\l)N_n(\l)$,  
where 
\[
N_n(\l)= (VG_0(\l))^{n-1}V G(\l)V (G_0(\l)V)^{n}.  
\]
If we substitute this for $G(\l)V$ in the right of \refeq(high), we have  
\begin{align}
W_>= \Psi(H_0)^2 + \sum_{j=1}^{2n} (-1)^j \W_j \Psi(H_0)^2 -\tilde \W_{2n+1}, \\
\tilde \W_{2n+1}=\frac{1}{i\pi} \int_0^\infty G_0(\l)N_n (G_0(\l)-G_0(-\l)) \tilde \Psi(\l) d\l, 
\end{align} 
where $\tilde \Phi(\l)=\l \Psi(\l^2)^2$.  
The operators $\Psi(H_0)$ and $\W_1, \ldots, \W_{2n}$ are bounded in $L^p$ for any $1\leq p \leq \infty$ 
by virtue of \reflm(wr1). 
We show that, if $n$ is large enough, the integral kernel 
\[ 
\tilde \W_{2n+1}(x,y)= \int_0^\infty \la N_n(\l)(G_0(\l)-G_0(-\l))\d_y, G_0(-\l)\d_x \ra \l \Psi^2(\l^2)d\l, 
\]
of $\tilde\W_{2n+1}$ is admissible. We define $\tilde G_{0}(\l,z,x) = e^{-i\l|x|} G_{0}(\l,x-z)$ 
and $\p(z,x)=|x-z|-|x|$ as previously.  
\begin{lemma} Let $j=0,1,2, \ldots$. We have for $|\l| \geq 1$ that 
\begin{equation}\lbeq(bG) 
\left|\left(\frac{\pa}{\pa \l}\right)^j \tilde G_0(\l,z,x)\right| 
\leq C_j\left(\frac{\az^j}{|x-z|^{m-2}}+ \frac{ \l^{\frac{m-3}2}\az^j}{|x-z|^{\frac{m-1}2}}\right) .
\end{equation}
\end{lemma}
\begin{proof} Differentiate $\tilde G_0^{(j)}(\l,z,x)$ by using Leibniz's formula. The result is 
a linear combination over $(\a,\b)$ such that $\a+\b=j$ of 
\[ 
\frac{e^{i\l\p(z,x)}\p(z,x)^\a }{|x-z|^{m-2-\b}}
\int^\infty_0 e^{-t}t^{\n-\frac12}\left(\frac{t}2-i\l|x-z|\right)^{\n-\frac12-\b}dt.
\]
Since $|\p(z,x)|^\a \leq \az^j$ for $0\leq\a\leq j$ and 
$|z-x|\leq |\frac{t}2-i\l|z-x|| \leq (t +\l|z-x|)$ when $|\l|\geq 1$, \refeq(bG) follows. 
\end{proof}
Define $T_{\pm}(\l, x,y)= \la  N_n (\l) \tilde G_{0}(\pm \l, \cdot,y), \tilde G_{0}(- \l, \cdot,x) \ra$ 
so that
\[
\tilde \W_{2n+1}(x,y)= \frac1{\pi i}\int^\infty_0 \left(e^{i\l(|x|+|y|)} T_{+}(\l, x,y)
-e^{i\l(|x|-|y|)} T_{-}(\l, x,y) \right) \tilde \Psi(\l)  d\l. \lbeq(ow)
\]
The following lemma may be 
proved by repeating line by line the proof of Lemma 3.14 of [I] by using \refeq(bG) and \reflm(Ag).
\begin{lemma} Let $0\leq s \leq \frac{m+2}{2}$. For sufficiently large $n$, we have 
\begin{equation}\lbeq(tdiff)
\left|\left(\frac{\pa}{\pa \l}\right)^s T_{\pm}(\l, x,y)\right|\leq C_{ns} \l^{-3} \ax^{-\frac{m-1}2} 
\ay^{-\frac{m-1}2}
\end{equation}
\end{lemma} 
We then integrate by parts $0\leq s \leq (m+2)/2$ times to obtain  
\[\br{l}
\ds \int^\infty_0 e^{i\l(|x|\pm |y|)}T_{\pm}(\l, x,y) \tilde \Psi(\l) d\l \\
\ds =\frac1{(|x|\pm |y|)^s}\int^\infty_0 e^{i\l(|x|\pm |y|)} 
\ds \left(\frac{\pa}{\pa \l}\right)^s\left(T_{\pm}(\l, x,y) \tilde \Psi(\l) \right) d\l
\er
\]
and estimate the right hand side by using \refeq(tdiff). We obtain  
\[
|\tilde \W_{n+1}(x,y)| \leq C \sum_{\pm} \la |x| \pm |y|\ra^{-\frac{m+2}{2}}\ax^{-\frac{m-1}{2}}\ay^{-\frac{m-1}{2}}
\]
and $\tilde \W_{n+1}(x,y)$ is admissible. \refprp(1) follows.  

\section{Completion of proof of Theorem}

To complete the proof of \refth(even) we have only to prove the continuity of $W$ in Sobolev spaces. 
We prove this for the case $1<p<\infty$ only. For the cases $p=1$ and $p=\infty$, we may apply 
without any change the proof presented in Section 4 of \cite{Y-d3} for odd dimensional cases 
where we estimated the multiple commutators $[p_{i_1}, [p_{i_2}, [ \cdots, [p_{i_\ell}, W_\pm] \cdots] ] ]$. 
We use the following two lemmas.  

\begin{lemma}\lblm(cut) Let $1<p<\infty$ and $|V(x)|\leq C <\infty$. Then for large negative $\l$,  
$R(\l)\in \Bb(L^p(\R^m), W^{2,p}(\R^m))$ and $R(\l)^\frac12\in \Bb(L^p(\R^m), W^{1,p}(\R^m))$.  
\end{lemma}
\begin{proof} We first remark that $H$ is bounded from below and $R(\l)^\frac12$ is well defined 
bounded operator in $\Hg$ for large negative $\l$ and 
$R(\l)^\frac12\in \Bb(L^p, W^{1,p})$ means that $R(\l)^\frac12$ defined on $L^2 \cap L^p$ 
can be extended to such an operator. For $\l<0$, we have $\|R_0(\l)\|_{p,p} \leq C |\l|^{-1}$ and 
$\|\nabla R_0(\l)\|_{p,p} \leq C_p |\l|^{-\frac12}$. It follows that, for large negative $\l$, 
$1+ R_0(\l)V$ is an isomorphism of $L^p$, $R(\l)=(1+ R_0(\l)V)^{-1}R_0(\l)$ also in $L^p$ 
and $\|R(\l)\|_{p,p} \leq C |\l|^{-1}$. Hence, the resolvent equation is also valid in $L^p$, 
\begin{equation}\lbeq(7-1)
R(\l)=R_0(\l) - R_0 (\l) V R(\l),  
\end{equation}
and this implies $R(\l) \in \Bb(L^p, W^{2,p})$. It also follows that the integral in  
\[
\nabla R(\l)^{\frac12} =\nabla R_0(\l)^{\frac12}- C 
\int^\infty_0 \m^{-\frac12}\nabla R_0(\l-\m)^{-1}V R(\l-\m) d\m 
\]
converges in the norm of $\Bb(L^p)$ and $\nabla R(\l)^{\frac12}$ is bounded in $L^p(\R^m)$. 
\end{proof}

\begin{lemma}\lblm(resolve)  Let $1<p<\infty$ and $n=1,2, \ldots$. Then, for 
large negative $\l$ the following statements are satisfied: 

\noindent {\rm (1)} 
Let $|\pa^\a V(x)|\leq C_\a$ for $|\a|\leq 2(n-1)$. Then, $R(\l)^n \in \Bb(L^p, W^{2n,p})$. 

\noindent 
{\rm (2)} Let $|\pa^\a V(x)|\leq C_\a$ for $|\a|\leq 2n-1$. Then, $R(\l)^n \in \Bb(W^{1,p}, W^{2n+1,p})$.
\end{lemma}
\begin{proof} We first prove (1) by induction on $n$. If $n=1$, (1) is contained in \reflm(cut). 
Let $n \geq 2$ and suppose that (1) is already proved for 
smaller values of $n$. By virtue of \refeq(7-1), 
\begin{equation}\lbeq(7-2)
R(\l)^{n}=R_0(\l) R(\l)^{n-1}- R_0(\l)VR(\l)^{n-1} R(\l) .
\end{equation}
By the assumption on $V$ and the induction hypothesis 
$R(\l)^{n-1}, VR(\l)^{n-1}  \in \Bb(L^p, W^{2n,p})$ and (1) follows since $R_0(\l)$ maps 
$W^{2n,p}$ to $W^{2n+2,p}$ boundedly.  

\noindent
We next prove (2). Let $n=1$ first. Then, in \refeq(7-1), $R_0(\l) \in \Bb(W^{1,p}, W^{3,p})$ 
and $ V R(\l) \in \Bb(W^{1,p})$ by (1) for $n=1$ and the assumption on $V$. Hence (1) holds for 
$n=1$. Let $n \geq 2$ and suppose that (2) is already proved for smaller values of $n$. 
Then in \refeq(7-2), $ R(\l)^{n-1}\in \Bb(W^{1,p}, W^{2n-1,p})$ by the induction hypothesis, 
and $VR(\l)^{n-1} R(\l) \in \Bb(W^{1,p}, W^{2n-1,p})$ also by the assumption on $V$ and \reflm(cut). 
Since $R_0(\l) \in \Bb(W^{2n-1,p}, W^{2n+1,p})$, (2) follows. 
\end{proof} 

By intertwing property we have for sufficient laerge negative $\l$
\[
R(\l)^n W_\pm = W_\pm R_0(\l)^n, \quad R(\l)^{n+\frac12} W_\pm = W_\pm R_0(\l)^{n+\frac12}
\]
>From the first equation  and \reflm(resolve) (1) we see that, 
if $|\pa^\a V(x)|\leq C_\a$ for $|\a|\leq 2(n-1)$, $W_\pm \in \Bb(W^{2n,p}, W^{2n,p})$. 
Likewise from \reflm(cut) and \reflm(resolve) (2), we have 
$W_\pm \in \Bb(W^{2n+1,p}, W^{2n+1,p})$ if $|\pa^\a V(x)|\leq C_\a$ for $|\a|\leq 2n-1$. 
This completes the proof of \refth(even).


\end{document}